\documentclass[%
 aip,
 amsmath,amssymb,
 reprint,%
]{revtex4-2}

\usepackage{graphicx}
\usepackage{dcolumn}
\usepackage{bm}

\usepackage[utf8]{inputenc}
\usepackage[T1]{fontenc}
\usepackage{mathptmx}

\usepackage{braket}
\usepackage{amsmath,amssymb}
\usepackage{physics}
\usepackage{amsmath}
\usepackage{bm}
\usepackage{bbold}
\usepackage{braket}
\usepackage{xcolor}
\usepackage{placeins}
\usepackage{ulem}
\usepackage{natbib}

\begin{document}


\title[]{Semiclassical Dynamics in Wigner Phase Space I : Adiabatic Hybrid Wigner Dynamics}

\author{Shreyas Malpathak}
\altaffiliation[Current affiliation: ]{Department of Physical and Environmental Sciences, University of Toronto Scarborough, Toronto, Ontario M1C 1A4, Canada}
\author{Nandini Ananth}%
 \email{ananth@cornell.edu}
\affiliation{Department of Chemistry and Chemical Biology, Baker Laboratory, Cornell University,
Ithaca, 14853 NY, USA}%


\date{\today}

\begin{abstract}
The Wigner phase space formulation of quantum mechanics is a complete framework 
for quantum dynamic calculations that elegantly highlights connections with 
classical dynamics. In this series of two articles, building upon previous 
efforts, we derive the full hierarchy of approximate semiclassical (SC) 
dynamic methods for adiabatic and non-adiabatic problems in Wigner phase space. 
In paper I, focusing on adiabatic single surface processes, we derive the well-known 
Double Herman-Kluk (DHK) approximation for real-time correlation functions 
in Wigner phase space, and connect it to the Linearized SC (LSC) 
approximation through a stationary phase approximation. 
We exploit this relationship to introduce a new hybrid SC method, termed 
Adiabatic Hybrid Wigner Dynamics (AHWD) that allows for a few 
important `system' degrees of freedom (dofs) to be treated at the DHK level, 
while treating the rest of the dofs (the `bath') at the LSC level. 
AHWD is shown to accurately capture quantum interference effects in models 
of coupled oscillators, and the decoherence of vibrational probability 
density of a model I$_2$ Morse oscillator coupled to an Ohmic thermal bath. 
We show that AHWD significantly mitigates the sign-problem and employs 
a reduced dimensional prefactor bringing calculations of complex 
system-bath problems within the reach of SC methods. 
Paper II focuses on extending this hybrid SC dynamics to 
nonadiabatic processes.
\end{abstract}

\maketitle

\section{Introduction}

The importance of quantum dynamical effects in processes of chemical interest 
has received considerable attention in recent years. 
Their role has been investigated in a wide array of systems including in
chemical reaction rates,~\cite{Habershon2013,Suleimanov2016a,Lawrence2020} 
spectra of protonated water clusters~\cite{Vendrell2007a,Vendrell2007b,
Vendrell2007c,Yu2017,Schroder2022} and 
biomolecules~\cite{Conte2020c,Gabas2020,Botti2022,Botti2023,Moscato2023,Moscato2024} 
in the gas phase, as well as, water and ice in the 
condensed phase,~\cite{Benson2020,Althorpe2021a}  
chemistry at surfaces and 
interfaces,~\cite{Suleimanov2012,Liu2019,Jiang2021,Zhou2022,Shi2023} 
condensed phase proton transfer,~\cite{Zhang2020,Liu2021,Slocombe2022} 
vibrational decoherence,~\cite{Buchholz2012,Joutsuka2016,Qiang2024} 
and cavity-modified chemistry.~\cite{Triana2020,Fischer2021,Yang2021,
Li2022b,Mondal2022,Lindoy2023,Lieberherr2023} 
Despite tremendous progress in numerically exact quantum dynamic simulation 
methods like the Quasi-Adiabatic Path Integral (QUAPI),\cite{Makri1992} 
Hierarchy Equations of Motion (HEOM),\cite{Tanimura1989} and 
Multi-Configuration Time Dependent Hartree (MCTDH),\cite{Meyer1990,Manthe1992,Beck2000} 
their expensive scaling with respect to the number of degrees of freedom (dofs) 
limits their use for complex chemical problems.
To develop rigorous approximations to quantum dynamics that have favorable 
scaling with system size, it is important to employ approaches that exploit 
underlying connections with classical dynamics. 
Approximate quantum dynamic methods based on imaginary time path integrals, 
like Centroid Molecular Dynamics (CMD),
\cite{Cao1994a,Cao1994b,Cao1994c,Cao1994d,Cao1994e} 
Ring Polymer Molecular Dynamics,\cite{Craig2004,Craig2005} and 
Matsubara dynamics,\cite{Hele2015b,Hele2015a} along with their more 
recent extensions,\cite{Willatt2018,Trenins2018,Trenins2019,Benson2020,
Benson2021,Haggard2021,Fletcher2021,Althorpe2021a,Musil2022,Lawrence2023,Prada2023}
employ classical trajectories and have been successfully used to simulate
condensed phase chemistry.
However, these methods are most accurate for thermal processes and are not
suitable for simulations of systems where quantum interference effects play
a significant role.
With the aim of developing rigorous approximations to quantum dynamics while 
exploiting connections with underlying classical dynamics, in this sequence of 
articles we focus our attention on two distinct approaches \textemdash~quantum 
dynamics in Wigner phase space,\cite{Wigner1932,Hillery1984} 
and Semiclassical Initial Value Representation (SC-IVR)
methods.\cite{Miller2001a,Malpathak2022}

The phase space formulation of quantum mechanics offer an elegant framework that highlights quantum-classical correspondence.\cite{Berry1997,Wilkie1997a,Wilkie1997b} 
Although several different 
formulations of quantum mechanics in phase space exist,\cite{Cohen1966,Lee1995} 
we focus our attention on Wigner phase space.\cite{Wigner1932,Hillery1984} 
Here, quantum mechanical operators are transformed into functions of phase 
space variables $\bm{z}\equiv(\bm{q},\bm{p})$ using the Wigner 
transform,\cite{Wigner1932,Hillery1984} where $\bm{q}$ and $\bm{p}$ are the positions and momenta of the particles respectively. 
Quantum mechanical propagation involves propagating the Wigner transformed density operator, also known as the Wigner function. In stark contrast to classical dynamics in phase space, the propagation of the Wigner function cannot be performed using independent trajectories for general anharmonic potentials.\cite{Heller1976,Oliva2018} 
This is a hallmark of the non-locality of quantum mechanics, and makes exact propagation in Wigner phase space as expensive as other numerically exact approaches.
Attempts have been made to propagate the Wigner distribution using an ensemble of classical trajectories that interact with each other. 
For instance, the Entangled Trajectory Molecular Dynamics method 
makes use of a Gaussian ansatz for the density in phase space, 
and has been shown to capture tunneling rates accurately in model systems, but is prohibitively expensive for large multi-dimensional 
problems.\cite{Donoso2001,Donoso2003,Lopez2006} 

At high temperatures or for strong coupling to a dissipative environment, quantum coherence effects typically play a smaller 
role in the overall dynamics.
In these limits, approximating quantum dynamics with its classical 
counterpart can be reasonable, and is achieved 
by taking $\hbar \to 0$ as the classical limit of quantum propagation.
The classical limit of quantum dynamics in Wigner phase space 
is referred to as the Truncated Wigner approximation,\cite{Polkovnikov2010} 
and its counterpart in the SC-IVR framework for 
the calculation of correlation functions is the 
Linearized SC-Initial Value Representation 
(LSC-IVR)~\cite{Sun1998,Wang1998} 
or Linearized Path Integral approach.~\cite{Shi2003,Poulsen2003}
Efforts to go beyond the classical limit in Wigner phase space 
have explored adding `quantum corrections' using higher order terms in the Wigner Liouvillian,\cite{Liu2007,Liu2011a,Liu2011b,Liu2011c} however, approximations that employ independent trajectories are not 
based on a firm theoretical footing.

In parallel to the development of approximate methods in Wigner
phase space, a heirarchy of SC-IVR methods have been developed
that are derived from the quantum mechanical propagator using a stationary phase approximation.~\cite{Miller2001a} 
SC-IVR methods employ an ensemble of independent classical 
trajectories that can interfere with each other through a phase
corresponding to their classical action.~\cite{Miller2001a}
These methods have been shown to accurately capture a 
wide variety of quantum effects including zero-point energy, shallow 
tunneling, and interference and non-adiabatic effects in model 
systems.\cite{Miller2001a,Miller2009,Malpathak2022} 
A robust hierarchy of SC-IVR methods have been established, varying 
in accuracy and numerical cost with the double 
Herman-Kluk (DHK)-IVR method for approximating two-time correlation 
functions~\cite{Herman1984, Thoss2004} 
being amongst the more accurate methods in the hierarchy.
Unfortunately, DHK-IVR has found limited practical use since 
its application to large systems is limited by the cost of computing the SC prefactor and the dynamical sign problem.\cite{Miller2001a} 
The lowest accuracy approximations are classical-limit methods
like LSC-IVR and its variants, although even 
these methods have proven to be tremendously successful for adiabatic,~\cite{Shi2003c,BeingJ.Ka2005,Navrotskaya2006,Being2006,Being2006a,Vazquez2010,Liu2015} and nonadiabatic dynamic
simulations of population transfer in model systems,~\cite{Liu2020,Gao2020,Hu2022,Malpathak2024a} 
and on-the-fly dynamic simulations.\cite{Miyazaki2023} 
Other notable classical-limit variants include 
the Husimi-IVR,\cite{Antipov2015,Church2017,Malpathak2023} 
that propagates a Husimi distribution~\cite{Husimi1940} with classical trajectories and the Forward-Backward 
Semiclassical Dynamics that uses a modified Husimi distribution~\cite{Shao1999a,Shao1999b} 
However, these classical-limit SC methods cannot capture quantum interference effects and are typically accurate only for simulations of systems at high temperatures or in the condensed phase.

It is clear that moving beyond classical-limit SC without making the calculation prohibitively expensive for large systems requires
the development of a mixed quantum-classical SC framework where
a portion of the system can be described using a high-level SC theory like 
DHK-IVR while the rest of the system degrees of freedom are treated 
in the classical limit. 
The recently developed 
Mixed Quantum Classical-Initial Value Representation (MQC-IVR) offers 
such a framework, using a modified Filinov filtration scheme~\cite{Makri1987c,Makri1988b,Spanner2005}
to selectively filter phase contributions due to 
individual degrees of freedom.
\cite{Antipov2015,Church2017,Church2018,Church2019a,Malpathak2022} 
For a variety of applications, MQC-IVR successfully tunes 
between the quantum-limit 
DHK-IVR and classical-limit Husimi-IVR, and has been shown to 
successfully mitigate the sign problem.~\cite{Church2017}
However, recently, we showed that additional corrections are required 
to ensure that classical-limit MQC-IVR correctly reduces 
to Husimi-IVR for non-linear operator correlation functions,~\cite{Malpathak2022}
and perhaps more significantly, that Husimi-IVR is significantly less
accurate than its classical-limit counterpart LSC-IVR.~\cite{Malpathak2022,Lee1995} 

In a series of two papers, we introduce a new mixed quantum-classical 
SC approach, Hybrid Wigner Dynamics, that corresponds to treating the quantized subsystem
at the DHK-IVR level of theory while using 
LSC-IVR to capture the dynamics of the remaining system degrees of freedom. 
This manuscript, hereon referred to as paper I, introduces 
Adiabatic Hybrid Wigner Dynamics (AHWD) for the simulation of 
reactions on a single Born-Oppenheimer surface, 
and paper II~\cite{Malpathak2024c} introduces 
nonadiabatic HWD (NHWD) and demonstrates
different quantization schemes in the context of nonadiabatic dynamics.
Paper I starts with a new derivation of the DHK expression for two-time correlation functions 
in Wigner phase space in Sec.~\ref{sec:theory}, and 
shows that the LSC correlation function
can be derived from the Wigner-DHK framework via a stationary phase approximation. 
We use this result to derive a new hybrid approach, 
the AHWD method, and we describe the prefactor structure, 
trajectory dynamics, and discuss its relation to existing hybrid approaches. Sec.~\ref{sec:simulation} presents model 
systems used to demonstrate the AHWD approach, and provides simulation details. Sec.~\ref{sec:results} presents
the results from AHWD simulations and quantifies
the extent to which AHWD simulations can capture 
interference and decoherence effects. In Sec.~\ref{sec:conclusion}, we discuss some 
questions raised by developing an SC heirarchy 
in Wigner phase space, along with possible avenues 
for future work.

\begin{figure}
    \centering
    \includegraphics[width=0.49\textwidth]{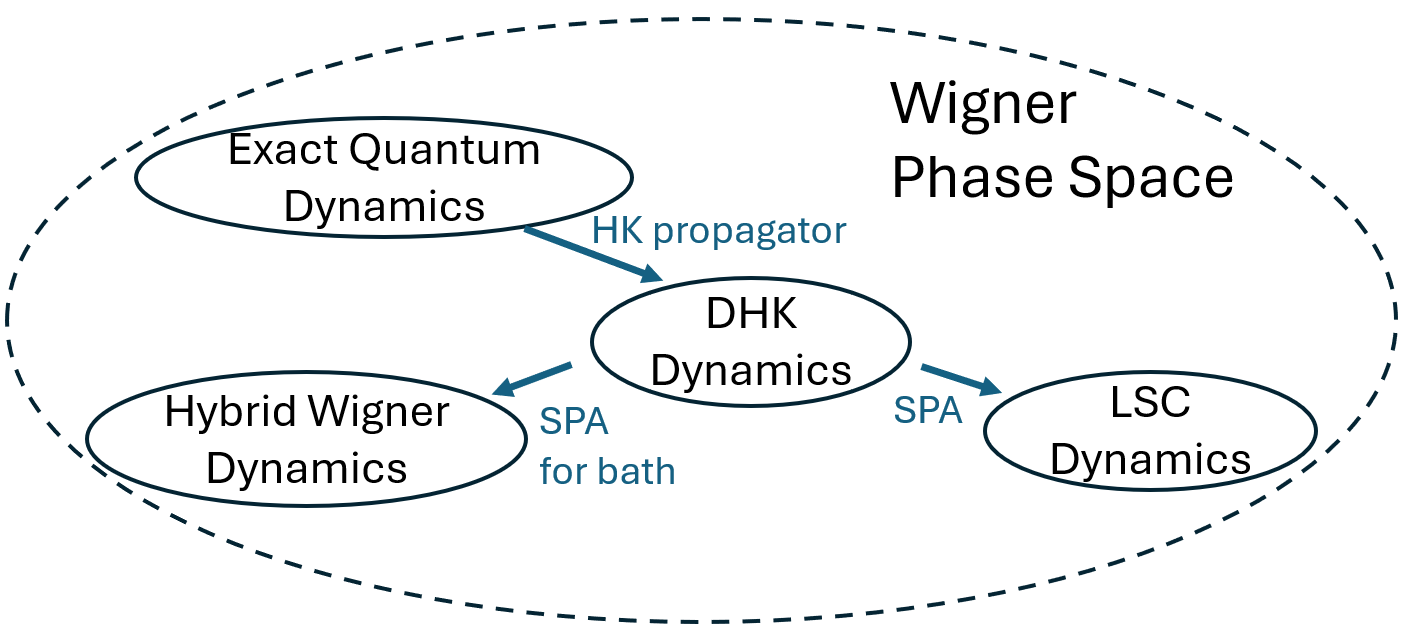}
    \caption{A schematic of the hierarchy of SC methods set in Wigner phase space set up in this manuscript. Refer to the text for details on the abbreviations used in the schematic.}
    \label{fig:wig_phase_space}
\end{figure}



 
\section{Theory} \label{sec:theory}

SC dynamic methods that go beyond the classical-limit do so by working in 
an extended phase space of forward-backward trajectories. Attempts to go beyond
the classical-limit in Wigner phase space include propagating 
Wigner functions using the Wigner transform of the semiclassical Van Vleck (VV) 
propagator,\cite{Rios2002,OzoriodeAlmeida2006a,Dittrich2006,Dittrich2010a,DeAlmeida2013,Lando2019} 
and deriving a propagator for Wigner functions based on the HK
and VV~\cite{Koda2015,Koda2016,Gottwald2018} 
propagators by re-interpreting 
the Wigner-Moyal equation as a Schrodinger equation in an extended phase space.
Here, we introduce an alternate path to the DHK expression for two-time correlation functions in Wigner phase space. The resulting expression has no advantages over the conventional DHK-IVR formulation in terms of accuracy or numerical implementation. 
However, it does offer a  direct connection to the LSC method through a stationary phase approximation. 
Exploiting this connection, we introduce the AHWD approach which partitions the problem into a relatively low-dimensional `system' that is treated at the DHK level of theory, and a high-dimensional `bath' that is treated at the LSC level of theory. Attempts at such a hybrid method have been 
made previously,\cite{Ovchinnikov1996,Sun1997c,Zhang2005,Grossmann2006,Koda2016,Church2019a} 
and we compare AHWD with previous attempts
in the Section~\ref{sec:hwd-a}.

\subsection{Herman Kluk Dynamics in Wigner Phase Space} \label{sec:w-dhk}
We start with the quantum-mechanical expression for a two-time correlation function,
\begin{align}
    C_{AB}(t) &= \frac{1}{\left(2\pi\hbar\right)^N}\int d\bm{z}\,[\hat{\rho}_A]_W(\bm{z})[\hat{B}(t)]_W(\bm{z}) 
    \label{eq:cab_exact}
\end{align}
where $\hat{\rho}_A \equiv \hat{\rho}\hat{A}$, $\hat{\rho}$ is the density operator,
$\hat A$ is evaluated at time zero. 
$\hat B(t)$ corresponds to the time-evolved Heisenberg operator $\hat B$, defined as  $\hat B(t) = K^{+}\hat{B}K^{-}$, where $K^{\pm}\equiv e^{\pm i\hat{H}t/\hbar}$ is the backward/forward time-evolution operator.
Here $N$ is the dimensionality of the system, $\bm{z}= (\bm{q},\bm{p})$ 
is the phase space coordinate, 
where $\bm{p}$ and $\bm{q}$ are the $N$-dimensional 
vectors for the momentum and position respectively, 
and $[.]_W$ denotes the Wigner transform of an operator, defined as,
\begin{align}
    [\hat{B}]_W(\bm{z})  = \int d\Delta \mel{\bm{q}+\frac{\Delta}{2}}{\hat{B}}{\bm{q}-\frac{\Delta}{2}}
    e^{-i\bm{p}\Delta/\hbar}.
    \label{eq:bwig}
\end{align}
Since $\hat B(t)$ is a product of three operators, its Wigner transform can be written using an identity,\cite{Koda2015}
\begin{align}
     [\hat{B}(t)]_W(\bm{z}) = \int d\bm{z}^{\prime}\, B_W(\bm{z}^{\prime}) F_{K^{+}K^{-}}(\bm{z},\bm{z}^{\prime}), \label{bt-wig}
\end{align}
where $F_{K^{+}K^{-}}(\bm{z},\bm{z}^{\prime})$  is defined as, 
\begin{align}
    F_{K^{+}K^{-}}(\bm{z},\bm{z}^{\prime}) &= \frac{1}{\left(2\pi\hbar\right)^N}\int d\bm{\Delta} \int d\bm{\Delta^{\prime}}\, e^{-i\left(\bm{p}^{T}\cdot\bm{\Delta}-{\bm{p}^{\prime}}^{T}\cdot\bm{\Delta^{\prime}}\right)/\hbar} \notag \\ 
    & \times \mel{\bm{q}+\frac{\bm{\Delta}}{2}}{K^{+}}{\bm{q^{\prime}}+\frac{\bm{\Delta^{\prime}}}{2}} \notag \\
    & \times \mel{\bm{q^{\prime}}-\frac{\bm{\Delta^{\prime}}}{2}}{K^{-}}{\bm{q}+\frac{\bm{\Delta}}{2}}. \label{eq:f_ker}
    \end{align}    
So far all manipulations have been exact. We now introduce the approximate HK time-evolution operator,\cite{Herman1984,Kay2005} 
\begin{align}
    e^{-i\hat{H}t/\hbar} & \approx \frac{1}{\left(2\pi\hbar\right)^N} \int \, d\bm{z}_0 \ket{\bm{z}_t} \mathcal{C}_t\left(\bm{z}_0\right) e^{iS_t(\bm{z}_0)/\hbar} \bra{\bm{z}_0}, \label{eq:hk-prop}
\end{align}
where $\ket{\bm{z}_0}$ is the time $t=0$ coherent state 
centered at $\bm{z}_0$ 
with a diagonal width matrix, $\bm{\gamma}$, defined as,
\begin{align}
    \braket{\bm{x}}{\bm{z}} = \text{det}\left(\frac{\bm{\gamma}}{\pi}\right)^{1/4} 
    e^{-\frac{1}{2}\left(\bm{x}-\bm{q}\right)^{T}.\bm{\gamma}.\left(\bm{x}-\bm{q}\right) 
    + i\bm{p}^{T}.\left(\bm{x}-\bm{q}\right)/\hbar}. 
    \label{eq:cs-def}
\end{align}
In Eq.~\eqref{eq:hk-prop}, $\mathcal{C}_t\left(\bm{z}_0\right)$ is the HK prefactor,
\begin{align}
    \mathcal{C}_{t}\left(\bm{z}_{0}\right) & = \text{det}\left|\frac{1}{2}\left[\bm{M}_{qq}  + \bm{M}_{pp}  
    - i\hbar\gamma\bm{M}_{qp}+\frac{i}{\hbar}\gamma^{-{1}}\bm{M}_{pq}\right]\right|^{1/2},
    \label{eq:hk_pre}
\end{align}
where $\bm{M}_{\alpha\beta}=\frac{\partial\alpha_t}{\partial\beta_0}$ are the monodromy matrix elements,
and $\bm{z}_t$, the center of the time-evolved coherent state $\ket{\bm{z}_t}$
obtained by propagating a classical trajectory under the classical Hamiltonian $H(\bm{z})$ 
for time $\textit{t}$ starting at $\bm{z}_0$, and $S_t(\bm{z}_0)$ is its classical action. 

To obtain the DHK approximation to the correlation function in Wigner phase 
space \textemdash~ which we will refer to as Wigner-DHK (WDHK) \textemdash~ 
we use the HK approximation to the propagator, Eq.~\eqref{eq:hk-prop}, for 
the forward propagator, and its complex conjugate for the backward propagator 
in Eq.~\eqref{eq:f_ker}, yielding 
\begin{align}
    C_{AB}^{\text{WDHK}}(t)  & = \frac{1}{\left(2\pi\hbar\right)^{2N}} \int d\bm{{z}}^{+}_0 \int d\bm{ z}_0^{-}\,\tilde{\rho}^{*}_{A}(\bm{{z}}_0^{+},\bm{z}_0^{-})  \notag \\
    & \times \tilde{\mathcal{C}}_t(\bm{{z}}_0^{+},\bm{z}_0^{-}) \tilde{B}(\bm{{z}}_t^{+},\bm{z}_t^{-}) e^{i\tilde{S}_t(\bm{{z}}_0^{+},\bm{z}_0^{-})/\hbar}. \label{eq:dhk-wig-fb}
\end{align}
Transforming to mean and difference variables, defined as, $\bm{\bar{z}} = \left(\bm{z}^{+} + \bm{z}^{-}\right)/2$ and $ \bm{\Delta z} = \bm{z}^{+} - \bm{z}^{-} $ respectively, we obtain 
\begin{align}
    C_{AB}^{\text{WDHK}}(t)  & = \frac{1}{\left(2\pi\hbar\right)^{2N}} \int d\bm{\bar{z}}_0 \int d\bm{\Delta z}_0\,\tilde{\rho}^{*}_{A}(\bm{\bar{z}}_0,\bm{\Delta z}_0)  \notag \\
    & \times \tilde{\mathcal{C}}_t(\bm{\bar{z}}_0,\bm{\Delta z}_0) \tilde{B}(\bm{\bar{z}}_t,\bm{\Delta z}_t) e^{i\tilde{S}_t(\bm{\bar{z}}_0,\bm{\Delta z}_0)/\hbar}, \label{eq:dhk-wig}
\end{align}
where, $\tilde{\rho}^{*}_{A}(\bm{\bar{z}}_0,\bm{\Delta z}_0)$ and $\tilde{B}(\bm{\bar{z}}_t,\bm{\Delta z}_t)$ are Wigner transforms smoothed by complex Gaussian functions,
\begin{align}
    \tilde{B}(\bm{\bar{z}}_t,\bm{\Delta z}_t) = \int d \bm{z}^{\prime} B_W(\bm{z}^{\prime}) g\left(\bm{z}^{\prime};\bm{\bar{z}}_t,\bm{\Delta z}_t\right), \label{eq:b_tilda}
\end{align}
$\tilde{\rho}^{*}_{A}(\bm{\bar{z}}_0,\bm{\Delta z}_0)$ is defined similarly, and the $*$ denotes complex conjugate. 
The complex Gaussian function is defined as,\footnote{ In Ref. \citenum{Gottwald2018} 
the determinant has a power of 1/4, not 1/2, and the definition of $\bm{\Gamma}$ is different. 
There, the Wigner transform of the density operator is not normalized, but instead the factor 
of $(2\pi\hbar)^{-N}$ that usually shows up due to normalization is divided equally into the 
Wigner transforms of $\hat{\rho}_{\hat{A}}$ and $\hat{B}$. 
This, taken together with the different definition of $\bm{\Gamma}$ accounts for the difference in power.}
\begin{align}
    g\left(\bm{z};\bm{\bar{z}},\bm{\Delta z}\right) & = \text{det}\left(\frac{\bm{\Gamma}}{\pi\hbar}\right)^{1/2} e^{-\frac{1}{\hbar}\left(\bm{z}-\bm{\bar{z}}\right)^{T}\cdot\bm{\Gamma}\cdot\left(\bm{z}-\bm{\bar{z}}\right)} \notag \\
    & \times e^{i\bm{\Delta z}^{T}\cdot\bm{J}^{T}\cdot\left(\bm{z}-\bm{\bar{z}}\right)/\hbar}, \label{eq:comp_gaus}
\end{align}
where $\bm{\Gamma}$ is a $2N \times 2N$ dimensional width matrix defined as, 
\begin{align}
    \bm{\Gamma} & = \left(\begin{array}{cc}
    \bm{\gamma}\hbar & \mathbb{0} \\
    \mathbb{0} & \frac{1}{\hbar}\bm{\gamma}^{-1} \\
    \end{array}\right), & \text{and} & &  \bm{J} &= \left(\begin{array}{rr}
     \mathbb{0} & \mathbb{1} \\
    -\mathbb{1} & \mathbb{0} \\
    \end{array}\right), \label{eq:j_and_gamma}
\end{align}
is the  $2N \times 2N$ dimensional symplectic matrix. 
It is important to note that the derivation we have followed naturally leads to this form for
the matrix $\bm{\Gamma}$ such that $\text{det}(\bm{\Gamma})=1$. 
Previous derivations of WDHK allow for a general width matrix, with the special case of 
$\text{det}(\bm{\Gamma})=1$\footnote{In Ref.\citenum{Gottwald2018} 
the condition is written as $\text{det}(\bm{\Gamma})=4$. 
The difference between our expression and theirs stems from a factor of 2 that is different 
between the respective definitions of the width matrix $\bm{\gamma}$ in the coherent state 
$\ket{\bm{z}}$.} being equivalent to the conventional DHK expression taken into Wigner phase space.\cite{Gottwald2018}

The phase space functions $\tilde{\rho}^{*}_{A}$ and $\tilde{B}$ represent the 
operators $\hat{\rho}_{A}$ and $\hat{B}$, respectively, in the extended phase 
space of the forward-backward (or mean-difference) variables. 
Similar to a conventional DHK calculation, classical forward and backward 
trajectories $\bm{z}^{\pm}_t$ are generated by the classical Hamiltonian $H(\bm{z})$. 
The prefactor $\tilde{\mathcal{C}}_t$ is calculated using the monodromy matrices of 
the forward and backwards trajectories, and the action $\tilde{S}_t$ is calculated 
using their respective classical actions. 
A discussion on the interpretation of the phase space functions  
$\tilde{\rho}^{*}_{A}$ and $\tilde{B}$, and the relationship between the WDHK 
prefactor, $\tilde{\mathcal{C}}_t$, and the WDHK action, $\tilde{S}_t$, 
with their respective counterparts in the conventional DHK expression are 
provided in Appendix~\ref{app:wig_DHK}. 
Furthermore, it is important to note that the WDHK correlation function 
that we have derived here, Eq.~\eqref{eq:dhk-wig}, is \textit{completely equivalent} 
to the conventional DHK correlation function.\cite{Gottwald2018} Details are provided in Appendix~\ref{app:wig_DHK}. In Appendix~\ref{app:DHK_md} we present the WDHK correlation function in terms 
of mean and difference trajectories $\bm{\bar{z}}(t)$ and $\Delta\bm{z}(t)$, and their respective actions and monodromy matrices. 

The classical-limit of the mean-difference form of the WDHK correlation function in Eq.~\ref{eq:dhk-wig}, is the LSC correlation function. This is achieved by performing the integrals over $\bm{\xi}^T=\left(\bm{z},\bm{z}^{\prime},\bm{\Delta z}_0\right)^T$ using the stationary phase approximation (SPA) as detailed in Appendix~\ref{app:dhk-cl-lmt}. 
This is a powerful result: with LSC dynamics as the classical limit of WDHK, a complete hierarchy of systematic SC approximations can be derived in Wigner phase space. 
We note that in this manuscript, we frequently move between expressing quantities in either the forward backward variables $(\bm{z}^{+},\bm{z}^{-})$ or the mean-difference variables, $(\bm{\bar{z}},\bm{\Delta z})$, or, even a mix of the two representations depending on the context.

\subsection{Adiabatic Hybrid Wigner Dynamics} \label{sec:hwd-a}

To capture quantum effects beyond the classical limit at a reasonable computational cost, 
we can partition a high-dimensional Hamiltonian into a `system' which consists of the subset 
of dofs that are integral to an accurate description of the problem, and a `bath' subset
that alter the system dynamics, but comprises dofs that are not being directly interrogated
and are well described in the classical limit.
Similar strategies for adiabatic dynamics have been pursued,~\cite{Ovchinnikov1996,Sun1997c,Zhang2005,Grossmann2006,Koda2016} 
however, none of these approaches provides a uniform theoretical treatment across all dofs.
In particular, the WDHK framework has 
been previously used to obtain a mixed 
semiclassical-classical 
propagation scheme,~\cite{Koda2016} however
as we discuss later in this section, our approach has 
important differences from this work.

Consider the Hamiltonian,
\begin{align}
    H(\bm{z}) = \frac{1}{2}\bm{p}^T\cdot\bm{m}^{-1}\cdot\bm{p} + V(\bm{q}) \label{eq:H_ad}
\end{align}
where $\bm{p}$ and $\bm{q}$ are $N$-dimensional system-bath vectors for the momentum and position respectively, and $\bm{m}$ is the $N\times N$ diagonal mass matrix.  
We partition this Hamiltonian into system and bath parts assuming the kinetic energy is separable (generally true 
in Cartesian coordinates),
\begin{align}
   T({\bm p})= \sum_{x\in \{s, b\}} \frac{1}{2}\bm{p}_x^T\cdot\bm{m}_x^{-1}\cdot\bm{p}_x
\end{align}
and we express the potential energy as a sum,
\begin{align}
    V(\bm{q}) = V_s\left(\bm{q}_s\right) + V_b\left(\bm{q}_b\right) + V_{sb}\left(\bm{q}\right),
\end{align}
where the subscript $s$ denotes system and $b$ denotes bath dofs, the system and bath dimensionality are $N_s$ and $N_b$ respectively, with $N = N_s + N_b$, $V_{s/b}$ are the system/bath-only potentials depending only on $\bm{q}_{s/b}$ 
and $V_{sb}$ is the system-bath coupling that depends on all coordinates, $\bm q$.
No specific forms for the potential energy components are assumed, but as with all multiphysics
methods, we expect the HWD approach to work best when 
the system-bath coupling is small.
We introduce a system-bath basis, $\bm{z}_{sb}^T=(\bm{z}_s,\bm{z}_b)^T$, to represent the 
matrices,
\begin{align}
    \bm{\Gamma} & = \left(\begin{array}{cc}
     \bm{\Gamma}_s & \mathbb{0} \\
    \mathbb{0}  & \bm{\Gamma}_b\\ 
    \end{array}\right), & \text{and} & & \bm{J} = \left(\begin{array}{cc}
     \bm{J}_s & \mathbb{0} \\
    \mathbb{0}  & \bm{J}_b\\ 
    \end{array}\right), 
\end{align}
where,
\begin{align}
    \bm{\Gamma}_x & = \left(\begin{array}{cc}
     \bm{\gamma}_x \hbar & \mathbb{0} \\
    \mathbb{0}  & \frac{1}{\hbar}\bm{\gamma}_x^{-1}\\ 
    \end{array}\right), & \text{and} & & \bm{J}_x = \left(\begin{array}{cc}
     \mathbb{0}  & \mathbb{1}_x \\
    -\mathbb{1}_x & \mathbb{0}  \\ 
    \end{array}\right), \label{eq:jg_sb}
\end{align}
are $2N_x \times 2N_x$ dimensional block matrices, with $x \in \{s,b\}$. Furthermore the monodromy matrix blocks for the mean and difference trajectories, $\bm{\bar{M}}$, and $\bm{\Delta M}$, defined in Eq.~\eqref{eq:mbar_dM}, can be written in this basis as, 
\begin{align}
    \bm{\bar{M}} = \left(\begin{array}{cc}
       \bm{\bar{M}}_{ss} & \bm{\bar{M}}_{sb} \\
    \bm{\bar{M}}_{bs}  & \bm{\bar{M}}_{bb} \\ 
    \end{array}\right),
\end{align}
where $\bm{\bar{M}}_{xy} = \frac{\partial \bm{\bar{z}}_x(t)}{\partial \bm{\bar{z}}_y(0)}$, for $x,y \in \{s,b\}$ and similarly for $\bm{\Delta M}$.  

We start with the WDHK expression Eq.~\eqref{eq:dhk-wig} for the real-time correlation 
function and approximate the integral over just the bath dofs using an SPA as detailed
in Appendix~\ref{app:aHWD}. 
The resulting expression is the AHWD correlation function,
\begin{align}
    C_{AB}^{\text{AHWD}}(t) & = \frac{1}{\left(2\pi\hbar\right)^{2N_s + N_b}} 
    \int d\bm{z}_{0,s}^{\pm} \int d\bm{\bar{z}}_{0,b} \, \Tilde{{\rho}}_{A_s}^{*}(\bm{z}_{0,s}^{\pm}) 
    \Tilde{B_s}(\bm{z}_{t,s}^{\pm}) \notag \\
    & \times \tilde{\mathcal{C}}_{t}^{\text{AHWD}}(\bm{z}_{0,s}^{\pm},\bm{\bar{z}}_{0,b})
    e^{i\tilde{S}_t^{\text{AHWD}}(\bm{z}_{0,s}^{\pm},\bm{\bar{z}}_{0,b})/\hbar}\notag\\
    & \times \left[\hat{\rho}_{A_b}\right]_W\left(\bm{\bar{z}}_{0,b}\right)  
    \left[\hat{B}_b\right]_W\left(\bm{\bar{z}}_{t,b}\right).
    \label{eq:cab_ahwd}
\end{align}
The structure of the correlation function in Eq.~\ref{eq:cab_ahwd} 
that emerges is as follows:\\
1. As a consequence of the stationary phase treatment of the bath, the difference variables for the bath dofs are constrained to be zero, $\bm{\Delta z}_{t,b} =0$. 
As in an LSC calculation, the bath operators are then represented in mean-variable phase space by their respective Wigner transforms, $\left[\hat{\rho}_{A_b}\right]_W\left(\bm{\bar{z}}_{0,b}\right)$ and $ \left[\hat{B}_b\right]_W\left(\bm{\bar{z}}_{t,b}\right)$, in third line of Eq.~\ref{eq:cab_ahwd}.\\
2. The system operators are represented in the extended forward-backward phase space by their respective smoothed Wigner functions,  
$\Tilde{{\rho}}_{A_s}^{*}(\bm{z}_{0,s}^{\pm})$ and 
$\Tilde{B_s}(\bm{z}_{t,s}^{\pm})$, 
in the first line of Eq.~\eqref{eq:cab_ahwd}, with the corresponding prefactor and phase (action) terms 
in the second line.\\
3. The structure of the trajectories is such that system variables evolve with separate forward and backward trajectories, $\bm{z}_{t,s}^{\pm}$, whereas, only the mean variables, $\bm{\bar{z}_{t,b}}$, are evolved for the bath dofs. We emphasize here that trajectories are evolved 
under the \textit{\textbf{full}} potential that includes forces due to the system bath coupling, $V_{sb}(\bm{q})$. This is evident in the equations of motion.
\begin{align}
    \dot{\bm{q}}_s^{\pm} &= \bm{m}^{-1}_s \cdot\bm{p}_s^{\pm} \\
    \dot{\bm{\bar{q}}}_b &= \bm{m}^{-1}_b\cdot\bm{\bar{p}}_b \\
    \dot{\bm{p}}_s^{\pm} & = - \frac{V\left(\bm{q}_s^{\pm},\bm{\bar{q}}_b\right)}{d\bm{q}^{\pm}_s} \\
    \dot{\bm{\bar{p}}}_{b} & = -\frac{1}{2}\left[\frac{V\left(\bm{q}_s^{+},\bm{\bar{q}}_b\right)}{d\bm{\bar{q}}_b} + \frac{V\left(\bm{q}_s^{-},\bm{\bar{q}}_b\right)}{d\bm{\bar{q}}_b}\right], 
\end{align}
that conserve the Hamiltonian (not written in conjugate variables),
\begin{align}
    \bar{H}(\bm{z}_s^{\pm},\bm{\bar{z}}_b) & = \frac{1}{2}\bm{p}^{{+}^T}_s\cdot\bm{m}^{-1}_s\cdot\bm{p}^{+}_s + \frac{1}{2}\bm{p}^{{-}^T}_s\cdot\bm{m}^{-1}_s\cdot\bm{p}^{-}_s \notag \\ 
    & + \bm{\bar{p}}_b^T\cdot\bm{m}_b^{-1}\cdot\bm{\bar{p}}_b 
    + V\left(\bm{q}_s^+,\bm{\bar{q}}_b\right) + V\left(\bm{q}_s^-,\bm{\bar{q}}_b\right). \label{eq:AHWD_ham}
\end{align}

Consistent with DHK-like treatment for the system variables, the form of the AHWD prefactor in Eq.~\eqref{eq:cab_ahwd}, $\tilde{\mathcal{C}}_{t}^{\text{AHWD}}(\bm{z}_{0,s}^{\pm},\bm{\bar{z}}_{0,b})$, is the same as that of the DHK prefactor for just the system variables. The full expression is presented in Eq.~\eqref{eq:pref_ahwd}. In performing an SPA over only the bath degrees of freedom, 
we find that satisfying one of the stationary phase conditions
requires that we set the system-bath coupling, $V_{sb}(\bm{q})\to 0$ for the propagation of the monodromy matrix elements. Details of this requirement can be found in Appendices~\ref{app:aHWD} and ~\ref{ap:HWDbo_mono}. As a consequence, we find the prefactor can be obtained from propagating just the system-system block of the monodromy matrix elements using the equations of motion presented in Eq.~\eqref{eq:ahwd_mono_eom}. This represents a considerable reduction in computational cost relative to 
a full DHK simulation particulary when the number of system
dofs is small relative to the number of bath dofs. We note that several approximate methods to compute the DHK prefactor have been developed,~\cite{Gelabert2000d, Liberto2016b} 
and given that the AHWD prefactor has the same structure, these approximations can be leveraged as needed for high-dimensional system simulations.

Finally, the definition of the action in the AHWD correlation function, ${S}_t^{\text{AHWD}}(\bm{z}_{0,s}^{\pm},\bm{\bar{z}}_{0,b})$, is presented in Eq.~\eqref{eq:act_ahwd}. Owing to the coupled nature of the system and bath dynamics, the action is dependent on both the system and bath variables. However, the vanishing difference trajectory of the bath is expected to make the magnitude of the AHWD action considerably smaller than the full DHK action, especially in cases where the bath consists of a large number of dofs. As a result, the AHWD correlation function is expected to largely mitigate the sign-problem and be significantly easier to converge than a full DHK calculation, as discussed in the results section (Fig.~\ref{fig:Pr_error}).

We note in the AHWD correlation function, Eq.~\eqref{eq:cab_ahwd}, we have assumed that $\hat{B}\equiv \hat{B}_s \otimes \hat{B}_b$ can be factorized into system and bath operators, and the same for $\hat{\rho}_{A}$. While we expect that $\hat B$ will likely be a system-only operator for many processes of interest, we note that the factorization assumption is not necessary, we use it here only for clarity. 
In the general case when such a factorization is not possible, $\tilde{B}$, 
defined in Eq.~\eqref{eq:b_tilda} will need to be evaluated while enforcing 
the stationary phase condition. Finally, we note that 
using the definition of the smoothed Wigner functions, Eq.~\eqref{eq:b_tilda}, it 
can be shown that the AHWD correlation function is exact at time zero for all operators.

The AHWD method is very similar in spirit to other mixed semiclassical methods. For instance, AMQC-IVR also partitions the problem into system and bath components while treating the system with the `quantum limit' of the Filinov filter, $\bm{c}_s \to 0$ and treating the bath with its classical limit, $\bm{c}_b \to \infty$.\cite{Church2019a} However there are key differences between AHWD and AMQC. First, the bath dofs are treated within the classical Husimi framework in AMQC, which is not as accurate as the LSC employed here.~\cite{Malpathak2022,Lee1995}
Second, the trajectory structure in AMQC is such that the difference variable for the bath is only constrained to be zero at zero time, $\bm{z}_{0,b} = 0$, but can take non-zero values at later times, $\bm{z}_{t,b} \neq 0$ unlike the AHWD. 
And third, the AMQC expression involves a full system+bath dimensional prefactor. 
In the weak system-bath coupling, it has been suggested that a separable prefactor assumption can be made to reduce the dimensionality of the prefactor however it is significantly less accurate than the AHWD simplification as demonstrated in Sec.~\ref{sec:results} for coupled harmonic oscillators.

The AHWD method also differs from the mixed semiclassical-classical dynamics method in Wigner space developed by Koda.~\cite{Koda2016} Much like AMQC, but set in Wigner phase space, \citeauthor{Koda2016}'s method only restricts the difference variables for the bath to vanish at the initial time, $\bm{\Delta z}_{0,b} =0$, and requires a full system+bath dimensional prefactor.
We also compare AHWD to \citeauthor{Grossmann2006}'s SCHD method,\cite{Grossmann2006} which treats the system with the HK propagator and the bath with Thawed Gaussian Wavepacket Dynamics (TGWD).\cite{Heller1975} The TGWD method propagates Gaussian wavepackets using a single classical trajectory, while allowing it to accumulate a phase governed by the action of the trajectory, and also allowing the width of the wavepacket to vary. Within the SCHD framework, the bath dofs are restricted to start at the same initial phase space points for all trajectories in the ensemble, limiting its ability to sample the correct quantum mechanical distribution for the bath at time zero. In the AHWD method, the bath is sampled exactly at $t=0$ using its Wigner distribution. In addition, SCHD requires system-bath blocks of the monodromy matrix to be calculated, which can become expensive for large dimensional baths.


\section{Model Systems and Simulation Details}\label{sec:simulation}

Model I consists of two harmonic oscillators coupled bilinearly, 
one with a lighter mass, $m_1 = 1$ a.u., and the other with a 
heavier mass, $m_2 = 25$ a.u. The potential is, 
\begin{align}
    V\left(x_1,x_2\right) = \frac{1}{2}m_1\omega_1^2 x_1^2 + \frac{1}{2}m_2\omega_2^2 x_2^2 + kx_1x_2,
\end{align}
with the harmonic frequencies $\omega_1 = \sqrt{2}$ a.u., $\omega_2 = 1/3$ a.u. and varying coupling constants, $k = 0.5, \& \, 2$ a.u. The initial state is taken to be a direct product of coherent states, as defined in Eq.~\eqref{eq:cs-def}, in both $x_1$ and $x_2$ dofs, 
with widths $\gamma_j = m_j\omega_j/\hbar$ for $j \in \{1,2\}$ and centered at 
$x_{1_i} = x_{2_i} = 1.0$ a.u. with momenta $p_{1_i} = p_{2_i} = 0$ a.u.  
Both DHK and LSC are exact for a bilinearly coupled harmonic systems since a simple
normal mode transformation can be used to express the system in terms of uncoupled 
oscillators. In the uncoupled representation, it is trivial to show that AHWD will also
be exact, however, we use the coupled representation to test the accuracy of the AHWD 
and specifically the approximation that requires monodromy matrix elements between
the system and bath dofs to vanish. 

Model II is an an anharmonic oscillator bilinearly coupled to a heavy harmonic oscillator,
\begin{align}
    V\left(x_1,x_2\right) & = \frac{1}{2}m_1\omega_1^2 x_1^2 -0.1x_1^3 + 0.1x_1^4 \notag \\
    & + \frac{1}{2}m_2\omega_2^2 x_2^2 + kx_1x_2,
\end{align}
with the same masses, frequencies and initial state as in model I. For low coupling, $k =0.5$ a.u., and moderate coupling, $k =1.0$ a.u., strong quantum mechanical recurrences are observed in the average position of the light oscillator, $\langle \hat{x}_1(t) \rangle$. These are a hallmark of interference in an anharmonic system,\cite{Wang2009,LohoChoudhury2020} and cannot be captured by classical limit methods like LSC. For strong coupling, $k=2.0$ a.u., long-time recurrences are largely washed-out due to coupling to the bath.

\begin{figure}
    \centering
    \includegraphics[width=0.45\textwidth]{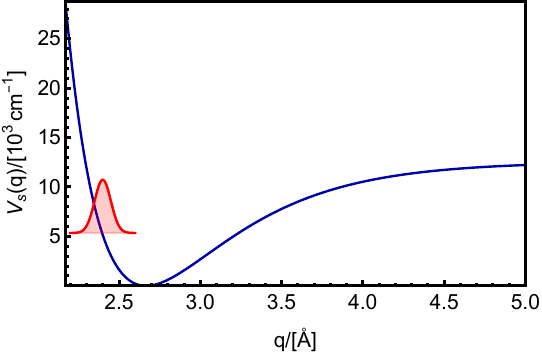}
    \caption{A schematic of the Morse oscillator potential $V_s(q)$ as function of the I$_2$ bond length $q$. Also shown in red is the initial coherent state wavepacket centered at $q_i = 2.4$\r{A} with an energy $E_i = 0.426 D_e$.}
    \label{fig:morse_osc}
\end{figure}

Model III is a Morse oscillator parameterized to describe an I$_2$ molecule, 
\begin{align}
    V_s(q) = D_e\left[1- e^{-\alpha (q-q_e)}\right]^2, 
\end{align}
where $D_e = 1.2547 \times 10^4$ cm$^{-1}$ is the dissociation energy, $q_e = 2.6663$ \r{A} is the equilibrium bond length of the I$_2$ molecule and $\alpha = 1.8576$ \r{A}$^{-1}$. Fig.~\ref{fig:morse_osc} plots this model Morse potential. The vibrational dof is 
coupled to a low frequency thermal bath which causes decoherence in the vibrational relaxation dynamics of the I$_2$ molecule. The Hamiltonian for the model is, 
\begin{align}
    H(z,\bm{Z}) & = \frac{p^2}{2\mu} + V_s(q)  + \sum_{j=1}^{N} \Biggl[\frac{P_j^2}{2m_j} \Biggr. \notag \\
    &  + \Biggl. \frac{1}{2}m_j\omega_j^2\left(Q_j + \frac{c_j}{m_j\omega_j^2}\left(q-q_e\right)\right)^2\Biggr],
\end{align}
where $z=(q,p)$ are the phase space coordinates for the vibrational dof of the I$_2$ molecule, with reduced mass $\mu=1.165 \times 10^5$ a.u., $Z_j = (Q_j,P_j)$ are the phase space coordinates for the $j^{th}$ bath mode, with mass $m_j = 1$ a.u., frequencies  $\omega_j$, and coupling constants $c_j$, for $j \in [1,N]$. Here we employ an Ohmic bath with spectral density,\cite{Leggett1987}
\begin{align}
     J_{bath}(\omega) = \eta\omega e^{-\omega/\omega_c}
\end{align}
with characteristic frequency, $\omega_c = 20$ cm$^{-1}$ and variable system-bath coupling strength $\eta$. The bath is discretized into $N$ modes as, 
\begin{align}
    J_{bath}(\omega) = \frac{\pi}{2}\sum_{j=1}^N \frac{c_j^2}{m_j\omega_j}\delta(\omega- \omega_j),
\end{align}
following the discretization protocal introduced by Craig and Manolopoulos.~\cite{Craig2005}
The initial state for the I$_2$ vibrational dof is chosen to be a coherent state $\ket{z_i;\gamma_i}$, as defined in Eq.~\eqref{eq:cs-def}, centered at $q_i = 2.4$ \r{A}, with momentum $p_i = 0$ a.u. and width $\gamma_i = \mu\omega_s/\hbar$, where $\omega_s = \alpha \sqrt{2 D_e/\mu} \approx 213.7$ cm$^{-1}$, is the harmonic frequency of the Morse oscillator. Fig.~\ref{fig:morse_osc} depicts a schematic of this initial coherent state centered at $q_i = 2.4$ \r{A}, with an initial energy $E_i = 0.426D_e$, set in the Morse oscillator potential $V_s(q)$. The bath is initially at thermal equilibrium at inverse temperature $\beta$, with the initial density $\hat{\rho}_b = e^{-\beta\hat{H}_b}/\text{Tr}[e^{-\beta\hat{H}_b}]$, where, 
\begin{align}
    H_b(\bm{Z}) & = \sum_{j=1}^{N} \Biggl[\frac{P_j^2}{2m_j} + \frac{1}{2}m_j\omega_j^2Q_j^2\Biggr],
\end{align}
is the Hamiltonian of the uncoupled bath. This model has previously been used to study quenching of interference effects in the vibrational probability density and the associated decoherence rates.\cite{Wang2001,Elran2004,Goletz2009,Sanz2014}

In models I and II we track the position of the light mode, $x_1(t) \equiv \langle \hat{x}_1(t) \rangle$, as a function of time.  For AHWD calculations, the light mode $x_1$ is chosen as the `system' and the heavy mode $x_2$ as the bath. Initial phase space points for the forward and backward trajectories of the system, $z_1^{\pm}(0)$, are sampled from the absolute magnitude of the coherent state matrix element, $\left|\braket{z_1^+(0)}{z_{1_i}}\braket{z_{1_i}}{z_1^-(0)}\right|$, whereas those for the mean trajectory of the bath, $\bar{z}_2(0)$, are sampled from the Wigner transform, $\left[\ketbra{z_{2_i}}{z_{2_i}}\right]_W(\bar{z}_2(0))$. For the former, we use Eq.~\eqref{eq:btilde_identity} that relates the smoothed Wigner transform of an operator and its coherent state matrix element. 

In model III we calculate the probability density of the Morse dof, $\mathbb{P}(q,t)\equiv \langle \delta(\hat{q}-q) \rangle_t$. The Morse dof $q$ is chosen as the system, and the harmonic oscillators $\bm{Q}$ represent the bath. As in the other models, initial phase space points for the forward and backward trajectories of the system, $z^{\pm}(0)$, are sampled from the absolute magnitude of the coherent state matrix element, $\left|\braket{z^+(0)}{z_{i}}\braket{z_{i}}{z^-(0)}\right|$, whereas those for the mean trajectory of the bath, $\bar{Z}(0)$, are sampled from the Wigner transform, $\left[e^{-\beta\hat{H}_b}\right]_W\left(\bar{Z}(0)\right)$. The probability density $\mathbb{P}(q,t)$ is normalized.

Trajectories are propagated under the classical Hamiltonian, Eq.~\eqref{eq:AHWD_ham}, using a fourth-order symplectic integrator\cite{Brewer1997}
with a time step of 0.01 a.u. for models I and II, and 2 fs for model III, to ensure total energy conservation. Trajectories breaking energy conservation, as indicated using $\left| 1 - E(t)/E(0)\right| \geq 10^{-5}$, or breaking the symplecticity condition, Eq.~\eqref{eq:ahwd_symp} for the system monodromy matrices within a tolerance of $10^{-3}$ are discarded. For all models $<1\%$ of total trajectories are discarded.

\section{Results and Discussion} \label{sec:results}
\subsection{Coupled Oscillator Models}

\begin{figure}
    \centering
    \includegraphics[width=0.45\textwidth]{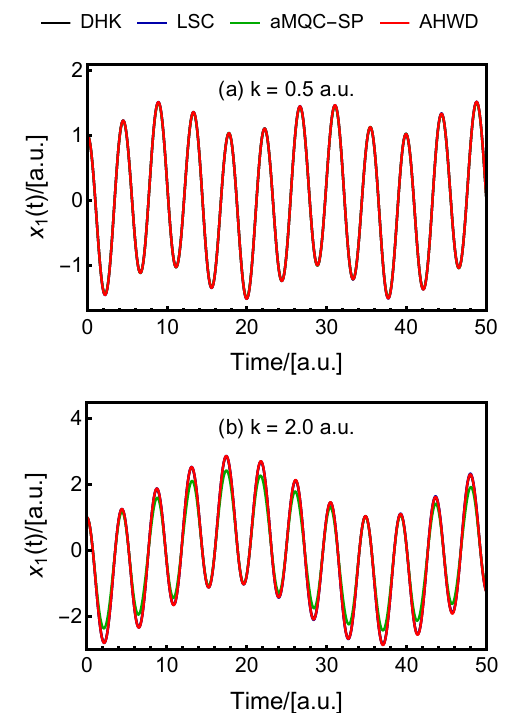}
    \caption{The average position of the light mode $\langle \hat{x}_1(t) \rangle$ is plotted as a function of time for model I for increasing coupling strengths, (a) $k = 0.5$ a.u., and (b) $k = 2.0$ a.u.. Both panels show results calculated using DHK (black), LSC (blue), AMQC-SP (green), and AHWD (red). DHK, LSC and AHWD  are exact for this model for both coupling strengths considered here, and overlap with each other. Unlike AHWD, AMQC-SP is not exact for the strong coupling case, $k = 2.0$ a.u.}
    \label{fig:xyt_2dho}
\end{figure}

Fig.~\ref{fig:xyt_2dho} plots the average position of the light mode $\langle \hat{x}_1(t) \rangle$ for model I for two coupling strengths, $k = 0.5$ and $2$ a.u. 
For the weak coupling case, $k = 0.5$ a.u., all the methods considered here, 
DHK, AHWD, AMQC with separable prefactor approximation (AMQC-SP) and LSC are exact. 
It is worth noting that both AHWD and AMQC-SP treat the heavy harmonic bath 
mode, $ \hat{x}_2 $, classically, and only require a prefactor for the light 
mode $\hat{x}_1$. However, for the strong coupling case, $k = 2$, AMQC-SP is no longer exact.
Owing to its more complete classical treatment of the bath mode,  $\bm{\Delta z}_{t,b} =0$, AHWD still predicts the exact result in agreement with DHK and LSC. 

\begin{figure}
    \centering
    \includegraphics[width=0.45\textwidth]{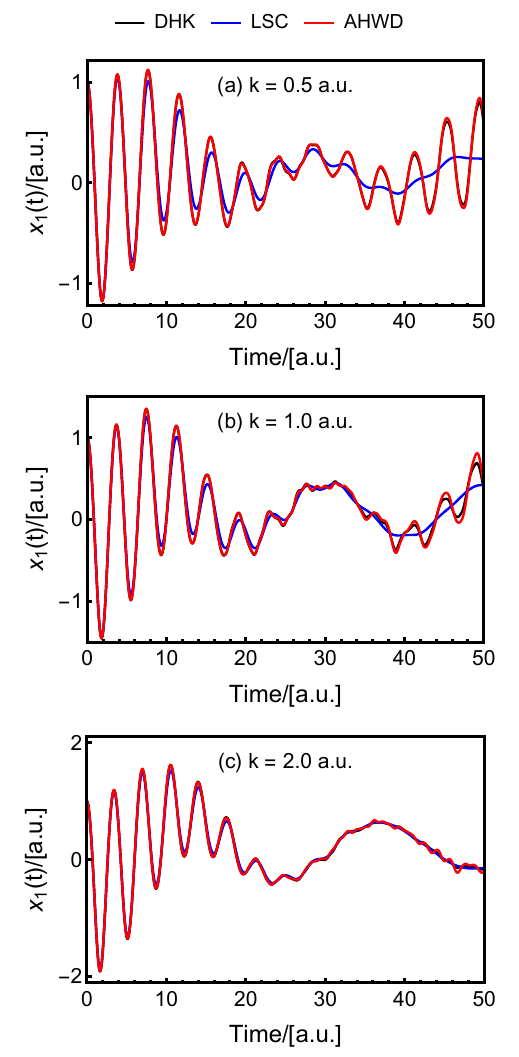}
    \caption{
    The average position of the light mode $\langle \hat{x}_1(t) \rangle$ is plotted as a function of time for model II for increasing coupling strengths, (a) $k = 0.5$ a.u., 
    (b) $k = 1.0$ a.u., and (c) $k=2.0$ a.u.
    All panels show results from DHK (black)
    compared with LSC (blue) and AHWD (red) simulations. For these systems, DHK and the exact quantum results are identical. AHWD and DHK results agree in all three cases shown, whereas LSC exhibits damped oscillations and agrees with DHK results only for the strong system-bath coupling cases in (c).} 
    
    \label{fig:xyt_hhb}
\end{figure}

\begin{figure}
    \centering
    \includegraphics[width=0.45\textwidth]{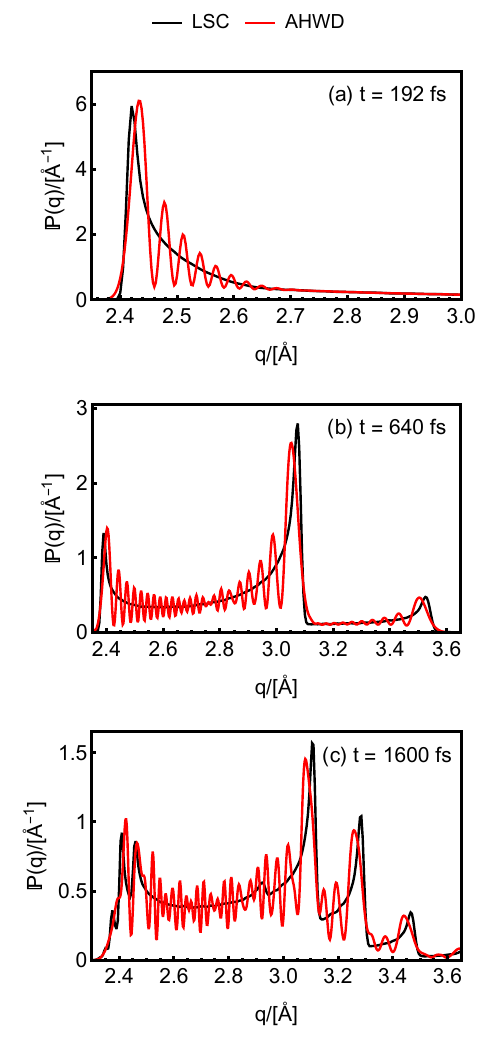}
    \caption{The probability density of the I$_2$ vibrational mode, $\mathbb{P}(q)$, for model 3 is plotted for coupling strength $\eta_{r} = 0.05$ and $T = 100$ K at (a) $t= 192$ fs, (b) $t = 640$ fs, and (c) $t = 1600$ fs. All panels show calculations using AHWD (red) and LSC (black).} 
    \label{fig:Pr_eta5}
\end{figure}

The anharmonic model II serves as a more stringent benchmark for AHWD. 
For all three coupling strengths, DHK is identical to the exact quantum results, and is able to capture long time recurrences in the oscillations of $\langle \hat{x}_1(t) \rangle$ for  weak coupling, $k=0.5$ a.u., and moderate coupling  $k=1.0$ a.u.. AHWD also accurately captures these long time recurrences in $\langle \hat{x}_1(t) \rangle$. Its accuracy is excellent when compared to DHK for the weak coupling case, and very good for the moderate coupling case (with slightly overestimated amplitudes at long time). Notably, AHWD results are obtained using $10^7$ trajectories, an order of magnitude fewer as compared to $10^8$ for DHK. LSC fails to capture long time recurrences for both these cases.
For strong coupling, $k=2.0$ a.u., the bath strongly affects system dynamics and long time recurrences are washed out. Since quantum interference effects are not required for an accurate description of the dynamics in this regime, LSC is fairly accurate. AHWD results are also in close agreement with DHK, however we see small amplitude oscillations emerge at longer times. We attribute this behavior in the strong system-bath coupling limit to the assumption of vanishing off-diagonal monodromy matrix blocks that is worse at longer times. This is further evidenced by the deterioration in unitarity and energy conservation of AHWD at long times, as presented in Appendix~\ref{app:HWD_long_time}.

To summarize our results for coupled oscillator models, we find 
AHWD accurately captures quantum mechanical recurrences in the weak and moderate coupling regime that are fairly difficult to simulate. For strong coupling where interference effects are washed out, and where mixed quantization is contra-indicated, using a classical limit method like LSC is preferred, with the accuracy of the AHWD method deteriorating at longer times.

\begin{figure}
    \centering
    \includegraphics[width=0.4\textwidth]{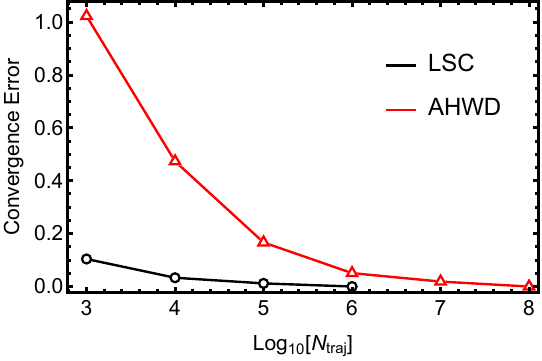}
    \caption{The convergence error, as defined in Eq.~\eqref{eq:conv-error}, is plotted as a function of the number of trajectories, $N_{traj}$, for both LSC (black) and AHWD (red) calculations of $\mathbb{P}(q)$ for T = 100K, $\eta_r = 0.05$, and $t = 1600$ fs. The plot demonstrates that AHWD calculations converge on increasing the number of trajectories and do not suffer from the dynamical sign problem.} 
    \label{fig:Pr_error}
\end{figure}

\subsection{Vibrational Relaxation in I$_2$}
Next we consider vibrational relaxation in model III. Fig.~\ref{fig:Pr_eta5} plots the probability density, $\mathbb{P}(q)$, for the reduced coupling strength $\eta_{r} \equiv \eta/\mu\omega_s = 0.05$ and $T = 100$ K. For this weak coupling and moderate temperature case, the time-evolved probability density calculated using AHWD has a highly oscillatory structure, which is a hallmark of quantum interference effects.\cite{Wang2001} 

\begin{figure*}
    \centering
    \includegraphics[width=0.8\textwidth]{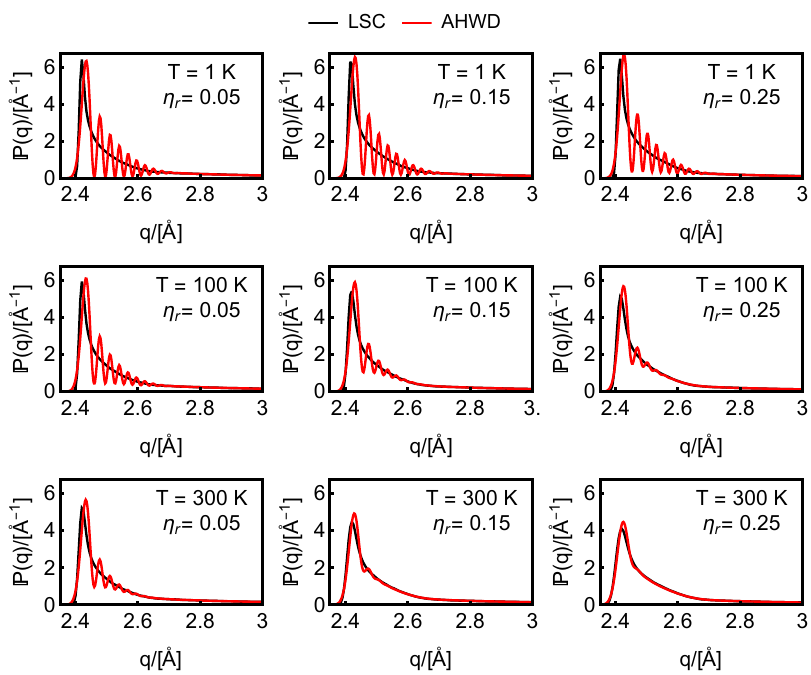}
    \caption{The probability density of the I$_2$ vibrational mode, $\mathbb{P}(q)$, for model 3 is plotted at $t= 192$ fs for various coupling strengths, $\eta_{r} = 0.05$ (left column), $\eta_{r} = 0.15$ (middle column) and $\eta_{r} = 0.25$ (right column) and various temperatures, $T = 1$ K (top row), $T = 100$ K (middle row), and $T = 300$ K (bottom row). All panels show calculations using AHWD (red) and LSC (black).} 
    \label{fig:Pr_t192}
\end{figure*}
AHWD captures these interference effects even after many vibrational periods at $t = 1600$ fs. In the AHWD method, the bath is treated in the classical limit and only monodromy matrices corresponding to the vibrational system dof need to be calculated, making such a calculation feasible for this model. A DHK calculation, which would also be able to account for the intricate oscillatory structure, would need the full system-bath prefactor, and moreover, would suffer from the dynamical sign problem for such a high dimensional system, necessitating the use of a phase filtering technique.\cite{Antipov2015,Church2017} We also note that LSC is unable to capture any oscillatory structure due to its lack of phase information and predicts a smooth, decohered vibrational probability distribution even at the shortest time presented here.

Next, we investigate the extent to which the AHWD approximation
mitigates the sign problem. We plot the convergence error 
between an unconverged calculation of the vibrational probability distribution against the converged calculations. The convergence error for $N_{traj}$ trajectories is defined as, 
\begin{align}
    \text{Convergence error} = \int_0^{\infty} dq \left|\mathbb{P}_{N_{traj}}(q) - \mathbb{P}_{ref}(q) \right|, \label{eq:conv-error}
\end{align}
where $\mathbb{P}_{ref}(q)$ is the converged calculation, with $10^8$ and $10^6$ trajectories for AHWD and LSC respectively. We plot the convergence error for $T=100$ K, $\eta_{eff} = 0.05$, and $t = 1600$ fs for both LSC and AHWD in Fig.~\ref{fig:Pr_error}. For both methods, the convergence error systematically decreases with increasing number of trajectories, demonstrating that, similar to LSC, AHWD calculations do not suffer from the sign problem here.

Finally, we use AHWD to study the effect of the temperature of the bath and the system-bath coupling strength on the decoherence of the vibrational distribution. Fig.~\ref{fig:Pr_t192} plots the probability density of the I$_2$ vibrational mode, $\mathbb{P}(q)$, at $t= 192$ fs for various coupling strengths, $\eta_{r} = 0.05$, $0.15$, and $0.25$, and various temperatures, $T = 1$, $100$, and $300$ K. Focusing on the effect of bath temperature for fixed system-bath coupling strength, AHWD calculations highlight that increasing the temperature of the bath results in more rapid decoherence, smoothing out the probability density. This effect is observed at all three system-bath coupling strengths studied here. Similarly, increasing the system bath coupling strength $\eta_r$ at a fixed temperature also results in more rapid decoherence. At $T=1$ K, this effect is not observed for the coupling strengths considered here. However, calculations using SCHD on this system have previously shown that increasing the coupling further to $\eta_r = 1.0$ results in effective damping of the oscillations in $\mathbb{P}(q)$ even at $T=0$ K.\cite{Goletz2009} Our analysis is in agreement with a similar analysis of the effect of both the system-bath coupling strength and the bath temperature on the decoherence of the vibrational probability density carried out previously using the Forward-Backward IVR method.\cite{Wang2001} It is interesting to note that the vibrational probability distribution predicted by 
LSC lacks oscillatory structure, consistent with the method's inability to describe interference effects. For the case of high temperature, $T =300$ K and strong system bath coupling $\eta_{eff} = 0.25$, the probability distribution predicted by LSC is in agreement with AHWD, suggesting that a classical limit calculation is only valid in regimes where strong decoherence from the bath washes out interference effects.

\section{Conclusions} \label{sec:conclusion}

In this paper, we introduce a complete hierarchy of SC methods in Wigner phase space. Starting with the exact quantum mechanical expression for the correlation function, we introduce a novel 
derivation of the DHK correlation function in Wigner phase space. We further demonstrate that the classical limit of this WDHK correlation function can be obtained using a stationary phase approximation and leads to the LSC approximation. 
Exploiting this connection with classical limit dynamics, the AHWD method for mixed quantization is derived. The resulting expression corresponds to describing important `system' dofs at the DHK level, while treating the bath within the LSC approximation. We demonstrate that the AHWD method can capture quantum interference effects in simple low-dimensional models. We further show that AHWD can accurately model system-bath dynamics through a study of the decoherence of the vibrational probability density of a model I$_2$ oscillator coupled to a thermal bath. Decoherence is shown to be facilitated by stronger coupling to the bath and by increasing the temperature of the bath, which is in agreement with previous studies.\cite{Wang2001,Goletz2009} It is worth noting that the AHWD framework requires only a system prefactor, which for large baths, drastically reduces the size of the the prefactor matrix when compared to DHK. Moreover the action in AHWD is expected to be smaller in magnitude as compared to DHK due to the linearization of the bath variables. This reduces the oscillatory nature of the phase term, helping to mitigate the sign problem. 

We foresee a variety of different directions for the development of the AHWD method. It can be extended to a continuum of bath dofs, which may have connections with existing SC approaches for studying dissipation and decoherence.\cite{Grossmann1995,Fiete2003,Pollak2007,Koch2008} Being able to treat large system-bath models while accurately capturing interference effects, AHWD is uniquely suited to the study of rate modification as a result of vibrational strong coupling.\cite{Lindoy2022,Sun2022,Lindoy2023} Lastly, to facilitate the use of AHWD for \textit{ab-initio} calculations, its performance with various existing approximations for the SC prefactor need to be tested.\cite{Gelabert2000d,Liberto2016b} These avenues will be pursued in future studies.

\begin{acknowledgments}
The authors thank Prof. Gregory Ezra for many stimulating discussions on quantum dynamics in phase space. S.M. acknowledges funding from the Cornell University Department of Chemistry and Chemical Biology.
\end{acknowledgments}
\appendix

\section{Details of the Wigner Double Herman Kluk Correlation Function} \label{app:wig_DHK}

As mentioned in the main text, the phase space functions $\tilde{\rho}^{*}_{A}$ and $\tilde{B}$ represent the operators $\hat{\rho}_{A}$ and $\hat{B}$, respectively, in the extended phase space of the forward-backward (or mean-difference) variables. To facilitate their interpretation, it helps to identify the mean or \textit{centre} variable $\bm{\bar{z}}$ and the (modified) difference, or \textit{chord} variable,  $\bm{J}\cdot\bm{\Delta z}$ as being conjugate bases in an extended phase space, analogous to the position and momentum bases in conventional phase space. Furthermore, the representation of an operator in centre phase space is the Wigner transform, also known as the centre function. Similarly, an operator represented in the chord phase space is the ``characteristic" function or the chord function.\cite{OzoriodeAlmeida2006a} Both these representations are completely equivalent and contain identical information. Now, note that for the case of a general width matrix $\bm{\Gamma}$ the complex Gaussian in Eq. \eqref{eq:comp_gaus} resembles a coherent state in the extended phase space centered at the centre variable $\bm{\bar{z}}$, with conjugate ``momentum" as the chord variable,  $\bm{J}\cdot\bm{\Delta z}$. In the limit $\bm{\Gamma} \to \infty$, the smoothing in Eq.~\eqref{eq:b_tilda} does nothing, yielding the Wigner transform of the operator, whereas in the limit $\bm{\Gamma} \to 0$, the smoothing is actually a Fourier transform that yields the chord function. For a finite width matrix $\bm{\Gamma}$, the smoothing results in an extended phase space function with finite width in both the centre and chord bases.

The WDHK prefactor $\tilde{\mathcal{C}}_t$ and action $\tilde{S}_t$ are related to their respective counterparts in the usual DHK expression as,
\begin{align}
    \tilde{\mathcal{C}}_t(\bm{z}^{+}_0,\bm{z}^{-}_0) & = \mathcal{C}_t\left(\bm{z}^{+}_0\right) \mathcal{C}^{*}_t\left(\bm{z}^{-}_0\right), \label{eq:pref-rel }\\
    \tilde{S}_t(\bm{z}^{+}_0,\bm{z}^{-}_0) & = -\bm{\bar{p}}^{T}_t\cdot\bm{\Delta q}_t + \bm{\bar{p}}^{T}_0\cdot\bm{\Delta q}_0 \notag \\ 
    & + S_t(\bm{z}^{+}_0) - S_t(\bm{z}^{-}_0). \label{eq:s_tilda}
\end{align}
Here, $S_t(\bm{z}^{\pm}_0)$ is the action of the forward/backward trajectory, $\bm{z}_{t}^{\pm}$. Finally,  we note that for $\bm{\Gamma}$ defined in Eq.~\eqref{eq:j_and_gamma}, the smoothed Wigner transform of an operator is related to its off-diagonal coherent state matrix element through the following identity, 
\begin{align}
    \mel{\bm{z}^{-}}{\hat{B}}{\bm{z}^{+}} = \tilde{B}(\bm{z}^{+},\bm{z}^{-}) e^{-i\bm{\bar{p}}^{T}\cdot\bm{\Delta q}/\hbar}. \label{eq:btilde_identity}
\end{align}
 
Along with Eqs.~\eqref{eq:pref-rel } and \eqref{eq:s_tilda}, it can be used to show that the WDHK expression, Eq.~\eqref{eq:dhk-wig}, is equivalent to the conventional DHK expression,\cite{Koda2015,Gottwald2018}
\begin{align}
    C_{AB}^{\text{DHK}}(t)  & =  \frac{1}{\left(2\pi\hbar\right)^{2N}} \int d\bm{z}^{+}_0 \int d\bm{z}^{-}_0\, \mel{\bm{z}^{+}_0}{\hat{\rho}_A}{\bm{z}^{-}_0} \mathcal{C}_t\left(\bm{z}^{+}_0\right) \notag \\
    & \times  \mathcal{C}^{*}_t\left(\bm{z}^{-}_0\right) \mel{\bm{z}^{-}_t}{\hat{B}}{\bm{z}^{+}_t} e^{i\left[S_t(\bm{z}^{+}_0) - S_t(\bm{z}^{-}_0)\right]/\hbar} . \label{dhk-hil}    
\end{align}

\section{Double Herman Kluk Dynamics in Mean-Difference Variables} \label{app:DHK_md}

Let us begin with the conventional Hamiltonian,
\begin{align}
    H(\bm{z}) = T(\bm{p}) + V(\bm{q}) = \frac{1}{2}\bm{p}^T\cdot\bm{m}^{-1}\cdot\bm{p} + V(\bm{q}) \label{eq:h_tpv}
\end{align}  where $\bm{m}$ is the diagonal mass matrix and the superscript $T$ indicates the transpose of a vector or matrix. This Hamiltonian propagates a set of conventional phase space variables $\bm{z}$ using Hamilton's equations,
\begin{align}
    \dot{\bm{z}} = \bm{J}\cdot\frac{\partial H}{\partial \bm{z}},
\end{align}
 and the classical action of the trajectory is defined as,
 \begin{align}
 S_t = \int_0^{t} d\tau \, \mathrm{L}(\bm{z_{\tau}}),
 \end{align} 
where the Lagrangian is $\mathrm{L}(\bm{z}) = T(\bm{p}) - V(\bm{q})$.  The monodromy matrix elements are then propagated using the equation of motion,
\begin{align}
    \dot{\bm{M}} = \bm{J}\cdot\frac{\partial^2 H}{\partial \bm{z}^2}\cdot\bm{M},
\end{align}
with $\bm{M}(0)=\mathbb{1}$ and following the symplecticity condition $\bm{M}^T\cdot\bm{J}\cdot\bm{M}=\bm{J}$.\cite{child2014} We now introduce an extended Hamiltonian, which we refer to as the ``sum"-Hamiltonian 
\begin{align}
    \bar{H}(\bm{z}^{\pm}) = H(\bm{z}^+) + H(\bm{z}^-). \label{eq:H_sum}
\end{align}
that generates a pair of trajectories: the ``forward" trajectory $\bm{z}^+(t)$ and the ``backward" trajectory $\bm{z}^-(t)$. Note that despite the name ``backward" \textemdash which is rooted in the trajectory representing the backward quantum propagator \textemdash  the extended Hamiltonian, Eq.~~\eqref{eq:H_sum} propagates $\bm{z}^-(t)$ forward in time.  The trajectories follow the equation of motion, 
\begin{align}
    \dot{\bm{z}}^{\pm} = \bm{\mathcal{J}}\cdot\frac{\partial \bar{H}}{\partial \bm{z}^{\pm}},
\end{align}
where $\bm{z}^{\pm^T} = (\bm{q}^{+},\bm{q}^{-},\bm{p}^{+},\bm{p}^{-})^T$ and $\mathcal{J}$ is a $4N \times 4N$ symplectic matrix . The forward and backward trajectories are thus individually propagated under $H(\bm{z}^+)$ and $H(\bm{z}^-)$ respectively, as expected. The action of this combined trajectory is just the sum of actions of the two trajectories, $\bar{S}_t(\bm{z}^{\pm}) = S_t(\bm{z}^+) + S_t(\bm{z}^-)$. We will come back to this aspect later.  The monodromy matrix for the extended Hamiltonian, $\mathcal{M}$, follows the equation of motion  $\dot{\mathcal{M}} = \mathcal{J}\cdot\frac{\partial^2 H}{\partial \bm{z}^{{\pm}^2}}\cdot\mathcal{M}$ when written in the basis $\bm{z}^{\pm^T}$. The symplecticity condition, $\mathcal{M}^T\cdot\mathcal{J}\cdot\mathcal{M}=\mathcal{J}$  is also obeyed. Note that when transformed into the basis $(\bm{z}^{+},\bm{z}^{-})^T$, we represent the monodromy matrix as $\mathcal{M}^{\pm}$, which stays block diagonal at all times,
\begin{align}
    \mathcal{M}^{\pm} & = \left(\begin{array}{cc}
    \bm{M}^{+} & \mathbb{0} \\
    \mathbb{0} & \bm{M}^{-} \\
    \end{array}\right),
\end{align}
where $\bm{M}^{+}$ and $\bm{M}^{-}$ are the monodromy matrices corresponding to the forward and backward trajectories respectively. It is important to note that the backward trajectory involved in the correlation function calculation is technically supposed to be run backward in time, but in practical implementations, it is run forward in time by exploiting the time reversal symmetry of classical mechanics. To account for its backward-in-time nature, the action of the forward-run backward trajectory is included with a negative sign, making the net forward-backward action  $S_t(\bm{z}^{+}_0) - S_t(\bm{z}^{-}_0)$ as opposed to $S_t(\bm{z}^{+}_0) + S_t(\bm{z}^{-}_0)$. Similarly, the inverse of the monodromy matrix of the forward-run backward trajectory is used to calculate the prefactor $\mathcal{C}^*_t(\bm{z}_0^{-})$.\cite{Church2017} These nuances can also be taken into account implicitly by working with another extended Hamiltonian, which we refer to as the ``difference"-Hamiltonian,
\begin{align}
    \Tilde{H}(\bm{z}^{\pm}) = H(\bm{z}^+) - H(\bm{z}^-). \label{eq:H_diff}
\end{align}
This Hamiltonian leads to similar equations of motion as the sum Hamiltonian, but has the action  $ S_t(\bm{z}^+) - S_t(\bm{z}^-)$, taking into account the backward-in-time nature of the backward trajectory.  Since both Hamiltonians have identical equations of motion for the phase space variables and the monodromy matrices, we stick with using the sum Hamiltonian because it offers simplicity in deriving conjugate variables when switching to mean-difference variables. However, we still properly account for the backward-in-time nature of the backward trajectory in the DHK prefactor, and by calculating the correct action $ S_t(\bm{z}^+) - S_t(\bm{z}^-)$. 

Next, we switch gears to mean and difference variables $\bm{\bar{z}}(t)$ and $\Delta\bm{z}(t)$. It is important to note that the mean and difference momenta are defined via their relation to the forward-backward momenta, $\bm{\bar{p}} = (\bm{p}^+ + \bm{p}^-)/2$ and $\Delta\bm{p} = \bm{p}^+ - \bm{p}^-$. Although $\bm{p}^{\pm}$ are the classical conjugate variables for $\bm{q}^{\pm}$ under the Hamiltonians $\bar{H}(\bm{z}^{\pm})$,  it is incorrect to assume that $(\bm{\bar{p}},\Delta\bm{p})$ will also be conjugate to  $(\bm{\bar{q}},\Delta\bm{q})$ when the same Hamiltonian is converted to mean and difference variables, $\bar{H}(\bm{\bar{z}},\Delta\bm{z})$. In view of this, we start by writing the Lagrangian $\bar{\mathrm{L}}$, corresponding to the sum Hamiltonian,
\begin{align}
    \bar{\mathrm{L}}(\bm{q}^{\pm},\dot{\bm{q}}^{\pm}) & =  \mathrm{L}(\bm{q}^+,\dot{\bm{q}}^+) + \mathrm{L}(\bm{q}^-,\dot{\bm{q}}^-) \notag \\
    &= T(\dot{\bm{q}}^{+}) - V(\bm{q}^+) + T(\dot{\bm{q}}^{-}) - V(\bm{q}^-) \notag \\
    & = \frac{1}{2}\dot{\bm{q}}^{{+}^T}\cdot\bm{m}\cdot\dot{\bm{q}}^{+} + \frac{1}{2}\dot{\bm{q}}^{{-}^T}\cdot\bm{m}\cdot\dot{\bm{q}}^{-} \notag \\
    & - V(\bm{q}^+) - V(\bm{q}^-), 
\end{align}
and converting it to mean and difference positions and velocities,
\begin{align}
    \bar{\mathrm{L}}\left(\bm{\bar{q}},\dot{\bm{\bar{q}}},\Delta\bm{q},\Delta\dot{\bm{q}}\right) & = \frac{1}{2}\dot{\bm{\bar{q}}}^T\cdot\bm{\bar{m}}\cdot\dot{\bm{\bar{q}}} + \frac{1}{2}\Delta\dot{\bm{q}}^T\cdot\Delta \bm{m}\cdot\Delta\dot{\bm{q}} \notag \\
    & - V\left(\bm{\bar{q}}+\frac{\Delta \bm{q}}{2}\right) - V\left(\bm{\bar{q}}-\frac{\Delta \bm{q}}{2}\right),
\end{align}
where $\bm{\bar{m}} = 2\bm{m}$ and $\Delta \bm{m} = \bm{m}/2$. The conjugate momenta to $(\bm{\bar{q}},\Delta\bm{q})$ can now be obtained as, 
\begin{align}
    \bm{\bar{\pi}} &= \frac{\partial \bar{\mathrm{L}}}{\partial \dot{\bm{\bar{q}}}} = \bm{\bar{m}}\cdot\dot{\bm{\bar{q}}} = 2 \bm{\bar{p}} \\
    \Delta \bm{\pi} & = \frac{\partial \bar{\mathrm{L}}}{\partial\Delta\dot{\bm{q}}} = \Delta \bm{m}\cdot\Delta\dot{\bm{q}} = \Delta \bm{p}/2.
\end{align} 
The Hamiltonian $\bar{H}(\bm{\bar{q}},\bm{\bar{\pi}},\Delta\bm{q},\Delta\bm{\pi})$ can be written in conjugate variables,
\begin{align}
    \bar{H}(\bm{\bar{q}},\bm{\bar{\pi}},\Delta\bm{q},\Delta\bm{\pi}) & = \bm{\bar{\pi}}^T\cdot\dot{\bm{\bar{q}}} + \Delta\bm{\pi}^T\cdot\Delta\dot{\bm{q}} - \bar{\mathrm{L}}(\bm{\bar{q}},\bm{\bar{\pi}},\Delta\bm{q},\Delta\bm{\pi}) \notag \\ 
    & = \frac{1}{2}\bm{\bar{\pi}}^T\cdot\bm{\bar{m}}^{-1}\cdot\bm{\bar{\pi}} + \frac{1}{2}\Delta\bm{\pi}^T\cdot\Delta \bm{m^{-1}}\cdot\Delta\bm{\pi} \notag \\
    &+ V\left(\bm{\bar{q}}+\frac{\Delta \bm{q}}{2}\right) + V\left(\bm{\bar{q}}-\frac{\Delta \bm{q}}{2}\right),
\end{align}

The equations of motion for the set of conjugate variables $(\bm{\bar{q}},\bm{\bar{\pi}},\Delta\bm{q},\Delta\bm{\pi})$ can be obtained from the Hamiltonian $\bar{H}(\bm{\bar{q}},\bm{\bar{\pi}},\Delta\bm{q},\Delta\bm{\pi})$, and then converted to the equations of motion for the ``standard" mean and difference variables $(\bm{\bar{z}},\Delta\bm{z})$, 
\begin{align}
    \dot{\bm{\bar{q}}} &= \bm{m}^{-1}\cdot\bm{\bar{p}} \label{eq:eom_qb} \\
    \Delta\dot{\bm{q}} &= \bm{m}^{-1}\cdot\Delta \bm{p} \\
    \dot{\bm{\bar{p}}} & = -\frac{1}{2}\left[\bm{V}^{\prime}\left(\bm{\bar{q}}+\frac{\Delta \bm{q}}{2}\right) + \bm{V}^{\prime}\left(\bm{\bar{q}}-\frac{\Delta \bm{q}}{2}\right)\right]  \label{eq:mean-force} \\
    \dot{\Delta \bm{p}} & = - \left[\bm{V}^{\prime}\left(\bm{\bar{q}}+\frac{\Delta \bm{q}}{2}\right) - \bm{V}^{\prime}\left(\bm{\bar{q}}-\frac{\Delta \bm{q}}{2}\right)\right]. \label{eq:diff-force}
\end{align}
Here, $\bm{V}^{\prime}\left(\bm{\bar{q}}\pm\frac{\Delta \bm{q}}{2}\right)$ represents $\left.\frac{dV(\bm{q})}{d\bm{q}}\right|_{\bm{q}=\bm{\bar{q}}\pm\frac{\Delta \bm{q}}{2}}$. Eqs.~\eqref{eq:mean-force}  and \eqref{eq:diff-force} highlight that the mean and difference trajectories, $\bm{\bar{z}}(t)$ and $\Delta \bm{z}(t)$ feel the mean and difference of the potentials evaluated at the forward and backward trajectory positions respectively. The Hamiltonian written in the standard mean and difference variables is, 
\begin{align}
    \bar{H}(\bm{\bar{z}},\Delta\bm{z}) & = \bm{\bar{p}}^T\cdot\bm{m}^{-1}\cdot\bm{\bar{p}} + \frac{1}{4}\Delta\bm{p}^T\cdot\bm{m}^{-1}\cdot\Delta\bm{p} \notag \\
    &
    + V\left(\bm{\bar{q}}+\frac{\Delta \bm{q}}{2}\right) + V\left(\bm{\bar{q}}-\frac{\Delta \bm{q}}{2}\right). \label{eq:H_ex_md}
\end{align}
Although this could've been obtained from Eq.~\eqref{eq:H_sum} by a simple coordinate transform, it was necessary to go through all these steps to obtain the correct equations of motion, since this Hamiltonian is \textbf{\textit{not}} written in conjugate variables and \textbf{\textit{cannot}} be used to directly obtain equations of motion using Hamilton's equations.  The forward-backward action can be written as, 
\begin{align}
    S_t(\bm{z}^{+}_0) - S_t(\bm{z}^{-}_0) & = \int_0^{\tau} d\tau \, \biggl\{\bm{\bar{p}}^T_{\tau}\cdot\bm{m}^{-1}\cdot\Delta\bm{p}_{\tau} \biggr. \notag \\
    & - \left. \left[V\left(\bm{\bar{q}}_{\tau}+\frac{\Delta \bm{q}_{\tau}}{2}\right) - V\left(\bm{\bar{q}}_{\tau}-\frac{\Delta \bm{q}_{\tau}}{2}\right) \right] \right\}. \label{eq:act_pm_md}
\end{align}
The monodromy matrix $\mathcal{M}_{cv}$, follows the equation of motion, $\dot{\mathcal{M}}_{cv} = \mathcal{J}\cdot\frac{\partial^2 H}{\partial \bm{z}_{cv}^2}\cdot\mathcal{M}_{cv}$ and the symplecticity condition $\mathcal{M}_{cv}^T\cdot\mathcal{J}\cdot\mathcal{M}_{cv}=\mathcal{J}$ . The subscript denotes that it is written in a basis of conjugate variables $\bm{z}_{cv}^T=\left(\bm{\bar{q}},\Delta\bm{q},\bm{\bar{\pi}},\Delta\bm{\pi}\right)^T$. When converted to a basis of standard mean-difference variables $\bm{z}_{md}^T = (\bm{\bar{z}},\Delta \bm{z})^T$, the monodromy matrix can be written as,
\begin{align}
    \mathcal{M}_{md} = \left(\begin{array}{cc}
    \bm{M}_{\bm{\bar{z}}\bm{\bar{z}}} & \bm{M}_{\bm{\bar{z}}\Delta\bm{z}} \\
    \bm{M}_{\Delta\bm{z}\bm{\bar{z}}}  &  \bm{M}_{\Delta\bm{z}\Delta\bm{z}}  \\
    \end{array}\right),
\end{align}
following the equation of motion,
\begin{align}
    \dot{\mathcal{M}}_{md} = \left(\begin{array}{cc}
    \mathcal{A} &  \frac{1}{4}\mathcal{B} \\
    \mathcal{B}  &  \mathcal{A} \\
    \end{array}\right) \cdot
    \mathcal{M}_{md}, \label{eq:mono_md_eom}
\end{align}
where, 
\begin{align}
   \mathcal{A} &=  \left(\begin{array}{cc}
     \mathbb{0} & \bm{m}^{-1}\\
    - \frac{1}{2}\bm{V}_+^{\prime\prime} & \mathbb{0} \\
    \end{array}\right), &  \mathcal{B} & = \left(\begin{array}{cc}
     \mathbb{0} & \mathbb{0}\\
    - \bm{V}_-^{\prime\prime} & \mathbb{0} \\
    \end{array}\right), \label{eq:mat_a&b}
\end{align}
$V_{\pm} \equiv V(\bm{q}^+) \pm V(\bm{q}^-)$ and $^{\prime\prime}$ indicates that it is the second derivative or Hessian matrix. The symplecticity condition becomes, 
\begin{align}
    \mathcal{M}_{md}^T \cdot \left(\begin{array}{cc}
     2\bm{J} & \mathbb{0} \\
    \mathbb{0}  & \frac{1}{2}\bm{J}\\
    \end{array}\right) \cdot \mathcal{M}_{md} =  \left(\begin{array}{cc}
     2\bm{J} & \mathbb{0} \\
    \mathbb{0}  & \frac{1}{2}\bm{J}\\
    \end{array}\right). \label{eq:sym_m-md}
\end{align}
The structure of the equation of motion Eq.~\eqref{eq:mono_md_eom} gives it a peculiar property.  If at time $t$, 
\begin{align}
    \bm{M}_{\bm{\bar{z}}\bm{\bar{z}}}(t) &= \bm{M}_{\Delta\bm{z}\Delta\bm{z}}(t), & \text{and} & & \bm{M}_{\bm{\bar{z}}\Delta\bm{z}}(t) &= \frac{1}{4}\bm{M}_{\Delta\bm{z}\bm{\bar{z}}}(t), \label{eq:mono_cond}
\end{align} 
then Eq.~\eqref{eq:mono_md_eom} implies that, 
\begin{align}
    \dot{\bm{M}}_{\bm{\bar{z}}\bm{\bar{z}}}(t) &= \dot{\bm{M}}_{\Delta\bm{z}\Delta\bm{z}}(t), & \text{and} & & \dot{\bm{M}}_{\bm{\bar{z}}\Delta\bm{z}}(t) &= \frac{1}{4}\dot{\bm{M}}_{\Delta\bm{z}\bm{\bar{z}}}(t). \label{eq:mono_cond_dot}
\end{align}
At $t=0$, $\mathcal{M}_{md}(0) = \mathbb{1}$, and Eq.~\eqref{eq:mono_cond} holds. Therefore, at $t=0$, Eq.\eqref{eq:mono_cond_dot} holds too. Taken together, these imply that Eq.~\eqref{eq:mono_cond} holds after an infinitesimal time-step, $t=\Delta t$. Thus, by induction, Eq.~\eqref{eq:mono_cond} holds for all time $t$, allowing us to write the monodromy matrix as,
\begin{align}
    \mathcal{M}_{md} = \left(\begin{array}{cc}
    \bm{\bar{M}} & \bm{\Delta M} \\
    4 \bm{\Delta M}  &  \bm{\bar{M}} \\
    \end{array}\right), \label{eq:mbar_dM}
\end{align}
where $\bm{\bar{M}} = \bm{M}_{\bm{\bar{z}}\bm{\bar{z}}} = \bm{M}_{\Delta\bm{z}\Delta\bm{z}}$ and $\bm{\Delta M} = \bm{M}_{\bm{\bar{z}}\Delta\bm{z}} = \frac{1}{4}\bm{M}_{\Delta\bm{z}\bm{\bar{z}}}$.  Thus, to propagate $\mathcal{M}_{md}$ we only need to propagate two $2N \times 2N$ dimensional matrices $\bm{\bar{M}}$ and $\bm{\Delta M}$, and not a full $4N \times 4N$ dimensional matrix. This is equivalent to propagating $\bm{M}^{\pm}$, and in fact it can be shown that, 
\begin{align}
    \bm{\bar{M}} & = \frac{1}{2}\left(\bm{M}^+ + \bm{M}^-\right), \label{eq:mb_dm1}
\end{align}
and, 
\begin{align}
    \bm{\Delta M} & = \frac{1}{4}\left(\bm{M}^+ - \bm{M}^-\right). \label{eq:mb_dm2}
\end{align}
Following Ref.~\citenum{Gottwald2018} the Wigner-DHK prefactor can also be written as,
\begin{align}
    \tilde{\mathcal{C}}_t(\bm{z}^{+}_0,\bm{z}^{-}_0) & = \mathcal{C}_t\left(\bm{z}^{+}_0\right) \mathcal{C}^{*}_t\left(\bm{z}^{-}_0\right) \notag \\ 
    & = \text{det}\left(2\bm{\Gamma}\right)^{-1/2}\text{det}\left[\frac{1}{2}\left(\bm{\Gamma}+i\bm{J}\right)\cdot\bm{M}^+\cdot\left(\mathbb{1}+i\bm{J}\cdot\bm{\Gamma}\right) \right. \notag \\
    & \left. + \frac{1}{2}\left(\bm{\Gamma}-i\bm{J}\right)\cdot\bm{M}^-\cdot\left(\mathbb{1}-i\bm{J}\cdot\bm{\Gamma}\right)\right]^{1/2}. \label{eq:dhkw_pref}
\end{align}
Inverting Eqs.~\eqref{eq:mb_dm1} and ~\eqref{eq:mb_dm2} and using them in Eq.~\eqref{eq:dhkw_pref} lets us write the prefactor as,
\begin{align}    \tilde{\mathcal{C}}_t(\bm{\bar{z}},\Delta\bm{z}) & = \text{det}\left(2\bm{\Gamma}\right)^{-1/2}\text{det}\left[\bm{\Gamma}\cdot\bm{\bar{M}} + 2i\bm{\Gamma}\cdot\bm{\Delta M}\cdot\bm{J}\cdot\bm{\Gamma} \right. \notag  \\
    & \left. + 2i\bm{J}\cdot\bm{\Delta M} - \bm{J}\cdot\bm{\bar{M}}\cdot\bm{J}\cdot\bm{\Gamma}  \right]^{1/2}. \label{eq:dhk_pref_md}
\end{align}
Finally, the symplecticity condition Eq.~\eqref{eq:sym_m-md} can be broken down into,
\begin{align}
    \bm{\bar{M}}^T\cdot\bm{J}\cdot\bm{\bar{M}} + 4\bm{\Delta M}^T\cdot\bm{J}\cdot\bm{\Delta M} & = \bm{J}, \label{eq:sym_mb_dm1}
\end{align}
and, 
\begin{align}
    \bm{\bar{M}}^T\cdot\bm{J}\cdot\bm{\Delta M} = \left(\bm{\bar{M}}^T\cdot\bm{J}\cdot\bm{\Delta M}\right)^T. \label{eq:sym_mb_dm2} 
\end{align}
The procedure outlined here to represent all the quantities required in a Wigner DHK calculation in terms of the mean and difference variables is restricted to Hamiltonians of the form in Eq.~\eqref{eq:h_tpv}. An equivalent procedure for the MMST Hamiltonian used for non-adiabatic SC dynamic calculations will be presented in Paper II.\cite{Malpathak2024c}

\section{Classical Limit of Herman Kluk Dynamics in Wigner Phase Space} \label{app:dhk-cl-lmt}
In this appendix, we show that the WDHK correlation function can be approximated using the stationary phase approximation (SPA) to yield LSC dynamics as the classical limit. This is expected since the LSC dynamics is the natural classical-limit dynamics in Wigner phase space. The stationary phase approximation to an integral is,
\begin{align}
    \int d\bm{\xi} f(\bm{\xi}) e^{i\phi(\bm{\xi})/\hbar} & \approx \sum_{\bm{\xi} = \bm{\xi}_{sp}} f\left(\bm{\xi}_{sp}\right) \notag \\
    & \times \text{det}\left[ \frac{1}{2\pi i \hbar} \bm{\phi}^{\prime\prime}\left( \bm{\xi}_{sp}\right)\right]^{-1/2} e^{i\phi(\bm{\xi}_{sp})/\hbar}.
\end{align}
We write the WDHK expression, Eq.~\eqref{eq:dhk-wig} as,
\begin{align}
    C_{AB}^{\text{WDHK}}(t)  & = \frac{1}{\left(2\pi\hbar\right)^{2N}} \int d\bm{\bar{z}}_0 \int d\bm{\Delta z}_0 \int d\bm{z}\int d\bm{z}^{\prime}\,\left[\hat{\rho}_{A}\right]_W(\bm{z}) \notag \\
    & \times g^{*}\left(\bm{z};\bm{\bar{z}}_0,\bm{\Delta z}_0\right) B_W(\bm{z}^{\prime})g\left(\bm{z}^{\prime};\bm{\bar{z}}_t,\bm{\Delta z}_t\right) \notag \\ & \times \tilde{\mathcal{C}}_t(\bm{\bar{z}}_0,\bm{\Delta z}_0) e^{i\tilde{S}_t(\bm{\bar{z}}_0,\bm{\Delta z}_0)/\hbar},
\end{align}
and approximate the integrals over $\bm{\xi}^T=\left(\bm{z},\bm{z}^{\prime},\bm{\Delta z}_0\right)^T$ using the SPA. The phase $\phi(\bm{\xi})$ in the SPA is identified as the action $\tilde{S}_t(\bm{\bar{z}}_0,\bm{\Delta z}_0)$ along with exponential contributions from $g^{*}\left(\bm{z};\bm{\bar{z}}_0,\bm{\Delta z}_0\right)$ and  $g\left(\bm{z}^{\prime};\bm{\bar{z}}_t,\bm{\Delta z}_t\right)$, 
\begin{align}
    \phi(\bm{\xi}) & = i\left(\bm{z}-\bm{\bar{z}}_0\right)^{T}\cdot\bm{\Gamma}\cdot\left(\bm{z}-\bm{\bar{z}}_0\right) - \bm{\Delta z}_0^{T}\cdot\bm{J}^{T}\cdot\left(\bm{z}-\bm{\bar{z}}_0\right) \notag \\
    & + i\left(\bm{z}^{\prime}-\bm{\bar{z}}_t\right)^{T}\cdot\bm{\Gamma}\cdot\left(\bm{z}^{\prime}-\bm{\bar{z}}_0\right) \notag \\
    &  + \bm{\Delta z}_t^{T}\cdot\bm{J}^{T}\cdot\left(\bm{z}^{\prime}-\bm{\bar{z}}_t\right)  + \tilde{S}_t(\bm{\bar{z}}_0,\bm{\Delta z}_0).
\end{align}
Note that other attempts to filter phase from the  conventional DHK expression have also successfully included real parts of Gaussians in the filtration before. \cite{Thoss2001a,Antipov2015} The first and second derivatives of the phase are,
\begin{align}
   \bm{\phi}^{\prime}(\bm{\xi})  =        \left(\begin{array}{cccc}
    2i\bm{\Gamma} & \mathbb{0} & \bm{J}^T & \mathbb{0} \\
   \mathbb{0} & 2i\bm{\Gamma} & \mathbb{0} & \bm{J} \\
  \bm{J} & \bm{F}^T & \mathbb{0} & \mathbb{0} 
\end{array}\right) \cdot
\left(\begin{array}{c}
\bm{z}-\bm{\bar{z}}_0 \\
\bm{z}^{\prime}-\bm{\bar{z}}_t  \\
    \bm{\Delta z}_0 \\ 
    \bm{\Delta z}_t
    \end{array}\right), \label{eq:phi_deriv}
\end{align}
and,
\begin{align}
   \bm{\phi}^{\prime\prime}(\bm{\xi}) = \left(\begin{array}{ccc}
    2i\bm{\Gamma} & \bm{0} & \bm{J}^T \\
    \bm{0} & 2i\bm{\Gamma} & \bm{F} \\
    \bm{J} & \bm{F}^T      & \bm{E} \\
    \end{array}\right), \label{eq:phi_hess}
\end{align}
respectively with $\bm{F} = \bm{J}\cdot\bm{\bar{M}}-2i\bm{\Gamma}\cdot\bm{\Delta M}$ and $\bm{E} = - \bm{F}^T\cdot\bm{\Delta M}$. The Hessian can be shown to be symmetric using the symplecticity condition, Eq.~\eqref{eq:sym_mb_dm2}. The stationary phase condition $\bm{\phi}^{\prime}(\bm{\xi}) = 0$ implies that,
\begin{align}
    \bm{z} &= \bm{\bar{z}}_0, & \bm{z}^{\prime} &= \bm{\bar{z}}_t,  & \text{and} &  & \bm{\Delta z}_t &= 0. \label{eq:sp_conditions} 
\end{align}
The first two conditions imply that the Gaussian smoothing of the Wigner transforms reduces to evaluating the Wigner transforms at the mean variables, where as the last condition implies that the difference trajectory is restricted to be zero! This is consistent with other classical limit SC methods where the difference trajectory is restricted to be zero. Under the stationary phase (SP) conditions, Eq.~\eqref{eq:sp_conditions}, the extended Hamiltonian, Eq.\eqref{eq:H_ex_md}, reverts back to the conventional Hamiltonian for the mean trajectory $H(\bm{\bar{z}})$, with the equations of motion, Eqs.~\eqref{eq:eom_qb} \& \eqref{eq:mean-force} reducing to conventional Hamilton's equations for the mean trajectory $\dot{\bm{\bar{z}}} = \bm{J}\cdot\frac{\partial H}{\partial \bm{\bar{z}}}$. Using Eq.~\eqref{eq:act_pm_md}, the action in Eq.~\eqref{eq:s_tilda} can be shown to vanish. Moreover, since the forward-backward trajectories coincide, their monodromy matrices coincide too, $\bm{M}^+ = \bm{M}^-$, implying $\bm{\Delta M}=0$. A similar conclusion can be drawn by inspecting the equations of motion for $\bm{\Delta M}$. This yields a simplification to the prefactor in Eq.~\eqref{eq:dhk_pref_md},
\begin{align}
    \tilde{\mathcal{C}}_t\left(\bm{\xi}_{sp}\right) = \text{det}\left(2\bm{\Gamma}\right)^{-1/2}\text{det}\left(\bm{\Gamma}\cdot\bm{\bar{M}}-\bm{J}\cdot\bm{\bar{M}}\cdot\bm{J}\cdot\bm{\Gamma}\right)^{1/2}. \label{eq:pref-cl-limit}
\end{align}
The last piece of the puzzle is the Hessian of the phase factor. Using standard techniques for simplification of determinants of block matrices\footnote{Refer to Eq.~A2 in Ref.\citenum{Antipov2015}.} to evaluate the determinant of Eq.~\eqref{eq:phi_hess}, and using the simplifications, $\bm{J}\cdot\bm{\Gamma}^{-1}\cdot\bm{J}^{T}= \bm{\Gamma}$, and $\text{det}(\bm{\bar{M}}) = \text{det}(\bm{\bar{M}}^{-1}) = 1$, it can be shown that, 
\begin{align}
    \text{det}\left[ \frac{1}{2\pi i \hbar} \bm{\phi}^{\prime\prime}\left( \bm{\xi}_{sp}\right)\right] & = (2\pi\hbar)^{-6N}2^{2N} \notag \\
    & \times \text{det}\left(\bm{\Gamma}\cdot\bm{\bar{M}}-\bm{J}\cdot\bm{\bar{M}}\cdot\bm{J}\cdot\bm{\Gamma}\right). 
\end{align}
Note that it involves the same determinant as the prefactor in Eq.~\eqref{eq:pref-cl-limit}, and they will cancel out. Assembling together all these pieces, the stationary phase approximation to WDHK correlation function can be shown to be,
\begin{align}
    C_{AB}^{\text{WDHK-SP}}(t)  & = \frac{1}{\left(2\pi\hbar\right)^{2N}} \int d\bm{\bar{z}}_0 \,\biggl[\left[\hat{\rho}_{A}\right]_W(\bm{z})  g^{*}\left(\bm{z};\bm{\bar{z}}_0,\bm{\Delta z}_0\right) \biggr.  \notag \\ 
    & \times B_W(\bm{z}^{\prime}) g\left(\bm{z}^{\prime};\bm{\bar{z}}_t,\bm{\Delta z}_t\right) \text{det}\left[ \frac{1}{2\pi i \hbar} \bm{\phi}^{\prime\prime}\left( \bm{\xi}_{sp}\right)\right]^{-1/2}   \notag \\ 
    & \times \biggl. \tilde{\mathcal{C}}_t(\bm{\bar{z}}_0,\bm{\Delta z}_0) 
    e^{i\tilde{S}_t(\bm{\bar{z}}_0,\bm{\Delta z}_0)/\hbar}\biggr]_{\bm{\xi} = \bm{\xi}_{sp}} \\
    & = \frac{1}{\left(2\pi\hbar\right)^{N}} \int d\bm{\bar{z}}_0 \,\left[\hat{\rho}_{A}\right]_W(\bm{\bar{z}}_0) B_W(\bm{\bar{z}}_t) \notag \\
    & =  C_{AB}^{LSC}(t).
\end{align}

\section{Derivation of Adiabatic Hybrid Wigner Dynamics} \label{app:aHWD}

\subsubsection{Stationary Phase Conditions}
To derive AHWD the bath dofs are approximated using the SPA. 
The integrals over $\bm{\xi}_b^T=\left(\bm{z}_b,\bm{z}^{\prime}_b,\bm{\Delta z}_{0,b}\right)^T$ in the WDHK expression Eq.~\eqref{eq:dhk-wig} are performed by SPA, while keeping the integrals over the system dofs and the mean bath variable $\bm{\bar{z}}_{0,b}$ exact. The relevant derivatives of the phase $\phi(\bm{\xi})$ can be extracted from Eq.~\eqref{eq:phi_deriv} as, 
\begin{align}
    \frac{\partial \phi}{\partial \bm{z}_b} & = 2i\bm{\Gamma}_b\cdot\left(\bm{z}-\bm{\bar{z}}_0\right)_b + \bm{J}_b^T\cdot\bm{\Delta z}_{0,b} = 0, \label{eq:HWD_ad_spc1} \\ 
     \frac{\partial \phi}{\partial \bm{z}_b^{\prime}} & = 2i\bm{\Gamma}_b\cdot\left(\bm{z}^{\prime}-\bm{\bar{z}}_t\right)_b + \bm{J}_b^T\cdot\bm{\Delta z}_{t,b} = 0,   \label{eq:HWD_ad_spc2} \\
     \frac{\partial \phi}{\partial \bm{\Delta z}_{0,b}} & = \bm{J}_b\cdot\left(\bm{z}-\bm{\bar{z}}_0\right)_b + \left[\bm{\bar{M}}^T\cdot\bm{J}^T\cdot\left(\bm{z}^{\prime}-\bm{\bar{z}}_t\right)\right]_b \notag \\
     & -2i\left[\bm{\Delta M}^T\cdot\bm{\Gamma}\cdot\left(\bm{z}^{\prime}-\bm{\bar{z}}_t\right)\right]_b = 0,  \label{eq:HWD_ad_spc3}
\end{align} 
where the subscript $b$ denotes the bath dof components of the vector and $\bm{\Gamma}_b$ and $\bm{J}_b$ have been defined earlier in Eq.~\eqref{eq:jg_sb}. Eqs.~\eqref{eq:HWD_ad_spc1} \& \eqref{eq:HWD_ad_spc2} imply,
\begin{align}
    \bm{z}_b &= \bm{\bar{z}}_{0,b}, & \bm{z}^{\prime}_b &= \bm{\bar{z}}_{t,b},  & \text{and} &  & \bm{\Delta z}_{t,b} &= 0. \label{eq:HWD_ad_sp_conditions} 
\end{align}
Here the last equality signifies that $\bm{\Delta z}_b=0$ at all times $t$ including $t=0$. It is interesting to note that these conditions are not sufficient for the equality in Eq.~\eqref{eq:HWD_ad_spc3} to hold, unlike in the classical limit case, where the conditions arising from the first two equations were sufficient to satisfy the third. As we shall see, this is a result of the coupled nature of the system-bath dynamics.  Let us examine the Hamiltonian and the equations of motion for the mean-difference trajectories, Eqs.~\eqref{eq:eom_qb}-\eqref{eq:diff-force} on imposing the SP conditions, Eq.~\eqref{eq:HWD_ad_sp_conditions}. The Hamiltonian becomes, 
\begin{align}
    \bar{H}(\bm{\bar{z}},\Delta\bm{z}_s) & = \bm{\bar{p}}^T\cdot\bm{m}^{-1}\cdot\bm{\bar{p}} + \frac{1}{4}\Delta\bm{p}_s^T\cdot\bm{m}_s^{-1}\cdot\Delta\bm{p}_s \notag \\
    &
    + V\left(\bm{q}_s^+,\bm{\bar{q}}_b\right) + V\left(\bm{q}_s^-,\bm{\bar{q}}_b\right), \label{eq:ham_HWD_ad}
\end{align}
and the equations of motion are,
\begin{align}
    \dot{\bm{\bar{q}}} &= \bm{m}^{-1}\cdot\bm{\bar{p}} \label{eq:eom_qb_HWD_ad} \\
    \Delta\dot{\bm{q}}_s &= \bm{m}^{-1}_s \cdot\Delta \bm{p}_s \\
    \dot{\bm{\bar{p}}} & = -\frac{1}{2}\left[\bm{V}^{\prime}\left(\bm{q}_s^+,\bm{\bar{q}}_b\right) + \bm{V}^{\prime}\left(\bm{q}_s^-,\bm{\bar{q}}_b\right)\right]  \label{eq:mean-force_HWD_ad} \\
    \dot{\Delta \bm{p}}_s & = - \left[\bm{V}^{\prime}\left(\bm{q}_s^+,\bm{\bar{q}}_b\right) - \bm{V}^{\prime}\left(\bm{q}_s^-,\bm{\bar{q}}_b\right)\right]_s. \label{eq:diff-force_HWD_ad}
\end{align}
Here we have reverted back to forward-backward variables to depict dependence in the potential for the sake of brevity. The potentials are evaluated with the forward/backward position for the system variable and the mean variable for the bath. These equations of motion highlight that the system still has mean and difference trajectories that feel a mean and difference force, but subject to the SP conditions, while the bath only has a mean trajectory that feels a mean force, with the difference trajectory fixed at zero. The action difference becomes, 
\begin{align}
     {S}_t(\bm{z}_{0,s}^{+},\bm{\bar{z}}_{0,b}) - {S}_t(\bm{z}_{0,s}^{-},\bm{\bar{z}}_{0,b})  & = \int_0^{\tau} d\tau \, \left\{\bm{\bar{p}}^T_{\tau,s}\cdot\bm{m}^{-1}_s\cdot\Delta\bm{p}_{\tau,s}\right. \notag \\
& -\left.\left[V\left(\bm{q}_{\tau,s}^+,\bm{\bar{q}}_{\tau,b}\right) - V\left(\bm{q}_{\tau,s}^-,\bm{\bar{q}}_{\tau,b}\right) \right] \right\}.\label{eq:act_pm_HWD_ad}
\end{align}
Using Eqs.~\eqref{eq:mbar_dM} and \eqref{eq:mono_md_eom}, the equations of motion for the monodromy matrix $\mathcal{M}_{md}$ can be broken down into equations of motion for $\bm{\bar{M}}$ and $\bm{\Delta M}$. These can be written as,
\begin{align}
    \dot{\bm{\bar{M}}} & = \mathcal{A}\cdot\bm{\bar{M}} + \mathcal{B}\cdot\bm{\Delta M}, \label{eq:mbar_dot}
\end{align}
and,
\begin{align}
    \bm{\Delta}\dot{\bm{M}} & = \mathcal{A}\cdot\bm{\Delta M} + \frac{1}{4}\mathcal{B}\cdot\bm{\bar{M}}, \label{eq:dm_dot}
\end{align}
respectively. No major simplifications to these can be obtained on imposing the SP conditions, Eq.~\eqref{eq:HWD_ad_sp_conditions}, apart from the potential being evaluated at the mean variable for the bath dofs. Consequently, both matrices $\bm{\bar{M}}$ and $\bm{\Delta M}$ need to be propagated in full. With this information in hand, let us inspect the third stationary phase condition, Eq.~\eqref{eq:HWD_ad_spc3} after imposing Eqs.~\eqref{eq:HWD_ad_sp_conditions},
\begin{align}
    \frac{\partial \phi}{\partial \bm{\Delta z}_{0,b}} & = \bm{\bar{M}}^T_{sb}\cdot\bm{J}^T_s\cdot\left(\bm{z}^{\prime}-\bm{\bar{z}}_t\right)_s  \notag  \\
    & - 2i\bm{\Delta M}^T_{sb}\cdot\bm{\Gamma}_s\cdot\left(\bm{z}^{\prime}-\bm{\bar{z}}_t\right)_s = 0. \label{eq:HWD_ad_spc3_simp}
\end{align}
In general, this condition is not satisfied because $\left(\bm{z}^{\prime}-\bm{\bar{z}}_t\right)_s \neq 0$ and also,  $\bm{\bar{M}}_{sb}$,  $\bm{\Delta M}_{sb} \neq 0$. Our analysis suggests that for a general potential with coupled system-bath dynamics, we \textit{could not} obtain SP conditions that satisfy all three SP equations. This is indicative that for such a potential, the system and bath dynamics are intricately coupled and approximating one without the other is not straightforward. However, it helps to note that for uncoupled system-bath problems, that is, if $V_{sb}(\bm{q}) = 0$ in the propagation of the monodromy matrix elements, it can be shown that $\bm{\bar{M}}_{sb}(t) =  \bm{\Delta M}_{sb}(t)= \bm{\bar{M}}_{bs}(t) =  \bm{\Delta M}_{bs}(t) = 0$, and Eq.~\eqref{eq:HWD_ad_spc3_simp} holds.  Details are provided in Appendix~\ref{ap:HWDbo_mono}. In light of this observation, we invoke a further approximation and set,
\begin{align}
    \bm{\bar{M}}_{sb}(t) =  \bm{\Delta M}_{sb}(t) = \bm{\bar{M}}_{bs}(t) =  \bm{\Delta M}_{bs}(t) = 0. \label{eq:msb_vanish}
\end{align}
This amounts to ignoring the system-bath coupling  $V_{sb}(\bm{q}) = 0$,  \textbf{only} in the propagation of the monodromy matrices, and \textbf{not} in the propagation of the phase space variables. With this approximation, all three SP equations can be satisfied with the SP conditions Eq.~\eqref{eq:HWD_ad_sp_conditions}. 

\subsubsection{Simplifying the Prefactor and SP Hessian}
Next, we focus our attention on simplifying the prefactor $\tilde{\mathcal{C}}_t$. Subject to the SP conditions, Eq.\eqref{eq:HWD_ad_sp_conditions} and vanishing off-diagonal monodromy matrix blocks, Eq.~\eqref{eq:msb_vanish}, the monodromy matrices  $\bm{\bar{M}}$ and $\bm{\Delta M}$ become block diagonal in the system-bath basis. This results in all the matrices appearing in the prefactor, Eq.~\eqref{eq:dhk_pref_md} being block diagonal in the system-bath basis, allowing to break the determinant into determinants of system and bath blocks,
\begin{align}
    \tilde{\mathcal{C}}_t(\bm{\bar{z}}_0,\bm{\Delta z}_{0,s}) = \tilde{\mathcal{C}}_{t,s}(\bm{\bar{z}}_0,\bm{\Delta z}_{0,s}) \tilde{\mathcal{C}}_{t,b}(\bm{\bar{z}}_0,\bm{\Delta z}_{0,s}),
\end{align}
where,
\begin{align}    
\tilde{\mathcal{C}}_{t,x}(\bm{\bar{z}}_0,\Delta\bm{z}_{0,s}) & = \text{det}\left(2\bm{\Gamma}_x\right)^{-1/2}\text{det}\left[\bm{\Gamma}_x\cdot\bm{\bar{M}}_{xx} \right. \notag  \\
  & + 2i\bm{\Gamma}_x\cdot \bm{\Delta M}_{xx}\cdot\bm{J}_x\cdot\bm{\Gamma}_x + 2i\bm{J}_x\cdot\bm{\Delta M}_{xx} \notag  \\
    & - \left. \bm{J}_x\cdot\bm{\bar{M}}_{xx}\cdot\bm{J}_x\cdot\bm{\Gamma}_x  \right]^{1/2},
\end{align}
and $x \in \{s,b\}$. Furthermore, as shown in Appendix~\ref{ap:HWDbo_mono} the vanishing system-bath coupling assumption yields $\bm{\Delta M}_{bb}=0$, which allows the bath prefactor to be simplified to,
\begin{align}    
\tilde{\mathcal{C}}_{t,b}(\bm{\bar{z}}_0,\Delta\bm{z}_{0,s}) & = \text{det}\left(2\bm{\Gamma}_b\right)^{-1/2}\text{det}\left[\bm{\Gamma}_b\cdot\bm{\bar{M}}_{bb} \right. \notag \\
& - \left. \bm{J}_b\cdot\bm{\bar{M}}_{bb}\cdot\bm{J}_b\cdot\bm{\Gamma}_b  \right]^{1/2}.
\end{align}
Lastly, we inspect the Hessian of the phase,
\begin{align}
   \bm{\phi}^{\prime\prime}(\bm{\xi}_b) = \left(\begin{array}{ccc}
    2i\bm{\Gamma}_b & \bm{0} & \bm{J}_b^T \\
    \bm{0} & 2i\bm{\Gamma}_b & \bm{F}_{bb} \\
    \bm{J}_b & \bm{F}_{bb}^T      & \bm{E}_{bb} \\
    \end{array}\right),
\end{align}
with $\bm{F}_{bb} = \bm{J}_b\cdot\bm{\bar{M}}_{bb}$ and $\bm{E}_{bb} = 0$ under the stationary phase conditions and the vanishing coupling approximation. Using  det$(\bm{\bar{M}}_{bb}) = 1$ and the symplecticity condition $\bm{\bar{M}}_{bb}^T\cdot\bm{J}_b\cdot\bm{\bar{M}}_{bb} = \bm{J}_b$, we can write,
\begin{align}
    \text{det}\left[ \frac{1}{2\pi i \hbar} \bm{\phi}^{\prime\prime}\left( \bm{\xi}_{b,sp}\right)\right] &= (2\pi\hbar)^{-6N_b}2^{2N_b} \text{det}\left(\bm{\Gamma}_b\cdot\bm{\bar{M}}_{bb} \right. \notag \\
    & - \left. \bm{J}_b\cdot\bm{\bar{M}}_{bb}\cdot\bm{J_b}\cdot\bm{\Gamma}_b\right),
\end{align}
and note that this Hessian will cancel the determinant in the bath prefactor.

\subsubsection{AHWD Correlation Function}
Assembling all the pieces after performing the SPA along with the vanishing coupling approximation for the monodromy matrix, we obtain the following expression for the AHWD correlation function, 
\begin{align}
    C_{AB}^{\text{AHWD}}(t)  & = \frac{1}{\left(2\pi\hbar\right)^{2N}}  \int d\bm{\bar{z}}_{0} \int d\bm{\Delta z}_{0,s}\,\biggl[\left[\hat{\rho}_{A}\right]_W(\bm{z}) B_W(\bm{z}^{\prime})   \biggr.  \notag \\ 
    & \times g^{*}\left(\bm{z};\bm{\bar{z}}_0,\bm{\Delta z}_0\right)  g\left(\bm{z}^{\prime};\bm{\bar{z}}_t,\bm{\Delta z}_t\right) \tilde{\mathcal{C}}_t(\bm{\bar{z}}_0,\bm{\Delta z}_0)   \notag \\ 
    & \times \biggl.  \text{det}\left[ \frac{1}{2\pi i \hbar} \bm{\phi}^{\prime\prime}\left( \bm{\xi}_{sp}\right)\right]^{-1/2} 
    e^{i\tilde{S}_t(\bm{\bar{z}}_0,\bm{\Delta z}_0)/\hbar}\biggr]_{\bm{\xi}_b = \bm{\xi}_{b,sp}} \notag \\
    & = \frac{1}{\left(2\pi\hbar\right)^{2N_s + N_b}} \int d\bm{z}_{0,s}^{\pm} \int d\bm{\bar{z}}_{0,b} \, \Tilde{{\rho}_{A_s}^{*}}(\bm{z_{0,s}^{\pm}}) \notag \\
    & \times \left[\hat{\rho}_{A_b}\right]_W\left(\bm{\bar{z}}_{0,b}\right)  \Tilde{B_s}(\bm{z_{t,s}^{\pm}}) \left[\hat{B}_b\right]_W\left(\bm{\bar{z}}_{t,b}\right) \notag \\  & \times \tilde{\mathcal{C}}_{t}^{\text{AHWD}}(\bm{z}_{0,s}^{\pm},\bm{\bar{z}}_{0,b})e^{i\tilde{S}_t^{\text{AHWD}}(\bm{z}_{0,s}^{\pm},\bm{\bar{z}}_{0,b})/\hbar} .
\end{align}
Here we have assumed that $\hat{B}\equiv \hat{B}_s \otimes \hat{B}_b$ can be factorized into system and bath operators, and the same for $\hat{\rho}_{A}$. However, this assumption is not required, we have only used it here to neatly factor the terms from both operators into system and bath parts. Moreover, for the sake of clarity we have  switched into forward-backward variables for the system, $\bm{z}_s^{\pm}$. The action becomes, 
\begin{align}
     & \tilde{S}_t^{\text{AHWD}}(\bm{z}_{0,s}^{\pm},\bm{\bar{z}}_{0,b}) \notag \\
     & = - \bm{\bar{p}}^{T}_{t,s}\cdot\bm{\Delta q}_{t,s} + \bm{\bar{p}}^{T}_{0,s}\cdot\bm{\Delta q}_{0,s}  \notag \\
    &   + \int_0^{\tau} d\tau \, \biggl\{ \frac{1}{2}\bm{p}^{+^T}_{\tau,s}\cdot\bm{m}^{-1}_s\cdot\bm{p}_{\tau,s}^{+^T} - \frac{1}{2}\bm{p}^{-^T}_{\tau,s}\cdot\bm{m}^{-1}_s\cdot\bm{p}_{\tau,s}^{-^T}  \biggr. \notag \\
& -\biggl.\left[V\left(\bm{q}_{\tau,s}^+,\bm{\bar{q}}_{\tau,b}\right) - V\left(\bm{q}_{\tau,s}^-,\bm{\bar{q}}_{\tau,b}\right) \right] \biggr\}, \label{eq:act_ahwd}
\end{align}
and the AHWD prefactor is,
\begin{align}
    \tilde{\mathcal{C}}_{t}^{\text{AHWD}}(\bm{z}_{0,s}^{\pm},\bm{\bar{z}}_{0,b}) & =  \text{det}\left[\frac{1}{2}\left(\bm{\Gamma}_s+i\bm{J}_s\right)\cdot\bm{M}_{ss}^+\cdot\left(\mathbb{1}_s+i\bm{J}_s\cdot\bm{\Gamma}_s\right) \right. \notag \\
    & \left. + \frac{1}{2}\left(\bm{\Gamma}_s-i\bm{J}_s\right)\cdot\bm{M}_{ss}^-\cdot\left(\mathbb{1}_s-i\bm{J}_s\cdot\bm{\Gamma}_s\right)\right]^{1/2} \notag \\
    & \times \text{det}\left(2\bm{\Gamma}_s\right)^{-1/2} \label{eq:pref_ahwd}.
\end{align}
When compared to Eq.~\eqref{eq:dhkw_pref}, it is evident that the form of the AHWD prefactor is the same as that of the DHK prefactor for just the system variables. The AHWD prefactor only needs the system-system block of the  monodromy matrix elements to be propagated using the equations of motion,
\begin{align}
    \dot{\bm{M}}^{\pm}_{ss}= \bm{J}_s\cdot\frac{\partial^2 \bar{H}(\bm{z}_s^{\pm},\bm{\bar{z}}_b)}{\partial \bm{z}^{\pm^2}_s}\cdot\bm{M}^{\pm}_{ss}, \label{eq:ahwd_mono_eom}
\end{align}
with $\bm{M}_{ss}^{\pm}(0)=\mathbb{1}$ and following the symplecticity condition
\begin{align}                   
\bm{M}_{ss}^{{\pm}^T}\cdot\bm{J}_{s}\cdot\bm{M}^{\pm}_{ss}=\bm{J}_{s}. \label{eq:ahwd_symp}
\end{align} 
Further details on the correlation function and its structure are presented in the main text.

\section{Equations of Motion for the Monodromy Matrices in AHWD Under Vanishing System-Bath Coupling} \label{ap:HWDbo_mono}

Consider the equations of motion for $\bm{\bar{M}}$ and $\bm{\Delta M}$, as given in Eqs.~\eqref{eq:mbar_dot} and \eqref{eq:dm_dot} respectively. If these are transformed to the system-bath basis, the assumption that the system-bath coupling vanishes, $V_{sb}(\bm{q}) = 0$, makes both $\bm{V}^{\prime\prime}_+$ and $\bm{V}^{\prime\prime}_-$  block diagonal, with the off-diagonal system-bath and bath-system blocks being zero. As a consequence, both $\mathcal{A}$ and $\mathcal{B}$ also become block diagonal: $\mathcal{A}_{sb}=\mathcal{A}_{bs} = 0$ and $\mathcal{B}_{sb}=\mathcal{B}_{bs} = 0$.  The equations of motion for the off-diagonal blocks of $\bm{\bar{M}}$ and $\bm{\Delta M}$ simplify to,
\begin{align}
    \dot{\bm{\bar{M}}}_{sb} & = \mathcal{A}_{ss}\cdot\bm{\bar{M}}_{sb} + \mathcal{B}_{ss}\cdot\bm{\Delta M}_{sb}, \notag \\
& \text{and,} \notag \\
\bm{\Delta}\dot{\bm{M}}_{sb} & = \mathcal{A}_{ss}\cdot\bm{\Delta M}_{sb} + \frac{1}{4}\mathcal{B}_{ss}\cdot\bm{\bar{M}}_{sb}, \label{eq:eom_msb}
\end{align}
with the time-dependence of the system-bath blocks being independent of the system-system, bath-bath and bath-system blocks. Similarly, the time-dependence of the bath-system blocks can be shown to only depend on the bath-system blocks themselves. This point greatly simplifies the dynamics when we note that  $\bm{\bar{M}}_{sb}(0) = \bm{\Delta M}_{sb}(0) =0$ .  Combined with the equation of motion Eq.~\eqref{eq:eom_msb}, this implies that $\bm{\bar{M}}_{sb}(t) = \bm{\Delta M}_{sb}(t)=0$. A similar argument can be made for the bath-system block. Thus, we have shown that $\bm{\bar{M}}$ and $\bm{\Delta M}$ stay block diagonal in the system-bath basis when the system-bath coupling is assumed to vanish in their equations of motion. 
Moreover, under the same assumption $V_{sb}(\bm{q}) = 0$, the bath-bath block of $\bm{V}^{\prime\prime}_-$ becomes, 
\begin{align}
    \frac{\partial^2 }{\partial \bm{\bar{q}}_b^2}\left[V\left(\bm{q}_s^+,\bm{\bar{q}}_b\right) - V\left(\bm{q}_s^-,\bm{\bar{q}}_b\right)\right] & = 0,
\end{align}
yielding,
\begin{align}
    \bm{\bar{M}}_{bb}(t) & = \bm{\bar{M}}_{bb}(0).e^{\mathcal{A}_{bb}t}, 
\end{align}
and, 
\begin{align}
    \bm{\Delta M}_{bb}(t) & = \bm{\Delta M}_{bb}(0).e^{\mathcal{A}_{bb}t}. 
\end{align}
Thus the equations of motion for $ \bm{\bar{M}}_{bb}(t)$ and $\bm{\Delta M}_{bb}(t)$ become independent of  each other. Noting that $\bm{\Delta M}_{bb}(0)= 0$, we get that $\bm{\Delta M}_{bb}(t)=0$, which is a further simplification. For the bath dofs, only $ \bm{\bar{M}}_{bb}(t)$ needs to be propagated,  resulting in det$(\bm{\bar{M}}_{bb}) = 1$ and the symplecticity condition $\bm{\bar{M}}_{bb}^T\cdot\bm{J}_b\cdot\bm{\bar{M}}_{bb} = \bm{J}_b$ . This simplification will be used to cancel the bath-prefactor with the Hessian from the SPA.  

\section{Long Time Accuracy of AHWD} \label{app:HWD_long_time}

\begin{figure}
    \centering
    \includegraphics[width=0.45\textwidth]{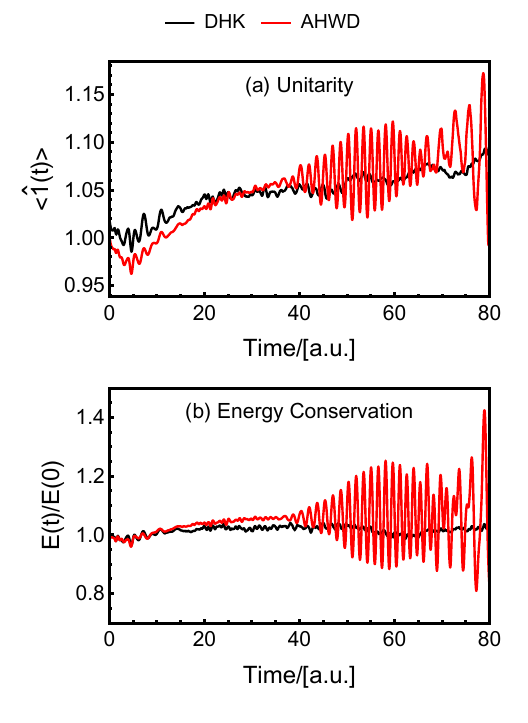}
    \caption{(a) $\langle \hat{1}(t) \rangle$, and (b)   $\langle \hat{E}(t)/\hat{E}(0) \rangle$ plotted as a function of time for model 2 with $k = 2$ a.u., calculated using DHK (black) and HWD (red). Quantities in both panels are expected to stay constant at 1 for all times under exact quantum dynamics.}
    \label{fig:HHB_longtime}
\end{figure}

Apart from the stationary phase approximation, the AHWD method also relies on assuming a vanishing system-bath coupling for the propagation of the monodromy matrices. Since the off-diagonal matrix elements of the monodromy matrix are zero at $t=0$, this assumption is valid then. However, its accuracy gets worse for longer times, especially for problems with strong system-bath coupling. Here, we track the unitarity of the AHWD method by calculating $\langle \hat{\mathbb{1}}(t) \rangle$. AHWD is not excepted to be exactly unitary, as even DHK is known to be only approximately unitary,\cite{Herman1986,Garashchuk1997,Harabati2004,Zhang2005,Tatchen2011}. Fig.~\ref{fig:HHB_longtime}(a) tracks the unitarity of AHWD and DHK as a function of time for model 2 for the strong coupling case, $k=2$ a.u. It is observed that the unitarity of AHWD drifts like that of DHK, but also exhibits oscillations at longer times. Moreover, we also track the expectation value of the energy, $E(t) \equiv \langle \hat{H}(t) \rangle$ for the same model. Since model 2 describes a closed system, the expectation value for the energy should stay constant over time for an exact calculation. However, as shown in Fig.~\ref{fig:HHB_longtime}(b), both DHK and AHWD do not exactly conserve energy, with AHWD predicting large high frequency oscillations in the energy at longer times. 

\bibliographystyle{apsrev4-2}
\bibliography{bibfile}

\providecommand{\noopsort}[1]{}\providecommand{\singleletter}[1]{#1}%
\begin{thebibliography}{147}%
\makeatletter
\providecommand \@ifxundefined [1]{%
 \@ifx{#1\undefined}
}%
\providecommand \@ifnum [1]{%
 \ifnum #1\expandafter \@firstoftwo
 \else \expandafter \@secondoftwo
 \fi
}%
\providecommand \@ifx [1]{%
 \ifx #1\expandafter \@firstoftwo
 \else \expandafter \@secondoftwo
 \fi
}%
\providecommand \natexlab [1]{#1}%
\providecommand \enquote  [1]{``#1''}%
\providecommand \bibnamefont  [1]{#1}%
\providecommand \bibfnamefont [1]{#1}%
\providecommand \citenamefont [1]{#1}%
\providecommand \href@noop [0]{\@secondoftwo}%
\providecommand \href [0]{\begingroup \@sanitize@url \@href}%
\providecommand \@href[1]{\@@startlink{#1}\@@href}%
\providecommand \@@href[1]{\endgroup#1\@@endlink}%
\providecommand \@sanitize@url [0]{\catcode `\\12\catcode `\$12\catcode
  `\&12\catcode `\#12\catcode `\^12\catcode `\_12\catcode `\%12\relax}%
\providecommand \@@startlink[1]{}%
\providecommand \@@endlink[0]{}%
\providecommand \url  [0]{\begingroup\@sanitize@url \@url }%
\providecommand \@url [1]{\endgroup\@href {#1}{\urlprefix }}%
\providecommand \urlprefix  [0]{URL }%
\providecommand \Eprint [0]{\href }%
\providecommand \doibase [0]{https://doi.org/}%
\providecommand \selectlanguage [0]{\@gobble}%
\providecommand \bibinfo  [0]{\@secondoftwo}%
\providecommand \bibfield  [0]{\@secondoftwo}%
\providecommand \translation [1]{[#1]}%
\providecommand \BibitemOpen [0]{}%
\providecommand \bibitemStop [0]{}%
\providecommand \bibitemNoStop [0]{.\EOS\space}%
\providecommand \EOS [0]{\spacefactor3000\relax}%
\providecommand \BibitemShut  [1]{\csname bibitem#1\endcsname}%
\let\auto@bib@innerbib\@empty
\bibitem [{\citenamefont {Habershon}\ \emph {et~al.}(2013)\citenamefont
  {Habershon}, \citenamefont {Manolopoulos}, \citenamefont {Markland},\ and\
  \citenamefont {Miller}}]{Habershon2013}%
  \BibitemOpen
  \bibfield  {author} {\bibinfo {author} {\bibfnamefont {S.}~\bibnamefont
  {Habershon}}, \bibinfo {author} {\bibfnamefont {D.~E.}\ \bibnamefont
  {Manolopoulos}}, \bibinfo {author} {\bibfnamefont {T.~E.}\ \bibnamefont
  {Markland}},\ and\ \bibinfo {author} {\bibfnamefont {T.~F.}\ \bibnamefont
  {Miller}},\ }\href {https://doi.org/10.1146/annurev-physchem-040412-110122}
  {\bibfield  {journal} {\bibinfo  {journal} {Annual Review of Physical
  Chemistry}\ }\textbf {\bibinfo {volume} {64}},\ \bibinfo {pages} {387}
  (\bibinfo {year} {2013})}\BibitemShut {NoStop}%
\bibitem [{\citenamefont {Suleimanov}\ \emph {et~al.}(2016)\citenamefont
  {Suleimanov}, \citenamefont {Aoiz},\ and\ \citenamefont
  {Guo}}]{Suleimanov2016a}%
  \BibitemOpen
  \bibfield  {author} {\bibinfo {author} {\bibfnamefont {Y.~V.}\ \bibnamefont
  {Suleimanov}}, \bibinfo {author} {\bibfnamefont {F.~J.}\ \bibnamefont
  {Aoiz}},\ and\ \bibinfo {author} {\bibfnamefont {H.}~\bibnamefont {Guo}},\
  }\href {https://doi.org/10.1021/acs.jpca.6b07140} {\bibfield  {journal}
  {\bibinfo  {journal} {The Journal of Physical Chemistry A}\ }\textbf
  {\bibinfo {volume} {120}},\ \bibinfo {pages} {8488} (\bibinfo {year}
  {2016})}\BibitemShut {NoStop}%
\bibitem [{\citenamefont {Lawrence}\ and\ \citenamefont
  {Manolopoulos}(2020)}]{Lawrence2020}%
  \BibitemOpen
  \bibfield  {author} {\bibinfo {author} {\bibfnamefont {J.~E.}\ \bibnamefont
  {Lawrence}}\ and\ \bibinfo {author} {\bibfnamefont {D.~E.}\ \bibnamefont
  {Manolopoulos}},\ }\href {https://doi.org/10.1039/C9FD00084D} {\bibfield
  {journal} {\bibinfo  {journal} {Faraday Discussions}\ }\textbf {\bibinfo
  {volume} {221}},\ \bibinfo {pages} {9} (\bibinfo {year} {2020})}\BibitemShut
  {NoStop}%
\bibitem [{\citenamefont {Vendrell}\ \emph
  {et~al.}(2007{\natexlab{a}})\citenamefont {Vendrell}, \citenamefont {Gatti},\
  and\ \citenamefont {Meyer}}]{Vendrell2007a}%
  \BibitemOpen
  \bibfield  {author} {\bibinfo {author} {\bibfnamefont {O.}~\bibnamefont
  {Vendrell}}, \bibinfo {author} {\bibfnamefont {F.}~\bibnamefont {Gatti}},\
  and\ \bibinfo {author} {\bibfnamefont {H.-D.}\ \bibnamefont {Meyer}},\ }\href
  {https://doi.org/10.1002/anie.200702201} {\bibfield  {journal} {\bibinfo
  {journal} {Angewandte Chemie (International ed. in English)}\ }\textbf
  {\bibinfo {volume} {46}},\ \bibinfo {pages} {6918} (\bibinfo {year}
  {2007}{\natexlab{a}})}\BibitemShut {NoStop}%
\bibitem [{\citenamefont {Vendrell}\ \emph
  {et~al.}(2007{\natexlab{b}})\citenamefont {Vendrell}, \citenamefont {Gatti},
  \citenamefont {Lauvergnat},\ and\ \citenamefont {Meyer}}]{Vendrell2007b}%
  \BibitemOpen
  \bibfield  {author} {\bibinfo {author} {\bibfnamefont {O.}~\bibnamefont
  {Vendrell}}, \bibinfo {author} {\bibfnamefont {F.}~\bibnamefont {Gatti}},
  \bibinfo {author} {\bibfnamefont {D.}~\bibnamefont {Lauvergnat}},\ and\
  \bibinfo {author} {\bibfnamefont {H.-D.}\ \bibnamefont {Meyer}},\ }\href
  {https://doi.org/10.1063/1.2787588} {\bibfield  {journal} {\bibinfo
  {journal} {The Journal of Chemical Physics}\ }\textbf {\bibinfo {volume}
  {127}},\ \bibinfo {pages} {184302} (\bibinfo {year}
  {2007}{\natexlab{b}})}\BibitemShut {NoStop}%
\bibitem [{\citenamefont {Vendrell}\ \emph
  {et~al.}(2007{\natexlab{c}})\citenamefont {Vendrell}, \citenamefont {Gatti},\
  and\ \citenamefont {Meyer}}]{Vendrell2007c}%
  \BibitemOpen
  \bibfield  {author} {\bibinfo {author} {\bibfnamefont {O.}~\bibnamefont
  {Vendrell}}, \bibinfo {author} {\bibfnamefont {F.}~\bibnamefont {Gatti}},\
  and\ \bibinfo {author} {\bibfnamefont {H.-D.}\ \bibnamefont {Meyer}},\ }\href
  {https://doi.org/10.1063/1.2787596} {\bibfield  {journal} {\bibinfo
  {journal} {The Journal of Chemical Physics}\ }\textbf {\bibinfo {volume}
  {127}},\ \bibinfo {pages} {244301} (\bibinfo {year}
  {2007}{\natexlab{c}})}\BibitemShut {NoStop}%
\bibitem [{\citenamefont {Yu}\ and\ \citenamefont {Bowman}(2017)}]{Yu2017}%
  \BibitemOpen
  \bibfield  {author} {\bibinfo {author} {\bibfnamefont {Q.}~\bibnamefont
  {Yu}}\ and\ \bibinfo {author} {\bibfnamefont {J.~M.}\ \bibnamefont
  {Bowman}},\ }\href {https://doi.org/10.1021/jacs.7b05459} {\bibfield
  {journal} {\bibinfo  {journal} {Journal of the American Chemical Society}\
  }\textbf {\bibinfo {volume} {139}},\ \bibinfo {pages} {10984} (\bibinfo
  {year} {2017})}\BibitemShut {NoStop}%
\bibitem [{\citenamefont {Schröder}\ \emph {et~al.}(2022)\citenamefont
  {Schröder}, \citenamefont {Gatti}, \citenamefont {Lauvergnat}, \citenamefont
  {Meyer},\ and\ \citenamefont {Vendrell}}]{Schroder2022}%
  \BibitemOpen
  \bibfield  {author} {\bibinfo {author} {\bibfnamefont {M.}~\bibnamefont
  {Schröder}}, \bibinfo {author} {\bibfnamefont {F.}~\bibnamefont {Gatti}},
  \bibinfo {author} {\bibfnamefont {D.}~\bibnamefont {Lauvergnat}}, \bibinfo
  {author} {\bibfnamefont {H.-D.}\ \bibnamefont {Meyer}},\ and\ \bibinfo
  {author} {\bibfnamefont {O.}~\bibnamefont {Vendrell}},\ }\href
  {https://doi.org/10.1038/s41467-022-33650-w} {\bibfield  {journal} {\bibinfo
  {journal} {Nature Communications}\ }\textbf {\bibinfo {volume} {13}},\
  \bibinfo {pages} {6170} (\bibinfo {year} {2022})}\BibitemShut {NoStop}%
\bibitem [{\citenamefont {Conte}\ and\ \citenamefont
  {Ceotto}(2020)}]{Conte2020c}%
  \BibitemOpen
  \bibfield  {author} {\bibinfo {author} {\bibfnamefont {R.}~\bibnamefont
  {Conte}}\ and\ \bibinfo {author} {\bibfnamefont {M.}~\bibnamefont {Ceotto}},\
  }\bibinfo {title} {Semiclassical molecular dynamics for spectroscopic
  calculations}\ (\bibinfo  {publisher} {Wiley},\ \bibinfo {year} {2020})\ pp.\
  \bibinfo {pages} {595--628}\BibitemShut {NoStop}%
\bibitem [{\citenamefont {Gabas}\ \emph {et~al.}(2020)\citenamefont {Gabas},
  \citenamefont {Conte},\ and\ \citenamefont {Ceotto}}]{Gabas2020}%
  \BibitemOpen
  \bibfield  {author} {\bibinfo {author} {\bibfnamefont {F.}~\bibnamefont
  {Gabas}}, \bibinfo {author} {\bibfnamefont {R.}~\bibnamefont {Conte}},\ and\
  \bibinfo {author} {\bibfnamefont {M.}~\bibnamefont {Ceotto}},\ }\href
  {https://doi.org/10.1021/acs.jctc.0c00127} {\bibfield  {journal} {\bibinfo
  {journal} {Journal of Chemical Theory and Computation}\ }\textbf {\bibinfo
  {volume} {16}},\ \bibinfo {pages} {3476} (\bibinfo {year}
  {2020})}\BibitemShut {NoStop}%
\bibitem [{\citenamefont {Botti}\ \emph {et~al.}(2022)\citenamefont {Botti},
  \citenamefont {Aieta},\ and\ \citenamefont {Conte}}]{Botti2022}%
  \BibitemOpen
  \bibfield  {author} {\bibinfo {author} {\bibfnamefont {G.}~\bibnamefont
  {Botti}}, \bibinfo {author} {\bibfnamefont {C.}~\bibnamefont {Aieta}},\ and\
  \bibinfo {author} {\bibfnamefont {R.}~\bibnamefont {Conte}},\ }\href
  {https://doi.org/10.1063/5.0089720} {\bibfield  {journal} {\bibinfo
  {journal} {The Journal of Chemical Physics}\ }\textbf {\bibinfo {volume}
  {156}},\ \bibinfo {pages} {164303} (\bibinfo {year} {2022})}\BibitemShut
  {NoStop}%
\bibitem [{\citenamefont {Botti}\ \emph {et~al.}(2023)\citenamefont {Botti},
  \citenamefont {Ceotto},\ and\ \citenamefont {Conte}}]{Botti2023}%
  \BibitemOpen
  \bibfield  {author} {\bibinfo {author} {\bibfnamefont {G.}~\bibnamefont
  {Botti}}, \bibinfo {author} {\bibfnamefont {M.}~\bibnamefont {Ceotto}},\ and\
  \bibinfo {author} {\bibfnamefont {R.}~\bibnamefont {Conte}},\ }\href
  {https://doi.org/10.1021/acs.jpclett.3c02073} {\bibfield  {journal} {\bibinfo
   {journal} {The Journal of Physical Chemistry Letters}\ }\textbf {\bibinfo
  {volume} {14}},\ \bibinfo {pages} {8940} (\bibinfo {year}
  {2023})}\BibitemShut {NoStop}%
\bibitem [{\citenamefont {Moscato}\ \emph {et~al.}(2023)\citenamefont
  {Moscato}, \citenamefont {Gabas}, \citenamefont {Conte},\ and\ \citenamefont
  {Ceotto}}]{Moscato2023}%
  \BibitemOpen
  \bibfield  {author} {\bibinfo {author} {\bibfnamefont {D.}~\bibnamefont
  {Moscato}}, \bibinfo {author} {\bibfnamefont {F.}~\bibnamefont {Gabas}},
  \bibinfo {author} {\bibfnamefont {R.}~\bibnamefont {Conte}},\ and\ \bibinfo
  {author} {\bibfnamefont {M.}~\bibnamefont {Ceotto}},\ }\href
  {https://doi.org/10.1080/07391102.2023.2180435} {\bibfield  {journal}
  {\bibinfo  {journal} {Journal of Biomolecular Structure and Dynamics}\
  }\textbf {\bibinfo {volume} {41}},\ \bibinfo {pages} {14248} (\bibinfo {year}
  {2023})}\BibitemShut {NoStop}%
\bibitem [{\citenamefont {Moscato}\ \emph {et~al.}(2024)\citenamefont
  {Moscato}, \citenamefont {Mandelli}, \citenamefont {Bondanza}, \citenamefont
  {Lipparini}, \citenamefont {Conte}, \citenamefont {Mennucci},\ and\
  \citenamefont {Ceotto}}]{Moscato2024}%
  \BibitemOpen
  \bibfield  {author} {\bibinfo {author} {\bibfnamefont {D.}~\bibnamefont
  {Moscato}}, \bibinfo {author} {\bibfnamefont {G.}~\bibnamefont {Mandelli}},
  \bibinfo {author} {\bibfnamefont {M.}~\bibnamefont {Bondanza}}, \bibinfo
  {author} {\bibfnamefont {F.}~\bibnamefont {Lipparini}}, \bibinfo {author}
  {\bibfnamefont {R.}~\bibnamefont {Conte}}, \bibinfo {author} {\bibfnamefont
  {B.}~\bibnamefont {Mennucci}},\ and\ \bibinfo {author} {\bibfnamefont
  {M.}~\bibnamefont {Ceotto}},\ }\href {https://doi.org/10.1021/jacs.3c12700}
  {\bibfield  {journal} {\bibinfo  {journal} {Journal of the American Chemical
  Society}\ }\textbf {\bibinfo {volume} {146}},\ \bibinfo {pages} {8179}
  (\bibinfo {year} {2024})}\BibitemShut {NoStop}%
\bibitem [{\citenamefont {Benson}\ \emph {et~al.}(2020)\citenamefont {Benson},
  \citenamefont {Trenins},\ and\ \citenamefont {Althorpe}}]{Benson2020}%
  \BibitemOpen
  \bibfield  {author} {\bibinfo {author} {\bibfnamefont {R.~L.}\ \bibnamefont
  {Benson}}, \bibinfo {author} {\bibfnamefont {G.}~\bibnamefont {Trenins}},\
  and\ \bibinfo {author} {\bibfnamefont {S.~C.}\ \bibnamefont {Althorpe}},\
  }\href {https://doi.org/10.1039/C9FD00077A} {\bibfield  {journal} {\bibinfo
  {journal} {Faraday Discuss.}\ }\textbf {\bibinfo {volume} {221}},\ \bibinfo
  {pages} {350} (\bibinfo {year} {2020})}\BibitemShut {NoStop}%
\bibitem [{\citenamefont {Althorpe}(2021)}]{Althorpe2021a}%
  \BibitemOpen
  \bibfield  {author} {\bibinfo {author} {\bibfnamefont {S.~C.}\ \bibnamefont
  {Althorpe}},\ }\href {https://doi.org/10.1140/epjb/s10051-021-00155-2}
  {\bibfield  {journal} {\bibinfo  {journal} {The European Physical Journal B}\
  }\textbf {\bibinfo {volume} {94}},\ \bibinfo {pages} {155} (\bibinfo {year}
  {2021})}\BibitemShut {NoStop}%
\bibitem [{\citenamefont {Suleimanov}(2012)}]{Suleimanov2012}%
  \BibitemOpen
  \bibfield  {author} {\bibinfo {author} {\bibfnamefont {Y.~V.}\ \bibnamefont
  {Suleimanov}},\ }\href {https://doi.org/10.1021/jp302453z} {\bibfield
  {journal} {\bibinfo  {journal} {The Journal of Physical Chemistry C}\
  }\textbf {\bibinfo {volume} {116}},\ \bibinfo {pages} {11141} (\bibinfo
  {year} {2012})}\BibitemShut {NoStop}%
\bibitem [{\citenamefont {Liu}\ \emph {et~al.}(2019)\citenamefont {Liu},
  \citenamefont {Zhang}, \citenamefont {Li},\ and\ \citenamefont
  {Jiang}}]{Liu2019}%
  \BibitemOpen
  \bibfield  {author} {\bibinfo {author} {\bibfnamefont {Q.}~\bibnamefont
  {Liu}}, \bibinfo {author} {\bibfnamefont {L.}~\bibnamefont {Zhang}}, \bibinfo
  {author} {\bibfnamefont {Y.}~\bibnamefont {Li}},\ and\ \bibinfo {author}
  {\bibfnamefont {B.}~\bibnamefont {Jiang}},\ }\href
  {https://doi.org/10.1021/acs.jpclett.9b02570} {\bibfield  {journal} {\bibinfo
   {journal} {The Journal of Physical Chemistry Letters}\ }\textbf {\bibinfo
  {volume} {10}},\ \bibinfo {pages} {7475} (\bibinfo {year}
  {2019})}\BibitemShut {NoStop}%
\bibitem [{\citenamefont {Jiang}\ \emph {et~al.}(2021)\citenamefont {Jiang},
  \citenamefont {Tao}, \citenamefont {Kammler}, \citenamefont {Ding},
  \citenamefont {Wodtke}, \citenamefont {Kandratsenka}, \citenamefont
  {Miller},\ and\ \citenamefont {Bünermann}}]{Jiang2021}%
  \BibitemOpen
  \bibfield  {author} {\bibinfo {author} {\bibfnamefont {H.}~\bibnamefont
  {Jiang}}, \bibinfo {author} {\bibfnamefont {X.}~\bibnamefont {Tao}}, \bibinfo
  {author} {\bibfnamefont {M.}~\bibnamefont {Kammler}}, \bibinfo {author}
  {\bibfnamefont {F.}~\bibnamefont {Ding}}, \bibinfo {author} {\bibfnamefont
  {A.~M.}\ \bibnamefont {Wodtke}}, \bibinfo {author} {\bibfnamefont
  {A.}~\bibnamefont {Kandratsenka}}, \bibinfo {author} {\bibfnamefont {T.~F.}\
  \bibnamefont {Miller}},\ and\ \bibinfo {author} {\bibfnamefont
  {O.}~\bibnamefont {Bünermann}},\ }\href
  {https://doi.org/10.1021/acs.jpclett.0c02933} {\bibfield  {journal} {\bibinfo
   {journal} {The Journal of Physical Chemistry Letters}\ }\textbf {\bibinfo
  {volume} {12}},\ \bibinfo {pages} {1991} (\bibinfo {year}
  {2021})}\BibitemShut {NoStop}%
\bibitem [{\citenamefont {Zhou}\ \emph {et~al.}(2022)\citenamefont {Zhou},
  \citenamefont {Meng}, \citenamefont {Guo},\ and\ \citenamefont
  {Jiang}}]{Zhou2022}%
  \BibitemOpen
  \bibfield  {author} {\bibinfo {author} {\bibfnamefont {X.}~\bibnamefont
  {Zhou}}, \bibinfo {author} {\bibfnamefont {G.}~\bibnamefont {Meng}}, \bibinfo
  {author} {\bibfnamefont {H.}~\bibnamefont {Guo}},\ and\ \bibinfo {author}
  {\bibfnamefont {B.}~\bibnamefont {Jiang}},\ }\href
  {https://doi.org/10.1021/acs.jpclett.2c00593} {\bibfield  {journal} {\bibinfo
   {journal} {The Journal of Physical Chemistry Letters}\ }\textbf {\bibinfo
  {volume} {13}},\ \bibinfo {pages} {3450} (\bibinfo {year}
  {2022})}\BibitemShut {NoStop}%
\bibitem [{\citenamefont {Shi}\ \emph {et~al.}(2023)\citenamefont {Shi},
  \citenamefont {Schröder}, \citenamefont {Meyer}, \citenamefont {Peláez},
  \citenamefont {Wodtke}, \citenamefont {Golibrzuch}, \citenamefont
  {Schönemann}, \citenamefont {Kandratsenka},\ and\ \citenamefont
  {Gatti}}]{Shi2023}%
  \BibitemOpen
  \bibfield  {author} {\bibinfo {author} {\bibfnamefont {L.}~\bibnamefont
  {Shi}}, \bibinfo {author} {\bibfnamefont {M.}~\bibnamefont {Schröder}},
  \bibinfo {author} {\bibfnamefont {H.-D.}\ \bibnamefont {Meyer}}, \bibinfo
  {author} {\bibfnamefont {D.}~\bibnamefont {Peláez}}, \bibinfo {author}
  {\bibfnamefont {A.~M.}\ \bibnamefont {Wodtke}}, \bibinfo {author}
  {\bibfnamefont {K.}~\bibnamefont {Golibrzuch}}, \bibinfo {author}
  {\bibfnamefont {A.-M.}\ \bibnamefont {Schönemann}}, \bibinfo {author}
  {\bibfnamefont {A.}~\bibnamefont {Kandratsenka}},\ and\ \bibinfo {author}
  {\bibfnamefont {F.}~\bibnamefont {Gatti}},\ }\href
  {https://doi.org/10.1063/5.0176655} {\bibfield  {journal} {\bibinfo
  {journal} {The Journal of Chemical Physics}\ }\textbf {\bibinfo {volume}
  {159}},\ \bibinfo {pages} {194102} (\bibinfo {year} {2023})}\BibitemShut
  {NoStop}%
\bibitem [{\citenamefont {Zhang}\ \emph {et~al.}(2020)\citenamefont {Zhang},
  \citenamefont {Borrelli},\ and\ \citenamefont {Tanimura}}]{Zhang2020}%
  \BibitemOpen
  \bibfield  {author} {\bibinfo {author} {\bibfnamefont {J.}~\bibnamefont
  {Zhang}}, \bibinfo {author} {\bibfnamefont {R.}~\bibnamefont {Borrelli}},\
  and\ \bibinfo {author} {\bibfnamefont {Y.}~\bibnamefont {Tanimura}},\ }\href
  {https://doi.org/10.1063/5.0010580} {\bibfield  {journal} {\bibinfo
  {journal} {The Journal of Chemical Physics}\ }\textbf {\bibinfo {volume}
  {152}},\ \bibinfo {pages} {214114} (\bibinfo {year} {2020})}\BibitemShut
  {NoStop}%
\bibitem [{\citenamefont {Liu}\ \emph {et~al.}(2021)\citenamefont {Liu},
  \citenamefont {Yan}, \citenamefont {Xing},\ and\ \citenamefont
  {Shi}}]{Liu2021}%
  \BibitemOpen
  \bibfield  {author} {\bibinfo {author} {\bibfnamefont {Y.}~\bibnamefont
  {Liu}}, \bibinfo {author} {\bibfnamefont {Y.}~\bibnamefont {Yan}}, \bibinfo
  {author} {\bibfnamefont {T.}~\bibnamefont {Xing}},\ and\ \bibinfo {author}
  {\bibfnamefont {Q.}~\bibnamefont {Shi}},\ }\href
  {https://doi.org/10.1021/acs.jpcb.1c02851} {\bibfield  {journal} {\bibinfo
  {journal} {The Journal of Physical Chemistry B}\ }\textbf {\bibinfo {volume}
  {125}},\ \bibinfo {pages} {5959} (\bibinfo {year} {2021})}\BibitemShut
  {NoStop}%
\bibitem [{\citenamefont {Slocombe}\ \emph {et~al.}(2022)\citenamefont
  {Slocombe}, \citenamefont {Sacchi},\ and\ \citenamefont
  {Al-Khalili}}]{Slocombe2022}%
  \BibitemOpen
  \bibfield  {author} {\bibinfo {author} {\bibfnamefont {L.}~\bibnamefont
  {Slocombe}}, \bibinfo {author} {\bibfnamefont {M.}~\bibnamefont {Sacchi}},\
  and\ \bibinfo {author} {\bibfnamefont {J.}~\bibnamefont {Al-Khalili}},\
  }\href {https://doi.org/10.1038/s42005-022-00881-8} {\bibfield  {journal}
  {\bibinfo  {journal} {Communications Physics}\ }\textbf {\bibinfo {volume}
  {5}},\ \bibinfo {pages} {109} (\bibinfo {year} {2022})}\BibitemShut {NoStop}%
\bibitem [{\citenamefont {Buchholz}\ \emph {et~al.}(2012)\citenamefont
  {Buchholz}, \citenamefont {Goletz}, \citenamefont {Grossmann}, \citenamefont
  {Schmidt}, \citenamefont {Heyda},\ and\ \citenamefont
  {Jungwirth}}]{Buchholz2012}%
  \BibitemOpen
  \bibfield  {author} {\bibinfo {author} {\bibfnamefont {M.}~\bibnamefont
  {Buchholz}}, \bibinfo {author} {\bibfnamefont {C.-M.}\ \bibnamefont
  {Goletz}}, \bibinfo {author} {\bibfnamefont {F.}~\bibnamefont {Grossmann}},
  \bibinfo {author} {\bibfnamefont {B.}~\bibnamefont {Schmidt}}, \bibinfo
  {author} {\bibfnamefont {J.}~\bibnamefont {Heyda}},\ and\ \bibinfo {author}
  {\bibfnamefont {P.}~\bibnamefont {Jungwirth}},\ }\href
  {https://doi.org/10.1021/jp305084f} {\bibfield  {journal} {\bibinfo
  {journal} {The Journal of Physical Chemistry A}\ }\textbf {\bibinfo {volume}
  {116}},\ \bibinfo {pages} {11199} (\bibinfo {year} {2012})}\BibitemShut
  {NoStop}%
\bibitem [{\citenamefont {Joutsuka}\ \emph {et~al.}(2016)\citenamefont
  {Joutsuka}, \citenamefont {Thompson},\ and\ \citenamefont
  {Laage}}]{Joutsuka2016}%
  \BibitemOpen
  \bibfield  {author} {\bibinfo {author} {\bibfnamefont {T.}~\bibnamefont
  {Joutsuka}}, \bibinfo {author} {\bibfnamefont {W.~H.}\ \bibnamefont
  {Thompson}},\ and\ \bibinfo {author} {\bibfnamefont {D.}~\bibnamefont
  {Laage}},\ }\href {https://doi.org/10.1021/acs.jpclett.5b02637} {\bibfield
  {journal} {\bibinfo  {journal} {The Journal of Physical Chemistry Letters}\
  }\textbf {\bibinfo {volume} {7}},\ \bibinfo {pages} {616} (\bibinfo {year}
  {2016})}\BibitemShut {NoStop}%
\bibitem [{\citenamefont {Qiang}\ \emph {et~al.}(2024)\citenamefont {Qiang},
  \citenamefont {Zhou}, \citenamefont {Peng}, \citenamefont {Yu}, \citenamefont
  {Lu}, \citenamefont {Pan}, \citenamefont {Lu}, \citenamefont {Chen},
  \citenamefont {Lu}, \citenamefont {Zhang},\ and\ \citenamefont
  {Wu}}]{Qiang2024}%
  \BibitemOpen
  \bibfield  {author} {\bibinfo {author} {\bibfnamefont {J.}~\bibnamefont
  {Qiang}}, \bibinfo {author} {\bibfnamefont {L.}~\bibnamefont {Zhou}},
  \bibinfo {author} {\bibfnamefont {Y.}~\bibnamefont {Peng}}, \bibinfo {author}
  {\bibfnamefont {C.}~\bibnamefont {Yu}}, \bibinfo {author} {\bibfnamefont
  {P.}~\bibnamefont {Lu}}, \bibinfo {author} {\bibfnamefont {S.}~\bibnamefont
  {Pan}}, \bibinfo {author} {\bibfnamefont {C.}~\bibnamefont {Lu}}, \bibinfo
  {author} {\bibfnamefont {G.}~\bibnamefont {Chen}}, \bibinfo {author}
  {\bibfnamefont {R.}~\bibnamefont {Lu}}, \bibinfo {author} {\bibfnamefont
  {W.}~\bibnamefont {Zhang}},\ and\ \bibinfo {author} {\bibfnamefont
  {J.}~\bibnamefont {Wu}},\ }\href
  {https://doi.org/10.1103/PhysRevLett.132.103201} {\bibfield  {journal}
  {\bibinfo  {journal} {Physical Review Letters}\ }\textbf {\bibinfo {volume}
  {132}},\ \bibinfo {pages} {103201} (\bibinfo {year} {2024})}\BibitemShut
  {NoStop}%
\bibitem [{\citenamefont {Triana}\ \emph {et~al.}(2020)\citenamefont {Triana},
  \citenamefont {Hernández},\ and\ \citenamefont {Herrera}}]{Triana2020}%
  \BibitemOpen
  \bibfield  {author} {\bibinfo {author} {\bibfnamefont {J.~F.}\ \bibnamefont
  {Triana}}, \bibinfo {author} {\bibfnamefont {F.~J.}\ \bibnamefont
  {Hernández}},\ and\ \bibinfo {author} {\bibfnamefont {F.}~\bibnamefont
  {Herrera}},\ }\href {https://doi.org/10.1063/5.0009869} {\bibfield  {journal}
  {\bibinfo  {journal} {The Journal of Chemical Physics}\ }\textbf {\bibinfo
  {volume} {152}},\ \bibinfo {pages} {234111} (\bibinfo {year}
  {2020})}\BibitemShut {NoStop}%
\bibitem [{\citenamefont {Fischer}\ and\ \citenamefont
  {Saalfrank}(2021)}]{Fischer2021}%
  \BibitemOpen
  \bibfield  {author} {\bibinfo {author} {\bibfnamefont {E.~W.}\ \bibnamefont
  {Fischer}}\ and\ \bibinfo {author} {\bibfnamefont {P.}~\bibnamefont
  {Saalfrank}},\ }\href {https://doi.org/10.1063/5.0040853} {\bibfield
  {journal} {\bibinfo  {journal} {The Journal of Chemical Physics}\ }\textbf
  {\bibinfo {volume} {154}},\ \bibinfo {pages} {104311} (\bibinfo {year}
  {2021})}\BibitemShut {NoStop}%
\bibitem [{\citenamefont {Yang}\ and\ \citenamefont {Cao}(2021)}]{Yang2021}%
  \BibitemOpen
  \bibfield  {author} {\bibinfo {author} {\bibfnamefont {P.-Y.}\ \bibnamefont
  {Yang}}\ and\ \bibinfo {author} {\bibfnamefont {J.}~\bibnamefont {Cao}},\
  }\href {https://doi.org/10.1021/acs.jpclett.1c02210} {\bibfield  {journal}
  {\bibinfo  {journal} {The Journal of Physical Chemistry Letters}\ }\textbf
  {\bibinfo {volume} {12}},\ \bibinfo {pages} {9531} (\bibinfo {year}
  {2021})}\BibitemShut {NoStop}%
\bibitem [{\citenamefont {Li}\ \emph {et~al.}(2022)\citenamefont {Li},
  \citenamefont {Nitzan}, \citenamefont {Hammes-Schiffer},\ and\ \citenamefont
  {Subotnik}}]{Li2022b}%
  \BibitemOpen
  \bibfield  {author} {\bibinfo {author} {\bibfnamefont {T.~E.}\ \bibnamefont
  {Li}}, \bibinfo {author} {\bibfnamefont {A.}~\bibnamefont {Nitzan}}, \bibinfo
  {author} {\bibfnamefont {S.}~\bibnamefont {Hammes-Schiffer}},\ and\ \bibinfo
  {author} {\bibfnamefont {J.~E.}\ \bibnamefont {Subotnik}},\ }\href
  {https://doi.org/10.1021/acs.jpclett.2c00613} {\bibfield  {journal} {\bibinfo
   {journal} {The Journal of Physical Chemistry Letters}\ }\textbf {\bibinfo
  {volume} {13}},\ \bibinfo {pages} {3890} (\bibinfo {year}
  {2022})}\BibitemShut {NoStop}%
\bibitem [{\citenamefont {Mondal}\ \emph {et~al.}(2022)\citenamefont {Mondal},
  \citenamefont {Wang},\ and\ \citenamefont {Keshavamurthy}}]{Mondal2022}%
  \BibitemOpen
  \bibfield  {author} {\bibinfo {author} {\bibfnamefont {S.}~\bibnamefont
  {Mondal}}, \bibinfo {author} {\bibfnamefont {D.~S.}\ \bibnamefont {Wang}},\
  and\ \bibinfo {author} {\bibfnamefont {S.}~\bibnamefont {Keshavamurthy}},\
  }\href {https://doi.org/10.1063/5.0124085} {\bibfield  {journal} {\bibinfo
  {journal} {The Journal of Chemical Physics}\ }\textbf {\bibinfo {volume}
  {157}},\ \bibinfo {pages} {244109} (\bibinfo {year} {2022})}\BibitemShut
  {NoStop}%
\bibitem [{\citenamefont {Lindoy}\ \emph {et~al.}(2023)\citenamefont {Lindoy},
  \citenamefont {Mandal},\ and\ \citenamefont {Reichman}}]{Lindoy2023}%
  \BibitemOpen
  \bibfield  {author} {\bibinfo {author} {\bibfnamefont {L.~P.}\ \bibnamefont
  {Lindoy}}, \bibinfo {author} {\bibfnamefont {A.}~\bibnamefont {Mandal}},\
  and\ \bibinfo {author} {\bibfnamefont {D.~R.}\ \bibnamefont {Reichman}},\
  }\href {https://doi.org/10.1038/s41467-023-38368-x} {\bibfield  {journal}
  {\bibinfo  {journal} {Nat. Commun}\ }\textbf {\bibinfo {volume} {14}},\
  \bibinfo {pages} {2733} (\bibinfo {year} {2023})}\BibitemShut {NoStop}%
\bibitem [{\citenamefont {Lieberherr}\ \emph {et~al.}(2023)\citenamefont
  {Lieberherr}, \citenamefont {Furniss}, \citenamefont {Lawrence},\ and\
  \citenamefont {Manolopoulos}}]{Lieberherr2023}%
  \BibitemOpen
  \bibfield  {author} {\bibinfo {author} {\bibfnamefont {A.~Z.}\ \bibnamefont
  {Lieberherr}}, \bibinfo {author} {\bibfnamefont {S.~T.~E.}\ \bibnamefont
  {Furniss}}, \bibinfo {author} {\bibfnamefont {J.~E.}\ \bibnamefont
  {Lawrence}},\ and\ \bibinfo {author} {\bibfnamefont {D.~E.}\ \bibnamefont
  {Manolopoulos}},\ }\href {https://doi.org/10.1063/5.0156808} {\bibfield
  {journal} {\bibinfo  {journal} {The Journal of Chemical Physics}\ }\textbf
  {\bibinfo {volume} {158}},\ \bibinfo {pages} {234106} (\bibinfo {year}
  {2023})}\BibitemShut {NoStop}%
\bibitem [{\citenamefont {Makri}(1992)}]{Makri1992}%
  \BibitemOpen
  \bibfield  {author} {\bibinfo {author} {\bibfnamefont {N.}~\bibnamefont
  {Makri}},\ }\href {https://doi.org/10.1016/0009-2614(92)85654-S} {\bibfield
  {journal} {\bibinfo  {journal} {Chem. Phys. Lett.}\ }\textbf {\bibinfo
  {volume} {193}},\ \bibinfo {pages} {435} (\bibinfo {year}
  {1992})}\BibitemShut {NoStop}%
\bibitem [{\citenamefont {Tanimura}\ and\ \citenamefont
  {Kubo}(1989)}]{Tanimura1989}%
  \BibitemOpen
  \bibfield  {author} {\bibinfo {author} {\bibfnamefont {Y.}~\bibnamefont
  {Tanimura}}\ and\ \bibinfo {author} {\bibfnamefont {R.}~\bibnamefont
  {Kubo}},\ }\href {https://doi.org/10.1143/JPSJ.58.101} {\bibfield  {journal}
  {\bibinfo  {journal} {J. Phys. Soc. Japan}\ }\textbf {\bibinfo {volume}
  {58}},\ \bibinfo {pages} {101} (\bibinfo {year} {1989})}\BibitemShut
  {NoStop}%
\bibitem [{\citenamefont {Meyer}\ \emph {et~al.}(1990)\citenamefont {Meyer},
  \citenamefont {Manthe},\ and\ \citenamefont {Cederbaum}}]{Meyer1990}%
  \BibitemOpen
  \bibfield  {author} {\bibinfo {author} {\bibfnamefont {H.-D.}\ \bibnamefont
  {Meyer}}, \bibinfo {author} {\bibfnamefont {U.}~\bibnamefont {Manthe}},\ and\
  \bibinfo {author} {\bibfnamefont {L.}~\bibnamefont {Cederbaum}},\ }\href
  {https://doi.org/10.1016/0009-2614(90)87014-I} {\bibfield  {journal}
  {\bibinfo  {journal} {Chem. Phys. Lett.}\ }\textbf {\bibinfo {volume}
  {165}},\ \bibinfo {pages} {73} (\bibinfo {year} {1990})}\BibitemShut
  {NoStop}%
\bibitem [{\citenamefont {Manthe}\ \emph {et~al.}(1992)\citenamefont {Manthe},
  \citenamefont {Meyer},\ and\ \citenamefont {Cederbaum}}]{Manthe1992}%
  \BibitemOpen
  \bibfield  {author} {\bibinfo {author} {\bibfnamefont {U.}~\bibnamefont
  {Manthe}}, \bibinfo {author} {\bibfnamefont {H.-D.}\ \bibnamefont {Meyer}},\
  and\ \bibinfo {author} {\bibfnamefont {L.~S.}\ \bibnamefont {Cederbaum}},\
  }\href {https://doi.org/10.1063/1.463007} {\bibfield  {journal} {\bibinfo
  {journal} {J. Chem. Phys.}\ }\textbf {\bibinfo {volume} {97}},\ \bibinfo
  {pages} {3199} (\bibinfo {year} {1992})}\BibitemShut {NoStop}%
\bibitem [{\citenamefont {Beck}(2000)}]{Beck2000}%
  \BibitemOpen
  \bibfield  {author} {\bibinfo {author} {\bibfnamefont {M.}~\bibnamefont
  {Beck}},\ }\href {https://doi.org/10.1016/S0370-1573(99)00047-2} {\bibfield
  {journal} {\bibinfo  {journal} {Phys. Rep.}\ }\textbf {\bibinfo {volume}
  {324}},\ \bibinfo {pages} {1} (\bibinfo {year} {2000})}\BibitemShut {NoStop}%
\bibitem [{\citenamefont {Cao}\ and\ \citenamefont
  {Voth}(1994{\natexlab{a}})}]{Cao1994a}%
  \BibitemOpen
  \bibfield  {author} {\bibinfo {author} {\bibfnamefont {J.}~\bibnamefont
  {Cao}}\ and\ \bibinfo {author} {\bibfnamefont {G.~A.}\ \bibnamefont {Voth}},\
  }\href {https://doi.org/10.1063/1.467175} {\bibfield  {journal} {\bibinfo
  {journal} {The Journal of Chemical Physics}\ }\textbf {\bibinfo {volume}
  {100}},\ \bibinfo {pages} {5093} (\bibinfo {year}
  {1994}{\natexlab{a}})}\BibitemShut {NoStop}%
\bibitem [{\citenamefont {Cao}\ and\ \citenamefont
  {Voth}(1994{\natexlab{b}})}]{Cao1994b}%
  \BibitemOpen
  \bibfield  {author} {\bibinfo {author} {\bibfnamefont {J.}~\bibnamefont
  {Cao}}\ and\ \bibinfo {author} {\bibfnamefont {G.~A.}\ \bibnamefont {Voth}},\
  }\href {https://doi.org/10.1063/1.467176} {\bibfield  {journal} {\bibinfo
  {journal} {The Journal of Chemical Physics}\ }\textbf {\bibinfo {volume}
  {100}},\ \bibinfo {pages} {5106} (\bibinfo {year}
  {1994}{\natexlab{b}})}\BibitemShut {NoStop}%
\bibitem [{\citenamefont {Cao}\ and\ \citenamefont
  {Voth}(1994{\natexlab{c}})}]{Cao1994c}%
  \BibitemOpen
  \bibfield  {author} {\bibinfo {author} {\bibfnamefont {J.}~\bibnamefont
  {Cao}}\ and\ \bibinfo {author} {\bibfnamefont {G.~A.}\ \bibnamefont {Voth}},\
  }\href {https://doi.org/10.1063/1.468503} {\bibfield  {journal} {\bibinfo
  {journal} {The Journal of Chemical Physics}\ }\textbf {\bibinfo {volume}
  {101}},\ \bibinfo {pages} {6157} (\bibinfo {year}
  {1994}{\natexlab{c}})}\BibitemShut {NoStop}%
\bibitem [{\citenamefont {Cao}\ and\ \citenamefont
  {Voth}(1994{\natexlab{d}})}]{Cao1994d}%
  \BibitemOpen
  \bibfield  {author} {\bibinfo {author} {\bibfnamefont {J.}~\bibnamefont
  {Cao}}\ and\ \bibinfo {author} {\bibfnamefont {G.~A.}\ \bibnamefont {Voth}},\
  }\href {https://doi.org/10.1063/1.468399} {\bibfield  {journal} {\bibinfo
  {journal} {The Journal of Chemical Physics}\ }\textbf {\bibinfo {volume}
  {101}},\ \bibinfo {pages} {6168} (\bibinfo {year}
  {1994}{\natexlab{d}})}\BibitemShut {NoStop}%
\bibitem [{\citenamefont {Cao}\ and\ \citenamefont
  {Voth}(1994{\natexlab{e}})}]{Cao1994e}%
  \BibitemOpen
  \bibfield  {author} {\bibinfo {author} {\bibfnamefont {J.}~\bibnamefont
  {Cao}}\ and\ \bibinfo {author} {\bibfnamefont {G.~A.}\ \bibnamefont {Voth}},\
  }\href {https://doi.org/10.1063/1.468400} {\bibfield  {journal} {\bibinfo
  {journal} {The Journal of Chemical Physics}\ }\textbf {\bibinfo {volume}
  {101}},\ \bibinfo {pages} {6184} (\bibinfo {year}
  {1994}{\natexlab{e}})}\BibitemShut {NoStop}%
\bibitem [{\citenamefont {Craig}\ and\ \citenamefont
  {Manolopoulos}(2004)}]{Craig2004}%
  \BibitemOpen
  \bibfield  {author} {\bibinfo {author} {\bibfnamefont {I.~R.}\ \bibnamefont
  {Craig}}\ and\ \bibinfo {author} {\bibfnamefont {D.~E.}\ \bibnamefont
  {Manolopoulos}},\ }\href {https://doi.org/10.1063/1.1777575} {\bibfield
  {journal} {\bibinfo  {journal} {J. Chem. Phys.}\ }\textbf {\bibinfo {volume}
  {121}},\ \bibinfo {pages} {3368} (\bibinfo {year} {2004})}\BibitemShut
  {NoStop}%
\bibitem [{\citenamefont {Craig}\ and\ \citenamefont
  {Manolopoulos}(2005)}]{Craig2005}%
  \BibitemOpen
  \bibfield  {author} {\bibinfo {author} {\bibfnamefont {I.~R.}\ \bibnamefont
  {Craig}}\ and\ \bibinfo {author} {\bibfnamefont {D.~E.}\ \bibnamefont
  {Manolopoulos}},\ }\href {https://doi.org/10.1063/1.1850093} {\bibfield
  {journal} {\bibinfo  {journal} {J. Chem. Phys.}\ }\textbf {\bibinfo {volume}
  {122}},\ \bibinfo {pages} {084106} (\bibinfo {year} {2005})}\BibitemShut
  {NoStop}%
\bibitem [{\citenamefont {Hele}\ \emph
  {et~al.}(2015{\natexlab{a}})\citenamefont {Hele}, \citenamefont {Willatt},
  \citenamefont {Muolo},\ and\ \citenamefont {Althorpe}}]{Hele2015b}%
  \BibitemOpen
  \bibfield  {author} {\bibinfo {author} {\bibfnamefont {T.~J.~H.}\
  \bibnamefont {Hele}}, \bibinfo {author} {\bibfnamefont {M.~J.}\ \bibnamefont
  {Willatt}}, \bibinfo {author} {\bibfnamefont {A.}~\bibnamefont {Muolo}},\
  and\ \bibinfo {author} {\bibfnamefont {S.~C.}\ \bibnamefont {Althorpe}},\
  }\href {https://doi.org/10.1063/1.4916311} {\bibfield  {journal} {\bibinfo
  {journal} {The Journal of Chemical Physics}\ }\textbf {\bibinfo {volume}
  {142}},\ \bibinfo {pages} {134103} (\bibinfo {year}
  {2015}{\natexlab{a}})}\BibitemShut {NoStop}%
\bibitem [{\citenamefont {Hele}\ \emph
  {et~al.}(2015{\natexlab{b}})\citenamefont {Hele}, \citenamefont {Willatt},
  \citenamefont {Muolo},\ and\ \citenamefont {Althorpe}}]{Hele2015a}%
  \BibitemOpen
  \bibfield  {author} {\bibinfo {author} {\bibfnamefont {T.~J.~H.}\
  \bibnamefont {Hele}}, \bibinfo {author} {\bibfnamefont {M.~J.}\ \bibnamefont
  {Willatt}}, \bibinfo {author} {\bibfnamefont {A.}~\bibnamefont {Muolo}},\
  and\ \bibinfo {author} {\bibfnamefont {S.~C.}\ \bibnamefont {Althorpe}},\
  }\href {https://doi.org/10.1063/1.4921234} {\bibfield  {journal} {\bibinfo
  {journal} {The Journal of Chemical Physics}\ }\textbf {\bibinfo {volume}
  {142}},\ \bibinfo {pages} {191101} (\bibinfo {year}
  {2015}{\natexlab{b}})}\BibitemShut {NoStop}%
\bibitem [{\citenamefont {Willatt}\ \emph {et~al.}(2018)\citenamefont
  {Willatt}, \citenamefont {Ceriotti},\ and\ \citenamefont
  {Althorpe}}]{Willatt2018}%
  \BibitemOpen
  \bibfield  {author} {\bibinfo {author} {\bibfnamefont {M.~J.}\ \bibnamefont
  {Willatt}}, \bibinfo {author} {\bibfnamefont {M.}~\bibnamefont {Ceriotti}},\
  and\ \bibinfo {author} {\bibfnamefont {S.~C.}\ \bibnamefont {Althorpe}},\
  }\href {https://doi.org/10.1063/1.5004808} {\bibfield  {journal} {\bibinfo
  {journal} {The Journal of Chemical Physics}\ }\textbf {\bibinfo {volume}
  {148}},\ \bibinfo {pages} {102336} (\bibinfo {year} {2018})}\BibitemShut
  {NoStop}%
\bibitem [{\citenamefont {Trenins}\ and\ \citenamefont
  {Althorpe}(2018)}]{Trenins2018}%
  \BibitemOpen
  \bibfield  {author} {\bibinfo {author} {\bibfnamefont {G.}~\bibnamefont
  {Trenins}}\ and\ \bibinfo {author} {\bibfnamefont {S.~C.}\ \bibnamefont
  {Althorpe}},\ }\href {https://doi.org/10.1063/1.5038616} {\bibfield
  {journal} {\bibinfo  {journal} {The Journal of Chemical Physics}\ }\textbf
  {\bibinfo {volume} {149}},\ \bibinfo {pages} {14102} (\bibinfo {year}
  {2018})}\BibitemShut {NoStop}%
\bibitem [{\citenamefont {Trenins}\ \emph {et~al.}(2019)\citenamefont
  {Trenins}, \citenamefont {Willatt},\ and\ \citenamefont
  {Althorpe}}]{Trenins2019}%
  \BibitemOpen
  \bibfield  {author} {\bibinfo {author} {\bibfnamefont {G.}~\bibnamefont
  {Trenins}}, \bibinfo {author} {\bibfnamefont {M.~J.}\ \bibnamefont
  {Willatt}},\ and\ \bibinfo {author} {\bibfnamefont {S.~C.}\ \bibnamefont
  {Althorpe}},\ }\href {https://doi.org/10.1063/1.5100587} {\bibfield
  {journal} {\bibinfo  {journal} {The Journal of Chemical Physics}\ }\textbf
  {\bibinfo {volume} {151}},\ \bibinfo {pages} {054109} (\bibinfo {year}
  {2019})}\BibitemShut {NoStop}%
\bibitem [{\citenamefont {Benson}\ and\ \citenamefont
  {Althorpe}(2021)}]{Benson2021}%
  \BibitemOpen
  \bibfield  {author} {\bibinfo {author} {\bibfnamefont {R.~L.}\ \bibnamefont
  {Benson}}\ and\ \bibinfo {author} {\bibfnamefont {S.~C.}\ \bibnamefont
  {Althorpe}},\ }\href {https://doi.org/10.1063/5.0056829} {\bibfield
  {journal} {\bibinfo  {journal} {The Journal of Chemical Physics}\ }\textbf
  {\bibinfo {volume} {155}},\ \bibinfo {pages} {104107} (\bibinfo {year}
  {2021})}\BibitemShut {NoStop}%
\bibitem [{\citenamefont {Haggard}\ \emph {et~al.}(2021)\citenamefont
  {Haggard}, \citenamefont {Sadhasivam}, \citenamefont {Trenins},\ and\
  \citenamefont {Althorpe}}]{Haggard2021}%
  \BibitemOpen
  \bibfield  {author} {\bibinfo {author} {\bibfnamefont {C.}~\bibnamefont
  {Haggard}}, \bibinfo {author} {\bibfnamefont {V.~G.}\ \bibnamefont
  {Sadhasivam}}, \bibinfo {author} {\bibfnamefont {G.}~\bibnamefont
  {Trenins}},\ and\ \bibinfo {author} {\bibfnamefont {S.~C.}\ \bibnamefont
  {Althorpe}},\ }\href {https://doi.org/10.1063/5.0068250} {\bibfield
  {journal} {\bibinfo  {journal} {The Journal of Chemical Physics}\ }\textbf
  {\bibinfo {volume} {155}},\ \bibinfo {pages} {174120} (\bibinfo {year}
  {2021})}\BibitemShut {NoStop}%
\bibitem [{\citenamefont {Fletcher}\ \emph {et~al.}(2021)\citenamefont
  {Fletcher}, \citenamefont {Zhu}, \citenamefont {Lawrence},\ and\
  \citenamefont {Manolopoulos}}]{Fletcher2021}%
  \BibitemOpen
  \bibfield  {author} {\bibinfo {author} {\bibfnamefont {T.}~\bibnamefont
  {Fletcher}}, \bibinfo {author} {\bibfnamefont {A.}~\bibnamefont {Zhu}},
  \bibinfo {author} {\bibfnamefont {J.~E.}\ \bibnamefont {Lawrence}},\ and\
  \bibinfo {author} {\bibfnamefont {D.~E.}\ \bibnamefont {Manolopoulos}},\
  }\href {https://doi.org/10.1063/5.0076704} {\bibfield  {journal} {\bibinfo
  {journal} {The Journal of Chemical Physics}\ }\textbf {\bibinfo {volume}
  {155}},\ \bibinfo {pages} {231101} (\bibinfo {year} {2021})}\BibitemShut
  {NoStop}%
\bibitem [{\citenamefont {Musil}\ \emph {et~al.}(2022)\citenamefont {Musil},
  \citenamefont {Zaporozhets}, \citenamefont {Noé}, \citenamefont {Clementi},\
  and\ \citenamefont {Kapil}}]{Musil2022}%
  \BibitemOpen
  \bibfield  {author} {\bibinfo {author} {\bibfnamefont {F.}~\bibnamefont
  {Musil}}, \bibinfo {author} {\bibfnamefont {I.}~\bibnamefont {Zaporozhets}},
  \bibinfo {author} {\bibfnamefont {F.}~\bibnamefont {Noé}}, \bibinfo {author}
  {\bibfnamefont {C.}~\bibnamefont {Clementi}},\ and\ \bibinfo {author}
  {\bibfnamefont {V.}~\bibnamefont {Kapil}},\ }\bibfield  {journal} {\bibinfo
  {journal} {The Journal of Chemical Physics}\ }\textbf {\bibinfo {volume}
  {157}},\ \href {https://doi.org/10.1063/5.0120386} {10.1063/5.0120386}
  (\bibinfo {year} {2022})\BibitemShut {NoStop}%
\bibitem [{\citenamefont {Lawrence}\ \emph {et~al.}(2023)\citenamefont
  {Lawrence}, \citenamefont {Lieberherr}, \citenamefont {Fletcher},\ and\
  \citenamefont {Manolopoulos}}]{Lawrence2023}%
  \BibitemOpen
  \bibfield  {author} {\bibinfo {author} {\bibfnamefont {J.~E.}\ \bibnamefont
  {Lawrence}}, \bibinfo {author} {\bibfnamefont {A.~Z.}\ \bibnamefont
  {Lieberherr}}, \bibinfo {author} {\bibfnamefont {T.}~\bibnamefont
  {Fletcher}},\ and\ \bibinfo {author} {\bibfnamefont {D.~E.}\ \bibnamefont
  {Manolopoulos}},\ }\href {https://doi.org/10.1021/acs.jpcb.3c05028}
  {\bibfield  {journal} {\bibinfo  {journal} {The Journal of Physical Chemistry
  B}\ }\textbf {\bibinfo {volume} {127}},\ \bibinfo {pages} {9172} (\bibinfo
  {year} {2023})}\BibitemShut {NoStop}%
\bibitem [{\citenamefont {Prada}\ \emph {et~al.}(2023)\citenamefont {Prada},
  \citenamefont {Pós},\ and\ \citenamefont {Althorpe}}]{Prada2023}%
  \BibitemOpen
  \bibfield  {author} {\bibinfo {author} {\bibfnamefont {A.}~\bibnamefont
  {Prada}}, \bibinfo {author} {\bibfnamefont {E.~S.}\ \bibnamefont {Pós}},\
  and\ \bibinfo {author} {\bibfnamefont {S.~C.}\ \bibnamefont {Althorpe}},\
  }\href {https://doi.org/10.1063/5.0138250} {\bibfield  {journal} {\bibinfo
  {journal} {The Journal of Chemical Physics}\ }\textbf {\bibinfo {volume}
  {158}},\ \bibinfo {pages} {114106} (\bibinfo {year} {2023})}\BibitemShut
  {NoStop}%
\bibitem [{\citenamefont {Wigner}(1932)}]{Wigner1932}%
  \BibitemOpen
  \bibfield  {author} {\bibinfo {author} {\bibfnamefont {E.}~\bibnamefont
  {Wigner}},\ }\href {https://doi.org/10.1103/PhysRev.40.749} {\bibfield
  {journal} {\bibinfo  {journal} {Phys. Rev.}\ }\textbf {\bibinfo {volume}
  {40}},\ \bibinfo {pages} {749} (\bibinfo {year} {1932})}\BibitemShut
  {NoStop}%
\bibitem [{\citenamefont {Hillery}\ \emph {et~al.}(1984)\citenamefont
  {Hillery}, \citenamefont {O'Connell}, \citenamefont {Scully},\ and\
  \citenamefont {Wigner}}]{Hillery1984}%
  \BibitemOpen
  \bibfield  {author} {\bibinfo {author} {\bibfnamefont {M.}~\bibnamefont
  {Hillery}}, \bibinfo {author} {\bibfnamefont {R.}~\bibnamefont {O'Connell}},
  \bibinfo {author} {\bibfnamefont {M.}~\bibnamefont {Scully}},\ and\ \bibinfo
  {author} {\bibfnamefont {E.}~\bibnamefont {Wigner}},\ }\href
  {https://doi.org/10.1016/0370-1573(84)90160-1} {\bibfield  {journal}
  {\bibinfo  {journal} {Physics Reports}\ }\textbf {\bibinfo {volume} {106}},\
  \bibinfo {pages} {121} (\bibinfo {year} {1984})}\BibitemShut {NoStop}%
\bibitem [{\citenamefont {Miller}(2001)}]{Miller2001a}%
  \BibitemOpen
  \bibfield  {author} {\bibinfo {author} {\bibfnamefont {W.~H.}\ \bibnamefont
  {Miller}},\ }\href {https://doi.org/10.1021/JP003712K} {\bibfield  {journal}
  {\bibinfo  {journal} {J. Phys. Chem. A}\ }\textbf {\bibinfo {volume} {105}},\
  \bibinfo {pages} {2942} (\bibinfo {year} {2001})}\BibitemShut {NoStop}%
\bibitem [{\citenamefont {Malpathak}\ \emph {et~al.}(2022)\citenamefont
  {Malpathak}, \citenamefont {Church},\ and\ \citenamefont
  {Ananth}}]{Malpathak2022}%
  \BibitemOpen
  \bibfield  {author} {\bibinfo {author} {\bibfnamefont {S.}~\bibnamefont
  {Malpathak}}, \bibinfo {author} {\bibfnamefont {M.~S.}\ \bibnamefont
  {Church}},\ and\ \bibinfo {author} {\bibfnamefont {N.}~\bibnamefont
  {Ananth}},\ }\href {https://doi.org/10.1021/acs.jpca.2c03467} {\bibfield
  {journal} {\bibinfo  {journal} {J. Phys. Chem. A}\ }\textbf {\bibinfo
  {volume} {126}},\ \bibinfo {pages} {6359} (\bibinfo {year}
  {2022})}\BibitemShut {NoStop}%
\bibitem [{\citenamefont {Berry}\ and\ \citenamefont
  {Ziman}(1997)}]{Berry1997}%
  \BibitemOpen
  \bibfield  {author} {\bibinfo {author} {\bibfnamefont {M.~V.}\ \bibnamefont
  {Berry}}\ and\ \bibinfo {author} {\bibfnamefont {J.~M.}\ \bibnamefont
  {Ziman}},\ }\href {https://doi.org/10.1098/rsta.1977.0145} {\bibfield
  {journal} {\bibinfo  {journal} {Philosophical Transactions of the Royal
  Society of London. Series A, Mathematical and Physical Sciences}\ }\textbf
  {\bibinfo {volume} {287}},\ \bibinfo {pages} {237} (\bibinfo {year}
  {1997})}\BibitemShut {NoStop}%
\bibitem [{\citenamefont {Wilkie}\ and\ \citenamefont
  {Brumer}(1997{\natexlab{a}})}]{Wilkie1997a}%
  \BibitemOpen
  \bibfield  {author} {\bibinfo {author} {\bibfnamefont {J.}~\bibnamefont
  {Wilkie}}\ and\ \bibinfo {author} {\bibfnamefont {P.}~\bibnamefont
  {Brumer}},\ }\href {https://doi.org/10.1103/PhysRevA.55.27} {\bibfield
  {journal} {\bibinfo  {journal} {Physical Review A}\ }\textbf {\bibinfo
  {volume} {55}},\ \bibinfo {pages} {27} (\bibinfo {year}
  {1997}{\natexlab{a}})}\BibitemShut {NoStop}%
\bibitem [{\citenamefont {Wilkie}\ and\ \citenamefont
  {Brumer}(1997{\natexlab{b}})}]{Wilkie1997b}%
  \BibitemOpen
  \bibfield  {author} {\bibinfo {author} {\bibfnamefont {J.}~\bibnamefont
  {Wilkie}}\ and\ \bibinfo {author} {\bibfnamefont {P.}~\bibnamefont
  {Brumer}},\ }\href {https://doi.org/10.1103/PhysRevA.55.43} {\bibfield
  {journal} {\bibinfo  {journal} {Physical Review A}\ }\textbf {\bibinfo
  {volume} {55}},\ \bibinfo {pages} {43} (\bibinfo {year}
  {1997}{\natexlab{b}})}\BibitemShut {NoStop}%
\bibitem [{\citenamefont {Cohen}(1966)}]{Cohen1966}%
  \BibitemOpen
  \bibfield  {author} {\bibinfo {author} {\bibfnamefont {L.}~\bibnamefont
  {Cohen}},\ }\href {https://doi.org/10.1063/1.1931206} {\bibfield  {journal}
  {\bibinfo  {journal} {J. Math. Phys.}\ }\textbf {\bibinfo {volume} {7}},\
  \bibinfo {pages} {781} (\bibinfo {year} {1966})}\BibitemShut {NoStop}%
\bibitem [{\citenamefont {Lee}(1995)}]{Lee1995}%
  \BibitemOpen
  \bibfield  {author} {\bibinfo {author} {\bibfnamefont {H.-W.}\ \bibnamefont
  {Lee}},\ }\href {https://doi.org/10.1016/0370-1573(95)00007-4} {\bibfield
  {journal} {\bibinfo  {journal} {Phys. Rep.}\ }\textbf {\bibinfo {volume}
  {259}},\ \bibinfo {pages} {147} (\bibinfo {year} {1995})}\BibitemShut
  {NoStop}%
\bibitem [{\citenamefont {Heller}(1976)}]{Heller1976}%
  \BibitemOpen
  \bibfield  {author} {\bibinfo {author} {\bibfnamefont {E.~J.}\ \bibnamefont
  {Heller}},\ }\href {https://doi.org/10.1063/1.433238} {\bibfield  {journal}
  {\bibinfo  {journal} {J. Chem. Phys.}\ }\textbf {\bibinfo {volume} {65}},\
  \bibinfo {pages} {1289} (\bibinfo {year} {1976})}\BibitemShut {NoStop}%
\bibitem [{\citenamefont {Oliva}\ \emph {et~al.}(2018)\citenamefont {Oliva},
  \citenamefont {Kakofengitis},\ and\ \citenamefont {Steuernagel}}]{Oliva2018}%
  \BibitemOpen
  \bibfield  {author} {\bibinfo {author} {\bibfnamefont {M.}~\bibnamefont
  {Oliva}}, \bibinfo {author} {\bibfnamefont {D.}~\bibnamefont
  {Kakofengitis}},\ and\ \bibinfo {author} {\bibfnamefont {O.}~\bibnamefont
  {Steuernagel}},\ }\href {https://doi.org/10.1016/j.physa.2017.10.047}
  {\bibfield  {journal} {\bibinfo  {journal} {Phys. A}\ }\textbf {\bibinfo
  {volume} {502}},\ \bibinfo {pages} {201} (\bibinfo {year}
  {2018})}\BibitemShut {NoStop}%
\bibitem [{\citenamefont {Donoso}\ and\ \citenamefont
  {Martens}(2001)}]{Donoso2001}%
  \BibitemOpen
  \bibfield  {author} {\bibinfo {author} {\bibfnamefont {A.}~\bibnamefont
  {Donoso}}\ and\ \bibinfo {author} {\bibfnamefont {C.~C.}\ \bibnamefont
  {Martens}},\ }\href {https://doi.org/10.1103/PhysRevLett.87.223202}
  {\bibfield  {journal} {\bibinfo  {journal} {Phys. Rev. Lett.}\ }\textbf
  {\bibinfo {volume} {87}},\ \bibinfo {pages} {223202} (\bibinfo {year}
  {2001})}\BibitemShut {NoStop}%
\bibitem [{\citenamefont {Donoso}\ \emph {et~al.}(2003)\citenamefont {Donoso},
  \citenamefont {Zheng},\ and\ \citenamefont {Martens}}]{Donoso2003}%
  \BibitemOpen
  \bibfield  {author} {\bibinfo {author} {\bibfnamefont {A.}~\bibnamefont
  {Donoso}}, \bibinfo {author} {\bibfnamefont {Y.}~\bibnamefont {Zheng}},\ and\
  \bibinfo {author} {\bibfnamefont {C.~C.}\ \bibnamefont {Martens}},\ }\href
  {https://doi.org/10.1063/1.1597496} {\bibfield  {journal} {\bibinfo
  {journal} {J. Chem. Phys.}\ }\textbf {\bibinfo {volume} {119}},\ \bibinfo
  {pages} {5010} (\bibinfo {year} {2003})}\BibitemShut {NoStop}%
\bibitem [{\citenamefont {López}\ \emph {et~al.}(2006)\citenamefont {López},
  \citenamefont {Martens},\ and\ \citenamefont {Donoso}}]{Lopez2006}%
  \BibitemOpen
  \bibfield  {author} {\bibinfo {author} {\bibfnamefont {H.}~\bibnamefont
  {López}}, \bibinfo {author} {\bibfnamefont {C.~C.}\ \bibnamefont
  {Martens}},\ and\ \bibinfo {author} {\bibfnamefont {A.}~\bibnamefont
  {Donoso}},\ }\href {https://doi.org/10.1063/1.2222368} {\bibfield  {journal}
  {\bibinfo  {journal} {J. Chem. Phys.}\ }\textbf {\bibinfo {volume} {125}},\
  \bibinfo {pages} {154111} (\bibinfo {year} {2006})}\BibitemShut {NoStop}%
\bibitem [{\citenamefont {Polkovnikov}(2010)}]{Polkovnikov2010}%
  \BibitemOpen
  \bibfield  {author} {\bibinfo {author} {\bibfnamefont {A.}~\bibnamefont
  {Polkovnikov}},\ }\href {https://doi.org/10.1016/J.AOP.2010.02.006}
  {\bibfield  {journal} {\bibinfo  {journal} {Ann. Phys. (NY)}\ }\textbf
  {\bibinfo {volume} {325}},\ \bibinfo {pages} {1790} (\bibinfo {year}
  {2010})}\BibitemShut {NoStop}%
\bibitem [{\citenamefont {Sun}\ \emph {et~al.}(1998)\citenamefont {Sun},
  \citenamefont {Wang},\ and\ \citenamefont {Miller}}]{Sun1998}%
  \BibitemOpen
  \bibfield  {author} {\bibinfo {author} {\bibfnamefont {X.}~\bibnamefont
  {Sun}}, \bibinfo {author} {\bibfnamefont {H.}~\bibnamefont {Wang}},\ and\
  \bibinfo {author} {\bibfnamefont {W.~H.}\ \bibnamefont {Miller}},\ }\href
  {http://jcp.aip.org/jcp/copyright.jsp} {\bibfield  {journal} {\bibinfo
  {journal} {J. Chem. Phys.}\ }\textbf {\bibinfo {volume} {109}},\ \bibinfo
  {pages} {4190} (\bibinfo {year} {1998})}\BibitemShut {NoStop}%
\bibitem [{\citenamefont {Wang}\ \emph {et~al.}(1998)\citenamefont {Wang},
  \citenamefont {Sun},\ and\ \citenamefont {Miller}}]{Wang1998}%
  \BibitemOpen
  \bibfield  {author} {\bibinfo {author} {\bibfnamefont {H.}~\bibnamefont
  {Wang}}, \bibinfo {author} {\bibfnamefont {X.}~\bibnamefont {Sun}},\ and\
  \bibinfo {author} {\bibfnamefont {W.~H.}\ \bibnamefont {Miller}},\ }\href
  {https://doi.org/10.1063/1.476447} {\bibfield  {journal} {\bibinfo  {journal}
  {J. Chem. Phys.}\ }\textbf {\bibinfo {volume} {108}},\ \bibinfo {pages}
  {9726} (\bibinfo {year} {1998})}\BibitemShut {NoStop}%
\bibitem [{\citenamefont {Shi}\ and\ \citenamefont
  {Geva}(2003{\natexlab{a}})}]{Shi2003}%
  \BibitemOpen
  \bibfield  {author} {\bibinfo {author} {\bibfnamefont {Q.}~\bibnamefont
  {Shi}}\ and\ \bibinfo {author} {\bibfnamefont {E.}~\bibnamefont {Geva}},\
  }\href {https://doi.org/10.1063/1.1564814} {\bibfield  {journal} {\bibinfo
  {journal} {The Journal of Chemical Physics}\ }\textbf {\bibinfo {volume}
  {118}},\ \bibinfo {pages} {8173} (\bibinfo {year}
  {2003}{\natexlab{a}})}\BibitemShut {NoStop}%
\bibitem [{\citenamefont {Poulsen}\ \emph {et~al.}(2003)\citenamefont
  {Poulsen}, \citenamefont {Nyman},\ and\ \citenamefont
  {Rossky}}]{Poulsen2003}%
  \BibitemOpen
  \bibfield  {author} {\bibinfo {author} {\bibfnamefont {J.~A.}\ \bibnamefont
  {Poulsen}}, \bibinfo {author} {\bibfnamefont {G.}~\bibnamefont {Nyman}},\
  and\ \bibinfo {author} {\bibfnamefont {P.~J.}\ \bibnamefont {Rossky}},\
  }\href {https://doi.org/10.1063/1.1626631} {\bibfield  {journal} {\bibinfo
  {journal} {The Journal of Chemical Physics}\ }\textbf {\bibinfo {volume}
  {119}},\ \bibinfo {pages} {12179} (\bibinfo {year} {2003})}\BibitemShut
  {NoStop}%
\bibitem [{\citenamefont {Liu}\ and\ \citenamefont {Miller}(2007)}]{Liu2007}%
  \BibitemOpen
  \bibfield  {author} {\bibinfo {author} {\bibfnamefont {J.}~\bibnamefont
  {Liu}}\ and\ \bibinfo {author} {\bibfnamefont {W.~H.}\ \bibnamefont
  {Miller}},\ }\href {https://doi.org/10.1063/1.2743023} {\bibfield  {journal}
  {\bibinfo  {journal} {J. Chem. Phys.}\ }\textbf {\bibinfo {volume} {126}},\
  \bibinfo {pages} {234110} (\bibinfo {year} {2007})}\BibitemShut {NoStop}%
\bibitem [{\citenamefont {Liu}\ and\ \citenamefont
  {Miller}(2011{\natexlab{a}})}]{Liu2011a}%
  \BibitemOpen
  \bibfield  {author} {\bibinfo {author} {\bibfnamefont {J.}~\bibnamefont
  {Liu}}\ and\ \bibinfo {author} {\bibfnamefont {W.~H.}\ \bibnamefont
  {Miller}},\ }\href {https://doi.org/10.1063/1.3555273} {\bibfield  {journal}
  {\bibinfo  {journal} {J. Chem. Phys.}\ }\textbf {\bibinfo {volume} {134}},\
  \bibinfo {pages} {104101} (\bibinfo {year} {2011}{\natexlab{a}})}\BibitemShut
  {NoStop}%
\bibitem [{\citenamefont {Liu}\ and\ \citenamefont
  {Miller}(2011{\natexlab{b}})}]{Liu2011b}%
  \BibitemOpen
  \bibfield  {author} {\bibinfo {author} {\bibfnamefont {J.}~\bibnamefont
  {Liu}}\ and\ \bibinfo {author} {\bibfnamefont {W.~H.}\ \bibnamefont
  {Miller}},\ }\href {https://doi.org/10.1063/1.3555274} {\bibfield  {journal}
  {\bibinfo  {journal} {J. Chem. Phys.}\ }\textbf {\bibinfo {volume} {134}},\
  \bibinfo {pages} {104102} (\bibinfo {year} {2011}{\natexlab{b}})}\BibitemShut
  {NoStop}%
\bibitem [{\citenamefont {Liu}(2011)}]{Liu2011c}%
  \BibitemOpen
  \bibfield  {author} {\bibinfo {author} {\bibfnamefont {J.}~\bibnamefont
  {Liu}},\ }\href {https://doi.org/10.1063/1.3589406} {\bibfield  {journal}
  {\bibinfo  {journal} {J. Chem. Phys.}\ }\textbf {\bibinfo {volume} {134}},\
  \bibinfo {pages} {194110} (\bibinfo {year} {2011})}\BibitemShut {NoStop}%
\bibitem [{\citenamefont {Miller}(2009)}]{Miller2009}%
  \BibitemOpen
  \bibfield  {author} {\bibinfo {author} {\bibfnamefont {W.~H.}\ \bibnamefont
  {Miller}},\ }\href {https://doi.org/10.1021/jp809907p} {\bibfield  {journal}
  {\bibinfo  {journal} {The Journal of Physical Chemistry A}\ }\textbf
  {\bibinfo {volume} {113}},\ \bibinfo {pages} {1405} (\bibinfo {year}
  {2009})}\BibitemShut {NoStop}%
\bibitem [{\citenamefont {Herman}\ and\ \citenamefont
  {Kluk}(1984)}]{Herman1984}%
  \BibitemOpen
  \bibfield  {author} {\bibinfo {author} {\bibfnamefont {M.~F.}\ \bibnamefont
  {Herman}}\ and\ \bibinfo {author} {\bibfnamefont {E.}~\bibnamefont {Kluk}},\
  }\href {https://doi.org/10.1016/0301-0104(84)80039-7} {\bibfield  {journal}
  {\bibinfo  {journal} {Chem. Phys.}\ }\textbf {\bibinfo {volume} {91}},\
  \bibinfo {pages} {27} (\bibinfo {year} {1984})}\BibitemShut {NoStop}%
\bibitem [{\citenamefont {Thoss}\ and\ \citenamefont {Wang}(2004)}]{Thoss2004}%
  \BibitemOpen
  \bibfield  {author} {\bibinfo {author} {\bibfnamefont {M.}~\bibnamefont
  {Thoss}}\ and\ \bibinfo {author} {\bibfnamefont {H.}~\bibnamefont {Wang}},\
  }\href {https://doi.org/10.1146/annurev.physchem.55.091602.094429} {\bibfield
   {journal} {\bibinfo  {journal} {Annual Review of Physical Chemistry}\
  }\textbf {\bibinfo {volume} {55}},\ \bibinfo {pages} {299} (\bibinfo {year}
  {2004})}\BibitemShut {NoStop}%
\bibitem [{\citenamefont {Shi}\ and\ \citenamefont
  {Geva}(2003{\natexlab{b}})}]{Shi2003c}%
  \BibitemOpen
  \bibfield  {author} {\bibinfo {author} {\bibfnamefont {Q.}~\bibnamefont
  {Shi}}\ and\ \bibinfo {author} {\bibfnamefont {E.}~\bibnamefont {Geva}},\
  }\href {https://doi.org/10.1021/jp030497} {\bibfield  {journal} {\bibinfo
  {journal} {J. Phys. Chem. A}\ }\textbf {\bibinfo {volume} {107}},\ \bibinfo
  {pages} {9059} (\bibinfo {year} {2003}{\natexlab{b}})}\BibitemShut {NoStop}%
\bibitem [{\citenamefont {Being}\ \emph {et~al.}(2005)\citenamefont {Being},
  \citenamefont {Shi},\ and\ \citenamefont {Geva}}]{BeingJ.Ka2005}%
  \BibitemOpen
  \bibfield  {author} {\bibinfo {author} {\bibfnamefont {K.~J.}\ \bibnamefont
  {Being}}, \bibinfo {author} {\bibfnamefont {Q.}~\bibnamefont {Shi}},\ and\
  \bibinfo {author} {\bibfnamefont {E.}~\bibnamefont {Geva}},\ }\href
  {https://doi.org/10.1021/JP051223K} {\bibfield  {journal} {\bibinfo
  {journal} {J. Phys. Chem. A}\ }\textbf {\bibinfo {volume} {109}},\ \bibinfo
  {pages} {5527} (\bibinfo {year} {2005})}\BibitemShut {NoStop}%
\bibitem [{\citenamefont {Navrotskaya}\ and\ \citenamefont
  {Geva}(2006)}]{Navrotskaya2006}%
  \BibitemOpen
  \bibfield  {author} {\bibinfo {author} {\bibfnamefont {I.}~\bibnamefont
  {Navrotskaya}}\ and\ \bibinfo {author} {\bibfnamefont {E.}~\bibnamefont
  {Geva}},\ }\href {https://doi.org/10.1021/JP066243G} {\bibfield  {journal}
  {\bibinfo  {journal} {J. Phys. Chem. A}\ }\textbf {\bibinfo {volume} {111}},\
  \bibinfo {pages} {460} (\bibinfo {year} {2006})}\BibitemShut {NoStop}%
\bibitem [{\citenamefont {Being}\ and\ \citenamefont
  {Geva}(2006{\natexlab{a}})}]{Being2006}%
  \BibitemOpen
  \bibfield  {author} {\bibinfo {author} {\bibfnamefont {K.~J.}\ \bibnamefont
  {Being}}\ and\ \bibinfo {author} {\bibfnamefont {E.}~\bibnamefont {Geva}},\
  }\href {https://doi.org/10.1021/JP063907D} {\bibfield  {journal} {\bibinfo
  {journal} {J. Phys. Chem. A}\ }\textbf {\bibinfo {volume} {110}},\ \bibinfo
  {pages} {13131} (\bibinfo {year} {2006}{\natexlab{a}})}\BibitemShut {NoStop}%
\bibitem [{\citenamefont {Being}\ and\ \citenamefont
  {Geva}(2006{\natexlab{b}})}]{Being2006a}%
  \BibitemOpen
  \bibfield  {author} {\bibinfo {author} {\bibfnamefont {K.~J.}\ \bibnamefont
  {Being}}\ and\ \bibinfo {author} {\bibfnamefont {E.}~\bibnamefont {Geva}},\
  }\href {https://doi.org/10.1021/JP062363C} {\bibfield  {journal} {\bibinfo
  {journal} {J. Phys. Chem. A}\ }\textbf {\bibinfo {volume} {110}},\ \bibinfo
  {pages} {9555} (\bibinfo {year} {2006}{\natexlab{b}})}\BibitemShut {NoStop}%
\bibitem [{\citenamefont {V{\'{a}}zquez}\ \emph {et~al.}(2010)\citenamefont
  {V{\'{a}}zquez}, \citenamefont {Navrotskaya},\ and\ \citenamefont
  {Geva}}]{Vazquez2010}%
  \BibitemOpen
  \bibfield  {author} {\bibinfo {author} {\bibfnamefont {F.~X.}\ \bibnamefont
  {V{\'{a}}zquez}}, \bibinfo {author} {\bibfnamefont {I.}~\bibnamefont
  {Navrotskaya}},\ and\ \bibinfo {author} {\bibfnamefont {E.}~\bibnamefont
  {Geva}},\ }\href {https://doi.org/10.1021/JP1010499} {\bibfield  {journal}
  {\bibinfo  {journal} {J. Phys. Chem. A}\ }\textbf {\bibinfo {volume} {114}},\
  \bibinfo {pages} {5682} (\bibinfo {year} {2010})}\BibitemShut {NoStop}%
\bibitem [{\citenamefont {Liu}(2015)}]{Liu2015}%
  \BibitemOpen
  \bibfield  {author} {\bibinfo {author} {\bibfnamefont {J.}~\bibnamefont
  {Liu}},\ }\href {https://doi.org/10.1002/qua.24872} {\bibfield  {journal}
  {\bibinfo  {journal} {Int. J. Quantum Chem.}\ }\textbf {\bibinfo {volume}
  {115}},\ \bibinfo {pages} {657} (\bibinfo {year} {2015})}\BibitemShut
  {NoStop}%
\bibitem [{\citenamefont {Liu}\ \emph {et~al.}(2020)\citenamefont {Liu},
  \citenamefont {Gao}, \citenamefont {Lai}, \citenamefont {Mulvihill},\ and\
  \citenamefont {Geva}}]{Liu2020}%
  \BibitemOpen
  \bibfield  {author} {\bibinfo {author} {\bibfnamefont {Y.}~\bibnamefont
  {Liu}}, \bibinfo {author} {\bibfnamefont {X.}~\bibnamefont {Gao}}, \bibinfo
  {author} {\bibfnamefont {Y.}~\bibnamefont {Lai}}, \bibinfo {author}
  {\bibfnamefont {E.}~\bibnamefont {Mulvihill}},\ and\ \bibinfo {author}
  {\bibfnamefont {E.}~\bibnamefont {Geva}},\ }\href
  {https://doi.org/10.1021/acs.jctc.0c00177} {\bibfield  {journal} {\bibinfo
  {journal} {Journal of Chemical Theory and Computation}\ }\textbf {\bibinfo
  {volume} {16}},\ \bibinfo {pages} {4479} (\bibinfo {year}
  {2020})}\BibitemShut {NoStop}%
\bibitem [{\citenamefont {Gao}\ \emph {et~al.}(2020)\citenamefont {Gao},
  \citenamefont {Saller}, \citenamefont {Liu}, \citenamefont {Kelly},
  \citenamefont {Richardson},\ and\ \citenamefont {Geva}}]{Gao2020}%
  \BibitemOpen
  \bibfield  {author} {\bibinfo {author} {\bibfnamefont {X.}~\bibnamefont
  {Gao}}, \bibinfo {author} {\bibfnamefont {M.~A.~C.}\ \bibnamefont {Saller}},
  \bibinfo {author} {\bibfnamefont {Y.}~\bibnamefont {Liu}}, \bibinfo {author}
  {\bibfnamefont {A.}~\bibnamefont {Kelly}}, \bibinfo {author} {\bibfnamefont
  {J.~O.}\ \bibnamefont {Richardson}},\ and\ \bibinfo {author} {\bibfnamefont
  {E.}~\bibnamefont {Geva}},\ }\href {https://doi.org/10.1021/acs.jctc.9b01267}
  {\bibfield  {journal} {\bibinfo  {journal} {J. Chem. Theory Comput.}\
  }\textbf {\bibinfo {volume} {16}},\ \bibinfo {pages} {2883} (\bibinfo {year}
  {2020})}\BibitemShut {NoStop}%
\bibitem [{\citenamefont {Hu}\ and\ \citenamefont {Sun}(2022)}]{Hu2022}%
  \BibitemOpen
  \bibfield  {author} {\bibinfo {author} {\bibfnamefont {Z.}~\bibnamefont
  {Hu}}\ and\ \bibinfo {author} {\bibfnamefont {X.}~\bibnamefont {Sun}},\
  }\href {https://doi.org/10.1021/acs.jctc.2c00631} {\bibfield  {journal}
  {\bibinfo  {journal} {Journal of Chemical Theory and Computation}\ }\textbf
  {\bibinfo {volume} {18}},\ \bibinfo {pages} {5819} (\bibinfo {year}
  {2022})}\BibitemShut {NoStop}%
\bibitem [{\citenamefont {Malpathak}\ and\ \citenamefont
  {Ananth}(2024{\natexlab{a}})}]{Malpathak2024a}%
  \BibitemOpen
  \bibfield  {author} {\bibinfo {author} {\bibfnamefont {S.}~\bibnamefont
  {Malpathak}}\ and\ \bibinfo {author} {\bibfnamefont {N.}~\bibnamefont
  {Ananth}},\ }\href {https://doi.org/10.1021/acs.jpclett.3c03041} {\bibfield
  {journal} {\bibinfo  {journal} {The Journal of Physical Chemistry Letters}\
  }\textbf {\bibinfo {volume} {15}},\ \bibinfo {pages} {794} (\bibinfo {year}
  {2024}{\natexlab{a}})}\BibitemShut {NoStop}%
\bibitem [{\citenamefont {Miyazaki}\ and\ \citenamefont
  {Ananth}(2023)}]{Miyazaki2023}%
  \BibitemOpen
  \bibfield  {author} {\bibinfo {author} {\bibfnamefont {K.}~\bibnamefont
  {Miyazaki}}\ and\ \bibinfo {author} {\bibfnamefont {N.}~\bibnamefont
  {Ananth}},\ }\href {https://doi.org/10.1063/5.0163371/2912680} {\bibfield
  {journal} {\bibinfo  {journal} {The Journal of Chemical Physics}\ }\textbf
  {\bibinfo {volume} {159}},\ \bibinfo {pages} {124110} (\bibinfo {year}
  {2023})}\BibitemShut {NoStop}%
\bibitem [{\citenamefont {Antipov}\ \emph {et~al.}(2015)\citenamefont
  {Antipov}, \citenamefont {Ye},\ and\ \citenamefont {Ananth}}]{Antipov2015}%
  \BibitemOpen
  \bibfield  {author} {\bibinfo {author} {\bibfnamefont {S.~V.}\ \bibnamefont
  {Antipov}}, \bibinfo {author} {\bibfnamefont {Z.}~\bibnamefont {Ye}},\ and\
  \bibinfo {author} {\bibfnamefont {N.}~\bibnamefont {Ananth}},\ }\href
  {https://doi.org/10.1063/1.4919667} {\bibfield  {journal} {\bibinfo
  {journal} {J. Chem. Phys.}\ }\textbf {\bibinfo {volume} {142}},\ \bibinfo
  {pages} {184102} (\bibinfo {year} {2015})}\BibitemShut {NoStop}%
\bibitem [{\citenamefont {Church}\ \emph {et~al.}(2017)\citenamefont {Church},
  \citenamefont {Antipov},\ and\ \citenamefont {Ananth}}]{Church2017}%
  \BibitemOpen
  \bibfield  {author} {\bibinfo {author} {\bibfnamefont {M.~S.}\ \bibnamefont
  {Church}}, \bibinfo {author} {\bibfnamefont {S.~V.}\ \bibnamefont
  {Antipov}},\ and\ \bibinfo {author} {\bibfnamefont {N.}~\bibnamefont
  {Ananth}},\ }\href {https://doi.org/10.1063/1.4986645} {\bibfield  {journal}
  {\bibinfo  {journal} {J. Chem. Phys.}\ }\textbf {\bibinfo {volume} {146}},\
  \bibinfo {pages} {234104} (\bibinfo {year} {2017})}\BibitemShut {NoStop}%
\bibitem [{\citenamefont {Malpathak}\ and\ \citenamefont
  {Ananth}(2023)}]{Malpathak2023}%
  \BibitemOpen
  \bibfield  {author} {\bibinfo {author} {\bibfnamefont {S.}~\bibnamefont
  {Malpathak}}\ and\ \bibinfo {author} {\bibfnamefont {N.}~\bibnamefont
  {Ananth}},\ }\href {https://doi.org/10.1063/5.0133222} {\bibfield  {journal}
  {\bibinfo  {journal} {J. Chem. Phys.}\ }\textbf {\bibinfo {volume} {158}},\
  \bibinfo {pages} {104106} (\bibinfo {year} {2023})}\BibitemShut {NoStop}%
\bibitem [{\citenamefont {Husimi}(1940)}]{Husimi1940}%
  \BibitemOpen
  \bibfield  {author} {\bibinfo {author} {\bibfnamefont {K.}~\bibnamefont
  {Husimi}},\ }\href
  {https://doi.org/https://doi.org/10.11429/ppmsj1919.22.4_264} {\bibfield
  {journal} {\bibinfo  {journal} {Phys. Math. Soc. Jpn.}\ }\textbf {\bibinfo
  {volume} {22}},\ \bibinfo {pages} {264} (\bibinfo {year} {1940})}\BibitemShut
  {NoStop}%
\bibitem [{\citenamefont {Shao}\ and\ \citenamefont
  {Makri}(1999{\natexlab{a}})}]{Shao1999a}%
  \BibitemOpen
  \bibfield  {author} {\bibinfo {author} {\bibfnamefont {J.}~\bibnamefont
  {Shao}}\ and\ \bibinfo {author} {\bibfnamefont {N.}~\bibnamefont {Makri}},\
  }\href {https://doi.org/10.1021/jp991433v} {\bibfield  {journal} {\bibinfo
  {journal} {J. Phys. Chem. A}\ }\textbf {\bibinfo {volume} {103}},\ \bibinfo
  {pages} {7753} (\bibinfo {year} {1999}{\natexlab{a}})}\BibitemShut {NoStop}%
\bibitem [{\citenamefont {Shao}\ and\ \citenamefont
  {Makri}(1999{\natexlab{b}})}]{Shao1999b}%
  \BibitemOpen
  \bibfield  {author} {\bibinfo {author} {\bibfnamefont {J.}~\bibnamefont
  {Shao}}\ and\ \bibinfo {author} {\bibfnamefont {N.}~\bibnamefont {Makri}},\
  }\href {https://doi.org/10.1021/jp991837n} {\bibfield  {journal} {\bibinfo
  {journal} {J. Phys. Chem. A}\ }\textbf {\bibinfo {volume} {103}},\ \bibinfo
  {pages} {9479} (\bibinfo {year} {1999}{\natexlab{b}})}\BibitemShut {NoStop}%
\bibitem [{\citenamefont {Makri}\ and\ \citenamefont
  {Miller}(1987)}]{Makri1987c}%
  \BibitemOpen
  \bibfield  {author} {\bibinfo {author} {\bibfnamefont {N.}~\bibnamefont
  {Makri}}\ and\ \bibinfo {author} {\bibfnamefont {W.~H.}\ \bibnamefont
  {Miller}},\ }\href {https://doi.org/10.1016/0009-2614(87)80142-2} {\bibfield
  {journal} {\bibinfo  {journal} {Chem. Phys. Lett.}\ }\textbf {\bibinfo
  {volume} {139}},\ \bibinfo {pages} {10} (\bibinfo {year} {1987})}\BibitemShut
  {NoStop}%
\bibitem [{\citenamefont {Makri}\ and\ \citenamefont
  {Miller}(1988)}]{Makri1988b}%
  \BibitemOpen
  \bibfield  {author} {\bibinfo {author} {\bibfnamefont {N.}~\bibnamefont
  {Makri}}\ and\ \bibinfo {author} {\bibfnamefont {W.~H.}\ \bibnamefont
  {Miller}},\ }\href {https://doi.org/10.1063/1.455061} {\bibfield  {journal}
  {\bibinfo  {journal} {J. Chem. Phys.}\ }\textbf {\bibinfo {volume} {89}},\
  \bibinfo {pages} {2170} (\bibinfo {year} {1988})}\BibitemShut {NoStop}%
\bibitem [{\citenamefont {Spanner}\ \emph {et~al.}(2005)\citenamefont
  {Spanner}, \citenamefont {Batista},\ and\ \citenamefont
  {Brumer}}]{Spanner2005}%
  \BibitemOpen
  \bibfield  {author} {\bibinfo {author} {\bibfnamefont {M.}~\bibnamefont
  {Spanner}}, \bibinfo {author} {\bibfnamefont {V.~S.}\ \bibnamefont
  {Batista}},\ and\ \bibinfo {author} {\bibfnamefont {P.}~\bibnamefont
  {Brumer}},\ }\href {https://doi.org/10.1063/1.1854634} {\bibfield  {journal}
  {\bibinfo  {journal} {The Journal of Chemical Physics}\ }\textbf {\bibinfo
  {volume} {122}},\ \bibinfo {pages} {084111} (\bibinfo {year}
  {2005})}\BibitemShut {NoStop}%
\bibitem [{\citenamefont {Church}\ \emph {et~al.}(2018)\citenamefont {Church},
  \citenamefont {Hele}, \citenamefont {Ezra},\ and\ \citenamefont
  {Ananth}}]{Church2018}%
  \BibitemOpen
  \bibfield  {author} {\bibinfo {author} {\bibfnamefont {M.~S.}\ \bibnamefont
  {Church}}, \bibinfo {author} {\bibfnamefont {T.~J.~H.}\ \bibnamefont {Hele}},
  \bibinfo {author} {\bibfnamefont {G.~S.}\ \bibnamefont {Ezra}},\ and\
  \bibinfo {author} {\bibfnamefont {N.}~\bibnamefont {Ananth}},\ }\href
  {https://doi.org/10.1063/1.5005557} {\bibfield  {journal} {\bibinfo
  {journal} {J. Chem. Phys.}\ }\textbf {\bibinfo {volume} {148}},\ \bibinfo
  {pages} {102326} (\bibinfo {year} {2018})}\BibitemShut {NoStop}%
\bibitem [{\citenamefont {Church}\ and\ \citenamefont
  {Ananth}(2019)}]{Church2019a}%
  \BibitemOpen
  \bibfield  {author} {\bibinfo {author} {\bibfnamefont {M.~S.}\ \bibnamefont
  {Church}}\ and\ \bibinfo {author} {\bibfnamefont {N.}~\bibnamefont
  {Ananth}},\ }\href {https://doi.org/10.1063/1.5117160} {\bibfield  {journal}
  {\bibinfo  {journal} {J. Chem. Phys.}\ }\textbf {\bibinfo {volume} {151}},\
  \bibinfo {pages} {134109} (\bibinfo {year} {2019})}\BibitemShut {NoStop}%
\bibitem [{\citenamefont {Malpathak}\ and\ \citenamefont
  {Ananth}(2024{\natexlab{b}})}]{Malpathak2024c}%
  \BibitemOpen
  \bibfield  {author} {\bibinfo {author} {\bibfnamefont {S.}~\bibnamefont
  {Malpathak}}\ and\ \bibinfo {author} {\bibfnamefont {N.}~\bibnamefont
  {Ananth}},\ }\href {https://doi.org/https://arxiv.org/abs/2405.20499} {\
  (\bibinfo {year} {2024}{\natexlab{b}})},\ \bibinfo {note} {paper II of this
  series. arXiv ID: 2405.20499}\BibitemShut {NoStop}%
\bibitem [{\citenamefont {Rios}\ and\ \citenamefont
  {de~Almeida}(2002)}]{Rios2002}%
  \BibitemOpen
  \bibfield  {author} {\bibinfo {author} {\bibfnamefont {P.~P. d.~M.}\
  \bibnamefont {Rios}}\ and\ \bibinfo {author} {\bibfnamefont {A.~M.~O.}\
  \bibnamefont {de~Almeida}},\ }\href
  {https://doi.org/10.1088/0305-4470/35/11/307} {\bibfield  {journal} {\bibinfo
   {journal} {Journal of Physics A: Mathematical and General}\ }\textbf
  {\bibinfo {volume} {35}},\ \bibinfo {pages} {2609} (\bibinfo {year}
  {2002})},\ \Eprint {https://arxiv.org/abs/0111012} {arXiv:0111012 [math-ph]}
  \BibitemShut {NoStop}%
\bibitem [{\citenamefont {{Ozorio de Almeida}}\ and\ \citenamefont
  {Brodier}(2006)}]{OzoriodeAlmeida2006a}%
  \BibitemOpen
  \bibfield  {author} {\bibinfo {author} {\bibfnamefont {A.}~\bibnamefont
  {{Ozorio de Almeida}}}\ and\ \bibinfo {author} {\bibfnamefont
  {O.}~\bibnamefont {Brodier}},\ }\href
  {https://doi.org/10.1016/j.aop.2006.03.007} {\bibfield  {journal} {\bibinfo
  {journal} {Annals of Physics}\ }\textbf {\bibinfo {volume} {321}},\ \bibinfo
  {pages} {1790} (\bibinfo {year} {2006})}\BibitemShut {NoStop}%
\bibitem [{\citenamefont {Dittrich}\ \emph {et~al.}(2006)\citenamefont
  {Dittrich}, \citenamefont {Viviescas},\ and\ \citenamefont
  {Sandoval}}]{Dittrich2006}%
  \BibitemOpen
  \bibfield  {author} {\bibinfo {author} {\bibfnamefont {T.}~\bibnamefont
  {Dittrich}}, \bibinfo {author} {\bibfnamefont {C.}~\bibnamefont
  {Viviescas}},\ and\ \bibinfo {author} {\bibfnamefont {L.}~\bibnamefont
  {Sandoval}},\ }\href {https://doi.org/10.1103/PhysRevLett.96.070403}
  {\bibfield  {journal} {\bibinfo  {journal} {Physical Review Letters}\
  }\textbf {\bibinfo {volume} {96}},\ \bibinfo {pages} {070403} (\bibinfo
  {year} {2006})}\BibitemShut {NoStop}%
\bibitem [{\citenamefont {Dittrich}\ \emph {et~al.}(2010)\citenamefont
  {Dittrich}, \citenamefont {G{\'{o}}mez},\ and\ \citenamefont
  {Pach{\'{o}}n}}]{Dittrich2010a}%
  \BibitemOpen
  \bibfield  {author} {\bibinfo {author} {\bibfnamefont {T.}~\bibnamefont
  {Dittrich}}, \bibinfo {author} {\bibfnamefont {E.~A.}\ \bibnamefont
  {G{\'{o}}mez}},\ and\ \bibinfo {author} {\bibfnamefont {L.~A.}\ \bibnamefont
  {Pach{\'{o}}n}},\ }\href {https://doi.org/10.1063/1.3425881} {\bibfield
  {journal} {\bibinfo  {journal} {The Journal of Chemical Physics}\ }\textbf
  {\bibinfo {volume} {132}},\ \bibinfo {pages} {214102} (\bibinfo {year}
  {2010})},\ \Eprint {https://arxiv.org/abs/0911.3871} {arXiv:0911.3871}
  \BibitemShut {NoStop}%
\bibitem [{\citenamefont {de~Almeida}\ \emph {et~al.}(2013)\citenamefont
  {de~Almeida}, \citenamefont {Vallejos},\ and\ \citenamefont
  {Zambrano}}]{DeAlmeida2013}%
  \BibitemOpen
  \bibfield  {author} {\bibinfo {author} {\bibfnamefont {A.~M.~O.}\
  \bibnamefont {de~Almeida}}, \bibinfo {author} {\bibfnamefont {R.~O.}\
  \bibnamefont {Vallejos}},\ and\ \bibinfo {author} {\bibfnamefont
  {E.}~\bibnamefont {Zambrano}},\ }\href
  {https://doi.org/10.1088/1751-8113/46/13/135304} {\bibfield  {journal}
  {\bibinfo  {journal} {Journal of Physics A: Mathematical and Theoretical}\
  }\textbf {\bibinfo {volume} {46}},\ \bibinfo {pages} {135304} (\bibinfo
  {year} {2013})}\BibitemShut {NoStop}%
\bibitem [{\citenamefont {Lando}\ \emph {et~al.}(2019)\citenamefont {Lando},
  \citenamefont {Vallejos}, \citenamefont {Ingold},\ and\ \citenamefont
  {de~Almeida}}]{Lando2019}%
  \BibitemOpen
  \bibfield  {author} {\bibinfo {author} {\bibfnamefont {G.~M.}\ \bibnamefont
  {Lando}}, \bibinfo {author} {\bibfnamefont {R.~O.}\ \bibnamefont {Vallejos}},
  \bibinfo {author} {\bibfnamefont {G.-L.}\ \bibnamefont {Ingold}},\ and\
  \bibinfo {author} {\bibfnamefont {A.~M.~O.}\ \bibnamefont {de~Almeida}},\
  }\href {https://doi.org/10.1103/PhysRevA.99.042125} {\bibfield  {journal}
  {\bibinfo  {journal} {Physical Review A}\ }\textbf {\bibinfo {volume} {99}},\
  \bibinfo {pages} {042125} (\bibinfo {year} {2019})}\BibitemShut {NoStop}%
\bibitem [{\citenamefont {Koda}(2015)}]{Koda2015}%
  \BibitemOpen
  \bibfield  {author} {\bibinfo {author} {\bibfnamefont {S.-i.}\ \bibnamefont
  {Koda}},\ }\href {https://doi.org/10.1063/1.4938235} {\bibfield  {journal}
  {\bibinfo  {journal} {The Journal of Chemical Physics}\ }\textbf {\bibinfo
  {volume} {143}},\ \bibinfo {pages} {244110} (\bibinfo {year}
  {2015})}\BibitemShut {NoStop}%
\bibitem [{\citenamefont {Koda}(2016)}]{Koda2016}%
  \BibitemOpen
  \bibfield  {author} {\bibinfo {author} {\bibfnamefont {S.-i.}\ \bibnamefont
  {Koda}},\ }\href {https://doi.org/10.1063/1.4947041} {\bibfield  {journal}
  {\bibinfo  {journal} {The Journal of Chemical Physics}\ }\textbf {\bibinfo
  {volume} {144}},\ \bibinfo {pages} {154108} (\bibinfo {year}
  {2016})}\BibitemShut {NoStop}%
\bibitem [{\citenamefont {Gottwald}\ and\ \citenamefont
  {Ivanov}(2018)}]{Gottwald2018}%
  \BibitemOpen
  \bibfield  {author} {\bibinfo {author} {\bibfnamefont {F.}~\bibnamefont
  {Gottwald}}\ and\ \bibinfo {author} {\bibfnamefont {S.~D.}\ \bibnamefont
  {Ivanov}},\ }\href {https://doi.org/10.1016/j.chemphys.2018.02.009}
  {\bibfield  {journal} {\bibinfo  {journal} {Chemical Physics}\ }\textbf
  {\bibinfo {volume} {503}},\ \bibinfo {pages} {77} (\bibinfo {year}
  {2018})}\BibitemShut {NoStop}%
\bibitem [{\citenamefont {Ovchinnikov}\ and\ \citenamefont
  {Apkarian}(1996)}]{Ovchinnikov1996}%
  \BibitemOpen
  \bibfield  {author} {\bibinfo {author} {\bibfnamefont {M.}~\bibnamefont
  {Ovchinnikov}}\ and\ \bibinfo {author} {\bibfnamefont {V.~A.}\ \bibnamefont
  {Apkarian}},\ }\href {https://doi.org/10.1063/1.472959} {\bibfield  {journal}
  {\bibinfo  {journal} {J. Chem. Phys.}\ }\textbf {\bibinfo {volume} {105}},\
  \bibinfo {pages} {10312} (\bibinfo {year} {1996})}\BibitemShut {NoStop}%
\bibitem [{\citenamefont {Sun}\ and\ \citenamefont {Miller}(1997)}]{Sun1997c}%
  \BibitemOpen
  \bibfield  {author} {\bibinfo {author} {\bibfnamefont {X.}~\bibnamefont
  {Sun}}\ and\ \bibinfo {author} {\bibfnamefont {W.~H.}\ \bibnamefont
  {Miller}},\ }\href {https://doi.org/10.1063/1.473171} {\bibfield  {journal}
  {\bibinfo  {journal} {J. Chem. Phys.}\ }\textbf {\bibinfo {volume} {106}},\
  \bibinfo {pages} {916} (\bibinfo {year} {1997})}\BibitemShut {NoStop}%
\bibitem [{\citenamefont {Zhang}\ and\ \citenamefont
  {Pollak}(2005)}]{Zhang2005}%
  \BibitemOpen
  \bibfield  {author} {\bibinfo {author} {\bibfnamefont {S.}~\bibnamefont
  {Zhang}}\ and\ \bibinfo {author} {\bibfnamefont {E.}~\bibnamefont {Pollak}},\
  }\href {https://doi.org/10.1021/ct0499074} {\bibfield  {journal} {\bibinfo
  {journal} {Journal of Chemical Theory and Computation}\ }\textbf {\bibinfo
  {volume} {1}},\ \bibinfo {pages} {345} (\bibinfo {year} {2005})}\BibitemShut
  {NoStop}%
\bibitem [{\citenamefont {Grossmann}(2006)}]{Grossmann2006}%
  \BibitemOpen
  \bibfield  {author} {\bibinfo {author} {\bibfnamefont {F.}~\bibnamefont
  {Grossmann}},\ }\href {https://doi.org/10.1063/1.2213255} {\bibfield
  {journal} {\bibinfo  {journal} {J. Chem. Phys.}\ }\textbf {\bibinfo {volume}
  {125}},\ \bibinfo {pages} {014111} (\bibinfo {year} {2006})}\BibitemShut
  {NoStop}%
\bibitem [{\citenamefont {Kay}(2005)}]{Kay2005}%
  \BibitemOpen
  \bibfield  {author} {\bibinfo {author} {\bibfnamefont {K.~G.}\ \bibnamefont
  {Kay}},\ }\href {https://doi.org/10.1146/annurev.physchem.56.092503.141257}
  {\bibfield  {journal} {\bibinfo  {journal} {Annual Review of Physical
  Chemistry}\ }\textbf {\bibinfo {volume} {56}},\ \bibinfo {pages} {255}
  (\bibinfo {year} {2005})}\BibitemShut {NoStop}%
\bibitem [{Note1()}]{Note1}%
  \BibitemOpen
  \bibinfo {note} {In Ref. \protect \citenum {Gottwald2018} the determinant has
  a power of 1/4, not 1/2, and the definition of $\protect \bm {\Gamma }$ is
  different. There, the Wigner transform of the density operator is not
  normalized, but instead the factor of $(2\pi \protect \hbar )^{-N}$ that
  usually shows up due to normalization is divided equally into the Wigner
  transforms of $\protect \hat {\rho }_{\protect \hat {A}}$ and $\protect \hat
  {B}$. This, taken together with the different definition of $\protect \bm
  {\Gamma }$ accounts for the difference in power.}\BibitemShut {Stop}%
\bibitem [{Note2()}]{Note2}%
  \BibitemOpen
  \bibinfo {note} {In Ref.\protect \citenum {Gottwald2018} the condition is
  written as $\protect \text {det}(\protect \bm {\Gamma })=4$. The difference
  between our expression and theirs stems from a factor of 2 that is different
  between the respective definitions of the width matrix $\protect \bm {\gamma
  }$ in the coherent state $\ket {\protect \bm {z}}$.}\BibitemShut {Stop}%
\bibitem [{\citenamefont {Gelabert}\ \emph {et~al.}(2000)\citenamefont
  {Gelabert}, \citenamefont {Giménez}, \citenamefont {Thoss}, \citenamefont
  {Wang},\ and\ \citenamefont {Miller}}]{Gelabert2000d}%
  \BibitemOpen
  \bibfield  {author} {\bibinfo {author} {\bibfnamefont {R.}~\bibnamefont
  {Gelabert}}, \bibinfo {author} {\bibfnamefont {X.}~\bibnamefont {Giménez}},
  \bibinfo {author} {\bibfnamefont {M.}~\bibnamefont {Thoss}}, \bibinfo
  {author} {\bibfnamefont {H.}~\bibnamefont {Wang}},\ and\ \bibinfo {author}
  {\bibfnamefont {W.~H.}\ \bibnamefont {Miller}},\ }\href
  {https://doi.org/10.1021/jp0012451} {\bibfield  {journal} {\bibinfo
  {journal} {The Journal of Physical Chemistry A}\ }\textbf {\bibinfo {volume}
  {104}},\ \bibinfo {pages} {10321} (\bibinfo {year} {2000})}\BibitemShut
  {NoStop}%
\bibitem [{\citenamefont {Liberto}\ and\ \citenamefont
  {Ceotto}(2016)}]{Liberto2016b}%
  \BibitemOpen
  \bibfield  {author} {\bibinfo {author} {\bibfnamefont {G.~D.}\ \bibnamefont
  {Liberto}}\ and\ \bibinfo {author} {\bibfnamefont {M.}~\bibnamefont
  {Ceotto}},\ }\href {https://doi.org/10.1063/1.4964308} {\bibfield  {journal}
  {\bibinfo  {journal} {The Journal of Chemical Physics}\ }\textbf {\bibinfo
  {volume} {145}},\ \bibinfo {pages} {144107} (\bibinfo {year}
  {2016})}\BibitemShut {NoStop}%
\bibitem [{\citenamefont {Heller}(1975)}]{Heller1975}%
  \BibitemOpen
  \bibfield  {author} {\bibinfo {author} {\bibfnamefont {E.~J.}\ \bibnamefont
  {Heller}},\ }\href {https://doi.org/10.1063/1.430620} {\bibfield  {journal}
  {\bibinfo  {journal} {The Journal of Chemical Physics}\ }\textbf {\bibinfo
  {volume} {62}},\ \bibinfo {pages} {1544} (\bibinfo {year}
  {1975})}\BibitemShut {NoStop}%
\bibitem [{\citenamefont {Wang}\ and\ \citenamefont {Heller}(2009)}]{Wang2009}%
  \BibitemOpen
  \bibfield  {author} {\bibinfo {author} {\bibfnamefont {Z.-x.}\ \bibnamefont
  {Wang}}\ and\ \bibinfo {author} {\bibfnamefont {E.~J.}\ \bibnamefont
  {Heller}},\ }\href {https://doi.org/10.1088/1751-8113/42/28/285304}
  {\bibfield  {journal} {\bibinfo  {journal} {Journal of Physics A:
  Mathematical and Theoretical}\ }\textbf {\bibinfo {volume} {42}},\ \bibinfo
  {pages} {285304} (\bibinfo {year} {2009})}\BibitemShut {NoStop}%
\bibitem [{\citenamefont {{Loho Choudhury}}\ and\ \citenamefont
  {Gro{\ss}mann}(2020)}]{LohoChoudhury2020}%
  \BibitemOpen
  \bibfield  {author} {\bibinfo {author} {\bibfnamefont {S.}~\bibnamefont
  {{Loho Choudhury}}}\ and\ \bibinfo {author} {\bibfnamefont {F.}~\bibnamefont
  {Gro{\ss}mann}},\ }\href {https://doi.org/10.3390/condmat5010003} {\bibfield
  {journal} {\bibinfo  {journal} {Condensed Matter}\ }\textbf {\bibinfo
  {volume} {5}},\ \bibinfo {pages} {3} (\bibinfo {year} {2020})}\BibitemShut
  {NoStop}%
\bibitem [{\citenamefont {Leggett}\ \emph {et~al.}(1987)\citenamefont
  {Leggett}, \citenamefont {Chakravarty}, \citenamefont {Dorsey}, \citenamefont
  {Fisher}, \citenamefont {Garg},\ and\ \citenamefont {Zwerger}}]{Leggett1987}%
  \BibitemOpen
  \bibfield  {author} {\bibinfo {author} {\bibfnamefont {A.~J.}\ \bibnamefont
  {Leggett}}, \bibinfo {author} {\bibfnamefont {S.}~\bibnamefont
  {Chakravarty}}, \bibinfo {author} {\bibfnamefont {A.~T.}\ \bibnamefont
  {Dorsey}}, \bibinfo {author} {\bibfnamefont {M.~P.~A.}\ \bibnamefont
  {Fisher}}, \bibinfo {author} {\bibfnamefont {A.}~\bibnamefont {Garg}},\ and\
  \bibinfo {author} {\bibfnamefont {W.}~\bibnamefont {Zwerger}},\ }\href
  {https://doi.org/10.1103/RevModPhys.59.1} {\bibfield  {journal} {\bibinfo
  {journal} {Rev. Mod. Phys.}\ }\textbf {\bibinfo {volume} {59}},\ \bibinfo
  {pages} {1} (\bibinfo {year} {1987})}\BibitemShut {NoStop}%
\bibitem [{\citenamefont {Wang}\ \emph {et~al.}(2001)\citenamefont {Wang},
  \citenamefont {Thoss}, \citenamefont {Sorge}, \citenamefont {Gelabert},
  \citenamefont {Gim{\'{e}}nez},\ and\ \citenamefont {Miller}}]{Wang2001}%
  \BibitemOpen
  \bibfield  {author} {\bibinfo {author} {\bibfnamefont {H.}~\bibnamefont
  {Wang}}, \bibinfo {author} {\bibfnamefont {M.}~\bibnamefont {Thoss}},
  \bibinfo {author} {\bibfnamefont {K.~L.}\ \bibnamefont {Sorge}}, \bibinfo
  {author} {\bibfnamefont {R.}~\bibnamefont {Gelabert}}, \bibinfo {author}
  {\bibfnamefont {X.}~\bibnamefont {Gim{\'{e}}nez}},\ and\ \bibinfo {author}
  {\bibfnamefont {W.~H.}\ \bibnamefont {Miller}},\ }\href
  {https://doi.org/10.1063/1.1337802} {\bibfield  {journal} {\bibinfo
  {journal} {The Journal of Chemical Physics}\ }\textbf {\bibinfo {volume}
  {114}},\ \bibinfo {pages} {2562} (\bibinfo {year} {2001})}\BibitemShut
  {NoStop}%
\bibitem [{\citenamefont {Elran}\ and\ \citenamefont
  {Brumer}(2004)}]{Elran2004}%
  \BibitemOpen
  \bibfield  {author} {\bibinfo {author} {\bibfnamefont {Y.}~\bibnamefont
  {Elran}}\ and\ \bibinfo {author} {\bibfnamefont {P.}~\bibnamefont {Brumer}},\
  }\href {https://doi.org/10.1063/1.1766009} {\bibfield  {journal} {\bibinfo
  {journal} {The Journal of Chemical Physics}\ }\textbf {\bibinfo {volume}
  {121}},\ \bibinfo {pages} {2673} (\bibinfo {year} {2004})}\BibitemShut
  {NoStop}%
\bibitem [{\citenamefont {Goletz}\ and\ \citenamefont
  {Grossmann}(2009)}]{Goletz2009}%
  \BibitemOpen
  \bibfield  {author} {\bibinfo {author} {\bibfnamefont {C.-M.}\ \bibnamefont
  {Goletz}}\ and\ \bibinfo {author} {\bibfnamefont {F.}~\bibnamefont
  {Grossmann}},\ }\href {https://doi.org/10.1063/1.3157162} {\bibfield
  {journal} {\bibinfo  {journal} {The Journal of Chemical Physics}\ }\textbf
  {\bibinfo {volume} {130}},\ \bibinfo {pages} {244107} (\bibinfo {year}
  {2009})}\BibitemShut {NoStop}%
\bibitem [{\citenamefont {Sanz}(2014)}]{Sanz2014}%
  \BibitemOpen
  \bibfield  {author} {\bibinfo {author} {\bibfnamefont {A.}~\bibnamefont
  {Sanz}},\ }\href {https://doi.org/10.1139/cjc-2013-0399} {\bibfield
  {journal} {\bibinfo  {journal} {Canadian Journal of Chemistry}\ }\textbf
  {\bibinfo {volume} {92}},\ \bibinfo {pages} {168} (\bibinfo {year}
  {2014})}\BibitemShut {NoStop}%
\bibitem [{\citenamefont {Brewer}\ \emph {et~al.}(1997)\citenamefont {Brewer},
  \citenamefont {Hulme},\ and\ \citenamefont {Manolopoulos}}]{Brewer1997}%
  \BibitemOpen
  \bibfield  {author} {\bibinfo {author} {\bibfnamefont {M.~L.}\ \bibnamefont
  {Brewer}}, \bibinfo {author} {\bibfnamefont {J.~S.}\ \bibnamefont {Hulme}},\
  and\ \bibinfo {author} {\bibfnamefont {D.~E.}\ \bibnamefont {Manolopoulos}},\
  }\href {https://doi.org/10.1063/1.473532} {\bibfield  {journal} {\bibinfo
  {journal} {The Journal of Chemical Physics}\ }\textbf {\bibinfo {volume}
  {106}},\ \bibinfo {pages} {4832} (\bibinfo {year} {1997})}\BibitemShut
  {NoStop}%
\bibitem [{\citenamefont {Grossmann}(1995)}]{Grossmann1995}%
  \BibitemOpen
  \bibfield  {author} {\bibinfo {author} {\bibfnamefont {F.}~\bibnamefont
  {Grossmann}},\ }\href {https://doi.org/10.1063/1.470046} {\bibfield
  {journal} {\bibinfo  {journal} {The Journal of Chemical Physics}\ }\textbf
  {\bibinfo {volume} {103}},\ \bibinfo {pages} {3696} (\bibinfo {year}
  {1995})}\BibitemShut {NoStop}%
\bibitem [{\citenamefont {Fiete}\ and\ \citenamefont
  {Heller}(2003)}]{Fiete2003}%
  \BibitemOpen
  \bibfield  {author} {\bibinfo {author} {\bibfnamefont {G.~A.}\ \bibnamefont
  {Fiete}}\ and\ \bibinfo {author} {\bibfnamefont {E.~J.}\ \bibnamefont
  {Heller}},\ }\href {https://doi.org/10.1103/PhysRevA.68.022112} {\bibfield
  {journal} {\bibinfo  {journal} {Physical Review A}\ }\textbf {\bibinfo
  {volume} {68}},\ \bibinfo {pages} {022112} (\bibinfo {year}
  {2003})}\BibitemShut {NoStop}%
\bibitem [{\citenamefont {Pollak}(2007)}]{Pollak2007}%
  \BibitemOpen
  \bibfield  {author} {\bibinfo {author} {\bibfnamefont {E.}~\bibnamefont
  {Pollak}},\ }\bibfield  {journal} {\bibinfo  {journal} {The Journal of
  Chemical Physics}\ }\textbf {\bibinfo {volume} {127}},\ \href
  {https://doi.org/10.1063/1.2753151} {10.1063/1.2753151} (\bibinfo {year}
  {2007})\BibitemShut {NoStop}%
\bibitem [{\citenamefont {Koch}\ \emph {et~al.}(2008)\citenamefont {Koch},
  \citenamefont {Grossmann}, \citenamefont {Stockburger},\ and\ \citenamefont
  {Ankerhold}}]{Koch2008}%
  \BibitemOpen
  \bibfield  {author} {\bibinfo {author} {\bibfnamefont {W.}~\bibnamefont
  {Koch}}, \bibinfo {author} {\bibfnamefont {F.}~\bibnamefont {Grossmann}},
  \bibinfo {author} {\bibfnamefont {J.~T.}\ \bibnamefont {Stockburger}},\ and\
  \bibinfo {author} {\bibfnamefont {J.}~\bibnamefont {Ankerhold}},\ }\href
  {https://doi.org/10.1103/PhysRevLett.100.230402} {\bibfield  {journal}
  {\bibinfo  {journal} {Physical Review Letters}\ }\textbf {\bibinfo {volume}
  {100}},\ \bibinfo {pages} {230402} (\bibinfo {year} {2008})}\BibitemShut
  {NoStop}%
\bibitem [{\citenamefont {Lindoy}\ \emph {et~al.}(2022)\citenamefont {Lindoy},
  \citenamefont {Mandal},\ and\ \citenamefont {Reichman}}]{Lindoy2022}%
  \BibitemOpen
  \bibfield  {author} {\bibinfo {author} {\bibfnamefont {L.~P.}\ \bibnamefont
  {Lindoy}}, \bibinfo {author} {\bibfnamefont {A.}~\bibnamefont {Mandal}},\
  and\ \bibinfo {author} {\bibfnamefont {D.~R.}\ \bibnamefont {Reichman}},\
  }\href {https://doi.org/10.1021/ACS.JPCLETT.2C01521} {\bibfield  {journal}
  {\bibinfo  {journal} {The Journal of Physical Chemistry Letters}\ }\textbf
  {\bibinfo {volume} {13}},\ \bibinfo {pages} {6580} (\bibinfo {year}
  {2022})}\BibitemShut {NoStop}%
\bibitem [{\citenamefont {Sun}\ and\ \citenamefont {Vendrell}(2022)}]{Sun2022}%
  \BibitemOpen
  \bibfield  {author} {\bibinfo {author} {\bibfnamefont {J.}~\bibnamefont
  {Sun}}\ and\ \bibinfo {author} {\bibfnamefont {O.}~\bibnamefont {Vendrell}},\
  }\href {https://doi.org/10.1021/acs.jpclett.2c00974} {\bibfield  {journal}
  {\bibinfo  {journal} {The Journal of Physical Chemistry Letters}\ }\textbf
  {\bibinfo {volume} {13}},\ \bibinfo {pages} {4441} (\bibinfo {year}
  {2022})}\BibitemShut {NoStop}%
\bibitem [{\citenamefont {Child}(2014)}]{child2014}%
  \BibitemOpen
  \bibfield  {author} {\bibinfo {author} {\bibfnamefont {M.}~\bibnamefont
  {Child}},\ }\href@noop {} {\emph {\bibinfo {title} {Semiclassical Mechanics
  with Molecular Applications}}}\ (\bibinfo  {publisher} {Oxford University
  Press},\ \bibinfo {year} {2014})\BibitemShut {NoStop}%
\bibitem [{\citenamefont {Thoss}\ \emph {et~al.}(2001)\citenamefont {Thoss},
  \citenamefont {Wang},\ and\ \citenamefont {Miller}}]{Thoss2001a}%
  \BibitemOpen
  \bibfield  {author} {\bibinfo {author} {\bibfnamefont {M.}~\bibnamefont
  {Thoss}}, \bibinfo {author} {\bibfnamefont {H.}~\bibnamefont {Wang}},\ and\
  \bibinfo {author} {\bibfnamefont {W.~H.}\ \bibnamefont {Miller}},\ }\href
  {https://doi.org/10.1063/1.1359242} {\bibfield  {journal} {\bibinfo
  {journal} {J. Chem. Phys.}\ }\textbf {\bibinfo {volume} {114}},\ \bibinfo
  {pages} {9220} (\bibinfo {year} {2001})}\BibitemShut {NoStop}%
\bibitem [{Note3()}]{Note3}%
  \BibitemOpen
  \bibinfo {note} {Refer to Eq.~A2 in Ref.\protect \citenum
  {Antipov2015}.}\BibitemShut {Stop}%
\bibitem [{\citenamefont {Herman}(1986)}]{Herman1986}%
  \BibitemOpen
  \bibfield  {author} {\bibinfo {author} {\bibfnamefont {M.~F.}\ \bibnamefont
  {Herman}},\ }\href {https://doi.org/10.1063/1.451150} {\bibfield  {journal}
  {\bibinfo  {journal} {J. Chem. Phys.}\ }\textbf {\bibinfo {volume} {85}},\
  \bibinfo {pages} {2069} (\bibinfo {year} {1986})}\BibitemShut {NoStop}%
\bibitem [{\citenamefont {Garashchuk}\ \emph {et~al.}(1997)\citenamefont
  {Garashchuk}, \citenamefont {Grossmann},\ and\ \citenamefont
  {Tannor}}]{Garashchuk1997}%
  \BibitemOpen
  \bibfield  {author} {\bibinfo {author} {\bibfnamefont {S.}~\bibnamefont
  {Garashchuk}}, \bibinfo {author} {\bibfnamefont {F.}~\bibnamefont
  {Grossmann}},\ and\ \bibinfo {author} {\bibfnamefont {D.}~\bibnamefont
  {Tannor}},\ }\href {https://doi.org/10.1039/A607595I} {\bibfield  {journal}
  {\bibinfo  {journal} {J. Chem. Soc. Faraday Trans.}\ }\textbf {\bibinfo
  {volume} {93}},\ \bibinfo {pages} {781} (\bibinfo {year} {1997})}\BibitemShut
  {NoStop}%
\bibitem [{\citenamefont {Harabati}\ \emph {et~al.}(2004)\citenamefont
  {Harabati}, \citenamefont {Rost},\ and\ \citenamefont
  {Grossmann}}]{Harabati2004}%
  \BibitemOpen
  \bibfield  {author} {\bibinfo {author} {\bibfnamefont {C.}~\bibnamefont
  {Harabati}}, \bibinfo {author} {\bibfnamefont {J.~M.}\ \bibnamefont {Rost}},\
  and\ \bibinfo {author} {\bibfnamefont {F.}~\bibnamefont {Grossmann}},\ }\href
  {https://doi.org/10.1063/1.1630033} {\bibfield  {journal} {\bibinfo
  {journal} {J. Chem. Phys.}\ }\textbf {\bibinfo {volume} {120}},\ \bibinfo
  {pages} {26} (\bibinfo {year} {2004})}\BibitemShut {NoStop}%
\bibitem [{\citenamefont {Tatchen}\ \emph {et~al.}(2011)\citenamefont
  {Tatchen}, \citenamefont {Pollak}, \citenamefont {Tao},\ and\ \citenamefont
  {Miller}}]{Tatchen2011}%
  \BibitemOpen
  \bibfield  {author} {\bibinfo {author} {\bibfnamefont {J.}~\bibnamefont
  {Tatchen}}, \bibinfo {author} {\bibfnamefont {E.}~\bibnamefont {Pollak}},
  \bibinfo {author} {\bibfnamefont {G.}~\bibnamefont {Tao}},\ and\ \bibinfo
  {author} {\bibfnamefont {W.~H.}\ \bibnamefont {Miller}},\ }\href
  {https://doi.org/10.1063/1.3573566} {\bibfield  {journal} {\bibinfo
  {journal} {J. Chem. Phys.}\ }\textbf {\bibinfo {volume} {134}},\ \bibinfo
  {pages} {134104} (\bibinfo {year} {2011})}\BibitemShut {NoStop}%
\end{thebibliography}%

\end{document}